\documentclass{article}

\usepackage{cite,latexsym,amsmath,amssymb,tabularx,supertabular,bm}

\usepackage{graphicx}
\usepackage{color}

\usepackage[letterpaper]{geometry}
\newcommand{\tr}{{\rm tr}}               

\newcommand{\y}{Y}
\newcommand{\vrr}{{\bm r}}
\newcommand{\vdl}{{\bm r}_1}
\newcommand{\vdr}{{\bm r}_2}
\newcommand{\rr}{r}
\newcommand{\dl}{r_1}
\newcommand{\dr}{r_2}

\newcommand{\minus}{-}
\newcommand{\plus}{+}

\def\lsi{\raise0.3ex\hbox{$<$\kern-0.75em\raise-1.1ex\hbox{$\sim$}}}
\def\gsi{\raise0.3ex\hbox{$>$\kern-0.75em\raise-1.1ex\hbox{$\sim$}}}
\newcommand{\lsim}{\mathop{\lsi}}

\def\BLM{\hbox{\tiny BLM}}

\def\MSbar{\hbox{\tiny ${\overline{\rm MS}}$}}
\DeclareMathOperator{\F}{F}
\newcommand{\qFp}[5]{\sideset{_{#1}}{_{#2}}\F\left(\left.\genfrac{}{}{0pt}{}{#3}{#4}\right\vert #5\right)}

\usepackage{fancyheadings,calc,ifthen}

\newcounter{hours}\newcounter{minutes}
\newcommand{\printtime}{%
  \setcounter{hours}{\time/60}%
  \setcounter{minutes}{\time-\value{hours}*60}%
  \ifthenelse{\value{hours}<10}{0}{}\thehours:%
  \ifthenelse{\value{minutes}<10}{0}{}\theminutes}

\begin{document}

\hfill{\small\sf CERN-PH-TH/2006-170} \vspace{-.1cm}

\hfill{\small\sf Cavendish-HEP-06/21} \vspace{1cm}

\noindent{\Large\bf Running coupling and power corrections\\[0.15cm]
in nonlinear evolution at the high--energy limit}

  \vspace{0.5cm}
{\large
  { Einan Gardi$^{1,2}$,  Janne Kuokkanen$^3$,
          Kari Rummukainen$^{3,4}$, and Heribert Weigert$^5$}
}
          \\\vspace{0.2cm}\\
{\small
{$^1$ Cavendish Laboratory, University of Cambridge, J J
Thomson Avenue, Cambridge CB3 0HE, UK}\\
{$^2$ Department of Applied Mathematics \& Theoretical Physics,
Wilberforce Road, Cambridge CB3 0WA,~UK }\\
{$^3$ Department of Physical Sciences, University of Oulu, P.O. Box 3000, FI-90014 Oulu,
  Finland} \\
{$^4$ Department of Physics, Theory Division, CERN, CH-1211 Geneva, Switzerland}\\
{$^5$ Department of Physics, The Ohio State University, Columbus, OH 43210, USA} \\[5mm]
}

\vspace{.2cm}
\noindent\begin{center}
\begin{minipage}{.92\textwidth}
  {\sf A main feature of high-energy scattering in QCD is saturation in
    the number density of gluons. This phenomenon is described by non-linear
    evolution equations, JIMWLK and BK, which have been derived at leading
    logarithmic accuracy. In this paper we generalize this framework to
    include running coupling corrections to the evolution kernel. We develop a
    dispersive representation of the dressed gluon propagator in the
    background of Weisz\"acker Williams fields and use it to compute ${\cal
      O}(\beta_0^{n-1}\alpha_s^n)$ corrections to the kernel to all orders in
    perturbation theory. The resummed kernels present infrared-renormalon
    ambiguities, which are indicative of the form and importance of
    non-perturbative power corrections.  We investigate numerically the effect
    of the newly computed perturbative corrections as well as the power
    corrections on the evolution and find that at present energies
    they are both significant.
  }
\end{minipage}
\end{center}
\vspace{1cm}

\section{Introduction}
\label{sec:introduction}

Modern hadron collider experiments such as HERA, RHIC and especially
the forthcoming LHC operate at high enough energies to observe new
phenomenon associated with high gluon density. The principal
characteristics of high--energy QCD scattering are the following:
firstly, owing to Lorentz contraction, the configurations probed
appear to be \emph{frozen in time} compared to the natural time
scales of the interaction. Secondly, the number density of soft
gluons gets \emph{saturated} at densities of the order of $1/g^2$.
These features, which are usually referred to as the Color Glass
Condensate (CGC), are a consequence of the non-Abelian nature of the
interaction and of the fact that gluons are massless. These features
 are therefore unique to QCD.

At sufficiently high energy the dominant interaction between the
projectile and the target can be described by ensembles of
boost--enhanced field configurations, the ``frozen'' modes mentioned
above. This has been extensively explored in the context of the
McLerran--Venugopalan model~\cite{McLerran:1994ni, McLerran:1994ka,
McLerran:1994vd,
  Kovchegov:1996ty, Kovchegov:1997pc, Jalilian-Marian:1997xn, Lam:1999wu,
  Lam:2000nz, Lam:2001ax}. This description is tailored for
  asymmetric situations in which the field of, say, the
  target can be argued to be much
stronger than that of the projectile. High--energy scattering of a
virtual photon on a large nucleus is the prototype example of this
situation. Nonetheless, at sufficiently high energy
this generic picture is applicable to nucleus--nucleus scattering as well.

Energy dependence can be incorporated into this picture by taking
into account fluctuations that acquire properties of the previously
frozen modes as one increases the collision energy. The relevant
contributions are characterized by large logarithms $\ln(s)$ in the
total invariant energy $s$ in the collision. At low gluon densities,
or weak fields, the resummation of high--energy logarithms has been
formulated long ago as a \emph{linear evolution equation} for the
gluon distribution function, the BFKL equation~\cite{Bal-Lip, Lipat,
Kuraev:1976ge, Kuraev:1977fs, Lipat-2}. However, at high densities
the resummation of these logarithms leads instead to
\emph{non-linear evolution equations} for gauge field correlators.
These can be formulated as a functional evolution equation known as
the JIMWLK equation~\cite{Jalilian-Marian:1997xn,
Jalilian-Marian:1997jx,
  Jalilian-Marian:1997gr, Jalilian-Marian:1997dw, Jalilian-Marian:1998cb,
  Kovner:2000pt, Weigert:2000gi, Iancu:2000hn,Ferreiro:2001qy}, or
equivalently, as an infinite coupled hierarchy of evolution
equations for field correlators known as the Balitsky
hierarchy~\cite{Balitsky:1996ub,
  Balitsky:1997mk, Balitsky:1998ya}. A truncation of this hierarchy that
retains most of its essential properties is known as the BK
equation~\cite{Balitsky:1996ub, Balitsky:1997mk,
  Balitsky:1998ya,Kovchegov:1999yj, Kovchegov:1999ua}. This equation
  describes high--gluon--density dynamics in terms of dipole degrees of
  freedom. In Refs.~\cite{Kovchegov:1999yj, Kovchegov:1999ua} the BK equation has
been derived from Mueller's dipole
model~\cite{Mueller:1994rr,Mueller:1994jq,Mueller:1995gb,Chen:1995pa},
using nuclear enhancement as a tool to trace the dominant field
configurations. This extends the ideas of the dipole model beyond
the mere onset of saturation effects~\cite{Mueller:1996te}.

The most prominent feature of the solutions of these non-linear
evolution equations is the emergence of an energy--dependent
transverse correlation length $R_s$, or saturation scale $Q_s\sim
1/R_s$, which, asymptotically, encodes all the energy dependence of
the cross section. The saturation scale characterizes the transverse
momentum scales of radiated gluons that contribute to the evolution
at any given energy. Modes much softer than $Q_s$ decouple: the
number densities of soft gluons are saturated, so they remain
constant as the energy increases. Independently of the initial
condition, at sufficiently high energies the saturation scale
increases rapidly with the energy. Therefore, $Q_s$ can be
considered a hard scale: $Q_s\gg \Lambda$.

The possibility to describe saturation by perturbative evolution
equations is a highly non-trivial result, since the gauge field
involved is necessarily strong. The evolution equation is derived
perturbatively by expanding in small fluctuations on a strong
background field. This is justified a posteriori: having found that
soft modes do not contribute to the evolution, the equation is
perturbatively consistent. This kind of infrared stability is a
direct consequence of gluon saturation. It is therefore not shared
by the linear BFKL equation, which is instead afflicted by diffusion
into the infrared. As was beautifully illustrated in
Ref.~\cite{Golec-Biernat:2001if} it is the non-linearity of the
JIMWLK and BK equations that makes them infrared stable.
  The presence of the nonlinearities and hence $Q_s$ will also modify the
  r\^ole and influence of power corrections compared to the BFKL case
  discussed in Ref.~\cite{Anderson:1997bw}.

Despite these strengths, JIMWLK and BK evolution suffer from a
serious shortcoming: they are derived only at leading logarithmic
accuracy, i.e. at fixed coupling.  To partially compensate for this,
all recent studies of the evolution have included running--coupling
effects in some more or less ad hoc manner. There are several
reasons why running--coupling effects are essential:
\begin{itemize}
\item{} Running--coupling effects are known~\cite{Ciafaloni:1998gs,
    Fadin:1998py, Kovchegov:1998ae, Brodsky:1998kn, Ball:2005mj,
    Altarelli:1999vw, Ciafaloni:1999yw, Ciafaloni:2003rd, Thorne:2001nr} to
  provide a large part of the next--to--leading--order (NLO) corrections to
  the evolution in the low density limit, where the description matches onto
  the BFKL equation.
\item{} On the purely phenomenological side, for example in fits to HERA
  data~\cite{Iancu:2003ge}, running coupling (or more precisely a dependence
  of the coupling on a scale involved in a single emission step, see below) is
  essential to slow down evolution by reducing gluon emission from small
  objects.
\item{} Conceptually, it is understood that the evolution is dominated by
  scales of the order of $Q_s$. The non-linearity of the equation ensures that
  dipoles much larger than $1/Q_s$ are inherently suppressed through the
  evolution (the saturation mechanism). However, with strictly fixed coupling,
  dipoles \emph{much smaller} than $1/Q_s$ still contribute to the evolution.
  As soon as the coupling depends on the size of the emitting dipole such
  contributions are also suppressed through the
  evolution~\cite{Rummukainen:2003ns}.
\end{itemize}
Despite both the practical and conceptual importance of running--coupling
effects in the nonlinear evolution equations, there has been no derivation of
how they enter. All simulations done so far involved ad hoc prescriptions
for the scale of the coupling,
based on nothing more than educated guesswork.

In this paper we approach the problem on the more fundamental level.
We show that the JIMWLK and BK equations can indeed be derived
beyond the fixed coupling level. We find that the equations take a
similar form to the fixed coupling case, while their kernel changes
in a rather drastic way. For example, it does not naturally appear
as a single scale--dependent coupling times the LO
scale--invariant kernel. We explicitly compute running coupling
${\cal O}(\beta_0^{n-1}\alpha_s^n)$ corrections to the kernel to all
orders and resum them by means of Borel summation. We find that the
resummed kernel present infrared-renormalon ambiguities. These are
indicative of the form and importance of non-perturbative power
corrections.

In order to perform this calculation we develop a dispersive
representation of the dressed gluon propagator \emph{in the
background of Weisz\"acker--Williams fields}. This is a
generalization of the well--known dispersive representation of the
\emph{free} dressed gluon propagator, a technique that has been used
to compute running--coupling corrections and estimate power
corrections in a variety of applications, see
e.g.~\cite{Bigi:1994em,Smith:1994id,Beneke:1994qe,Ball:1995ni,Dokshitzer:1995qm,Grunberg:1998ix,Gardi:1999dq,Gardi:2000yh,Cacciari:2002xb,Gardi:2003iv}.

As in the BFKL case, the non-linear evolution equations are expected
to receive additional sub-leading corrections, which are not related
to the running of the coupling. In this paper we will not attempt to
include such corrections. Some steps in this direction on the level
of the BK equation have been taken in Ref.~\cite{Balitsky:2001mr},
or with an entirely different focus in Ref.~\cite{Gotsman:2004xb},
and can be combined with our treatment where desired.

The structure of the paper is as follows: in
Sec.~\ref{sec:jimwlk-equation} we give a short introduction to the
physics described by the JIMWLK equation, in order to establish the
key ideas and the notation that will be used in the rest of the
paper. For more detailed background we refer the reader to the
original literature~\cite{Jalilian-Marian:1997xn,
  Jalilian-Marian:1997jx, Jalilian-Marian:1997gr, Jalilian-Marian:1997dw,
  Jalilian-Marian:1998cb, Kovner:2000pt, Weigert:2000gi,
  Iancu:2000hn,Ferreiro:2001qy, Balitsky:1996ub, Balitsky:1997mk,
  Balitsky:1998ya,Kovchegov:1999yj, Kovchegov:1999ua} or review
articles~\cite{Iancu:2002xk, Iancu:2003xm, Weigert:2005us,
  Jalilian-Marian:2005jf}. In
Sec.~\ref{sec:evolution-equations} we briefly review the JIMWLK and
BK equations and recall those details of their derivation that are
needed in what follows. Sec.~\ref{sec:running-coupling-types} is
devoted to a discussion of running--coupling effects in JIMWLK and
BK evolution, contrasting what has been done previously with what we
want to achieve in this paper. This will be important also to
clarify the terminology used in the remainder of the paper.
Sec.~\ref{sec:deriv-jimwlk-running} extends the derivation of the
JIMWLK equation to the running--coupling case. It is divided into
four subsections: Sec.~\ref{sec:runn-coupl-disp} collects the tools
of the conventional dispersive technique for the calculation of
running--coupling corrections in the free field case.
Sec.~\ref{sec:dress-prop-backgr} generalizes these tools for use in
the presence of the Weizs\"acker-Williams background as needed in
the derivation of the JIMWLK equation. Sec.~\ref{sec:disp-runn-pres}
presents a re-derivation of the JIMWLK equation with a running
coupling. Next, in Sec.~\ref{sec:borel-rep-of-resummed-kernel} we
formulate the newly computed corrections to the kernel as an
all--order Borel sum. In Sec.~\ref{sec:perturbative-expansions} we
discuss the convergence of perturbation theory. In
Sec.~\ref{sec:power-corrections} we use the renormalon singularities
to determine the parametric form and the typical magnitude of power
corrections affecting the kernel. Finally, in
Sec.~\ref{sec:BK-numerics} we investigate numerically the effect of
running coupling and power corrections on the BK evolution as a
function of the saturation scale. In Sec.~\ref{sec:conclusions} we
summarize our conclusions.

\section{The physics of the JIMWLK equation}
\label{sec:jimwlk-equation}

The key points can be most easily understood in the context of deep
inelastic scattering (DIS) of leptons on protons or nuclei, where
$q$ and $p$ are the momenta of the virtual photon and the target
respectively. Here, two kinematic variables play a r\^ole: (1) the
deeply--spacelike momentum $q^2=-Q^2<0$ carried by the exchanged
photon. This scale defines the transverse resolution of the probe
and thereby the apparent size of the quarks and gluons encountered;
and (2) Bjorken $x:={Q^2}/(2 p\cdot q)$, which is inversely
proportional to the total energy $s=(p+q)^2$ in the collision:
$x\approx {Q^2}/{s}$. At high energy, the \emph{rapidity} $\y$ is
directly related to Bjorken $x$ via $\y=\ln(1/x)$. The rapidity is
the natural evolution variable since $\ln(1/x)\approx
\ln\left(s/Q^2\right)$ reflects the large hierarchy of scales in the
high--energy limit, which appears with increasing powers in
perturbation theory.

At large $Q^2$ with fixed $x$ there are well-established methods to treat such
a system based on the Operator Product Expansion (OPE), a short--distance
expansion in powers of $1/Q^2$. Since $Q^2$ also controls the apparent size of
the particles encountered, the OPE can be viewed as a small density expansion,
despite the fact that particle numbers, driven by large logarithms in $Q^2$,
increase in parallel with increasing resolution. As a consequence the
description at large $Q^2$ can be based entirely on single--particle
properties such as quark and gluon distribution functions: particle
correlations are not important. This restriction is a key ingredient of the
derivation of the Dokshitzer--Gribov--Lipatov--Altarelli--Parisi (DGLAP)
equations that describe the increase of particle numbers with $Q^2$ in this
domain.

Going to small $x$ at fixed $Q^2$, no matter how large, one ends up
in an entirely different domain, that of high densities, even if one
starts out in a dilute situation. As the energy increases BFKL
evolution keeps generating new particles (mostly gluons) which are
all of effective size of ${\cal O}(1/Q^2)$, and so the density keeps
increasing. Eventually  the density reaches a level where particle
correlations become
essential~\cite{Mueller:1994rr,Mueller:1994jq,Mueller:1995gb,Chen:1995pa,Mueller:1996te}
and a description in terms of distribution functions alone becomes
untenable.  Since the BFKL description is based on gluon
distributions, this is also the point where this evolution equation
ceases to be adequate. Appropriate degrees of freedom and more
general evolution equations are needed to describe the system beyond
this point. The most general of these existing to date are the
JIMWLK equation, or, the completely equivalent Balitsky hierarchies,
with their factorized truncation, the BK equation.

JIMWLK and BK equations are formulated in terms of path--ordered
exponentials, as defined in Eq.~(\ref{eq:U-via-b}) below, with paths
collinear to the projectile direction, which can be interpreted as
quark and gluon constituents of the projectile. The path--ordered
exponentials encode the fact that, owing to the high energy in the
collision, these constituents penetrate the target without being
deflected from their straight--line trajectories. The $ \gamma^* A$
cross section then reads
\begin{equation}
  \label{eq:dipole-cross}
  \sigma_{\mathrm{DIS}}(\y,Q^2) = \text{\cal Im}\
  \begin{minipage}[c]{3cm}
    \includegraphics[height=.8cm]{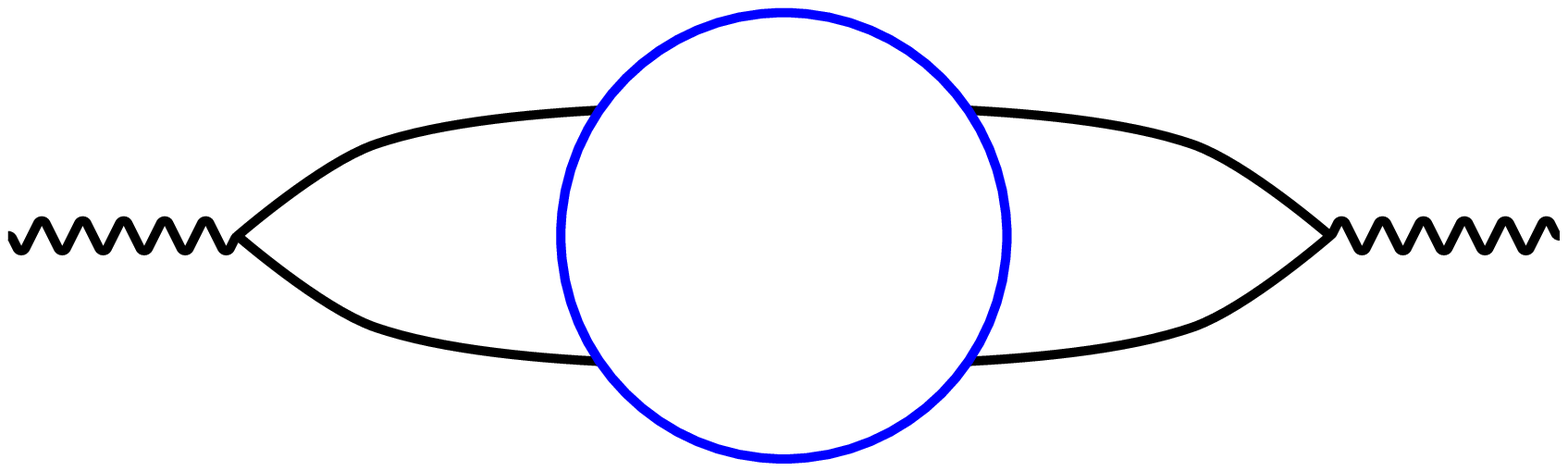}
  \end{minipage}
  =\int\!\!d^2 r
  \, \vert \psi\vert^2(r^2 Q^2) \hspace{.1cm}
  \int d^2b \ \left\langle
  \frac{\tr(1-U_{\bm{x}} U^\dagger_{\bm{y}})}{N_c}
  \right\rangle(\y)
\end{equation}
where $\bm{r}=\bm{x}-\bm{y}$ corresponds to the transverse size of a
given $q\Bar q$ dipole and $\bm{b}=(\bm{x}+\bm{y})/2$ to the impact
parameter of this dipole relative to the target. $q^2:=-Q^2$ is the
large spacelike momentum carried by the virtual photon. The square
of the $q\Bar q$ component of the photon wave function $\vert
\psi\vert^2(r^2 Q^2)$ describes the probability to find a $q\Bar q$
pair of size $r$ inside the virtual photon and can be calculated in
QED. It consists of a known combination of Bessel functions together
with an integral over longitudinal momentum fractions already
absorbed in the notation. The remaining factor, the expectation
value of $U$-operators is usually called the dipole cross section
$\sigma_{\text{dipole}}(\y,\bm{r})$ of the target in question. All
the properties of this interaction
--- details of the target wave function, gluon exchange between
the target and projectile etc. --- are encoded in this expectation
value. The leading--logarithmic corrections at small $x$ appear in
powers of $\alpha_s \ln(1/x)$. These corrections are resummed by the
JIMWLK equation.

The dipole operator $\Hat N_{\bm{x y}} := {\tr(1-U_{\bm{x}}
U^\dagger_{\bm{y}})}/{N_c}$ itself is naturally bounded between zero
and one. Typically, gluon densities grow towards large $r$, such
that the expectation value of $\Hat N_{\bm{x y}}$ interpolates
between $0$ for infinitesimally small dipoles, and $1$ for very
large ones. This encodes the idea of color transparency at short
distances and saturation at large distances where gluon densities
grow up to ${\cal O}(1/g^2)$. The length scale that characterizes
the transition between the two domains can be interpreted as the
correlation length $R_s$ of $U$-operators, or equivalently gluon
fields. The corresponding momentum scale $Q_s\propto 1/R_s$ is
usually called the saturation scale. Clearly, as more gluons are
generated in JIMWLK evolution towards small $x$, the correlation
length gets small and $Q_s$ increases.

One key feature that emerges for this evolution is that details about the
initial conditions are erased quickly and a universal scaling form of
correlators such as the dipole cross section is reached. From then on, all $x$
or $\y$ dependence is carried by the saturation scale $Q_s(\y)$. Such behavior
has been seen in HERA data (geometric scaling)~\cite{Golec-Biernat:1998js,
  Golec-Biernat:1999qd, Stasto:2000er} and has important consequences for RHIC
(disappearance of Cronin enhancement from mid to forward
rapidities)~\cite{Baier:2003hr, Kharzeev:2003wz,
  Jalilian-Marian:2003mf,Albacete:2003iq,Arsene:2004ux} and the
  LHC experiments where the energies are higher.

It had been noted early on in the context of the BK equation that a treatment
at the strictly leading--logarithmic level is insufficient: running--coupling
effects have a strong influence on the speed of evolution; quantities like the
evolution rate $\lambda(\y):=\partial_{\y} \ln Q_s^2(\y)$ are reduced by more
than $50\%$ if running--coupling effects are introduced (in some heuristic
way). Despite the explicit scale breaking introduced by the appearance of
$\Lambda_{\text{QCD}}$ in the running coupling, scaling of the dipole cross
section with $Q_s$ is retained to very good accuracy. In
Ref.~\cite{Rummukainen:2003ns} it was emphasized that running--coupling
effects are also conceptually important: only with running--coupling effects
included does the phase--space region active in the evolution center around
the physical scale $Q_s$. At strictly leading--logarithmic level --- in the
conformal limit for the evoluton kernel --- the evolution involves
short--distance\footnote{As explained above, the infrared is not a problem due
  to the presence of the correlation length $R_s$, which acts an an effective
  infrared cutoff.} contributions from more than 7 orders of magnitude away
from the physical scale.

Unlike the other evolution equations in QCD, such as DGLAP and BFKL, the
JIMWLK and BK equations have been derived only at leading logarithmic
accuracy. Only partial calculations of two--loop corrections to the BK equation
are available~\cite{Balitsky:2001mr} but they do not include any attempt to
determine the running of the coupling. For our purposes, the existing results
in the low--density limit for BFKL evolution are of limited use:
they offer no hint as for how to extend or extrapolate the
calculation into the high--density domain where non-linearities appear.
A direct calculation in the context of the JIMWLK and BK
equations is therefore necessary.

As announced in the introduction, in this paper we compute
running--coupling corrections to the JIMWLK and BK equations. In
Sec.~\ref{sec:evolution-equations} we briefly review the
fixed--coupling derivation of the JIMWLK equation, by considering
small fluctuations in a strong Weizs\"acker-Williams field that is
encoded in the eikonal factors of Eq.~\eqref{eq:dipole-cross}. This
derivation will then be generalized to the running--coupling case in
Sec.~\ref{sec:deriv-jimwlk-running} using a dispersive
representation of the dressed gluon propagator in such a background
field. This will not only enable us to calculate running--coupling
corrections, but also to explore non-perturbative effects in the
evolution.

\subsection{Evolution equations}
\label{sec:evolution-equations}

\subsubsection*{The JIMWLK equation and the Balitsky hierarchies}

The JIMWLK equation is a functional Fokker-Planck equation for the
statistical weight $Z_\y[U]$ defining the $\y$--dependent averaging
procedure $\langle\ldots\rangle(\y)$ that determines the expectation
value of operators $O[U]$ made of an arbitrary number of
path--ordered exponentials $U$, where
\begin{equation}
  \label{eq:U-via-b}
  U^{-1}_{\bm{x}} =
  P\exp\left\{
  ig \int\limits_{-\infty}^{\infty} dz^- \delta(z^-) \beta(\bm{x})
  \right\}
  \ .
\end{equation}
Such an average was already encountered above in Eq.~\eqref{eq:dipole-cross}
in the case of the dipole operator ${\tr(1-U_{\bm{x}}
  U^\dagger_{\bm{y}})}/{N_c}$.  To understand why these averages determine
virtually all cross sections at small~$x$, recall the origin of the
path--ordered exponentials (\ref{eq:U-via-b}): they encode the interaction of
fast moving quarks and gluons in the projectile wave function with the target
field. It is the high energy of the collision that allows the description of
this interaction in terms of the eikonal factors $U$, that at leading order
follow perfectly straight, lightlike worldlines. We have chosen a frame where
these trajectories extend along the minus light--cone direction at $x^+=0$. Each
particle is then characterized by the remaining coordinates, namely its
transverse location $\bm{x}$. The leading contribution comes from interaction
with the non-Abelian Weizs\"acker-Williams field of the target,~$A^+$. It
takes the form
\begin{equation}
  \label{eq:bgfield}
  A^+(x) =b^+(x^-,{\bm x}) +\delta A(x),
  \hspace{2cm}
  b^+(x^-,{\bm x}) = \delta(x^-)\beta(\bm{x})
\end{equation}
where $b^+$, or more specifically $\beta(\bm{x})$, is the single
leading degree-of-freedom, a strong field, while $\delta A$ is a
small fluctuation in which we will expand. The $\delta(x^-)$
reflects the lack of resolution in longitudinal direction: no
internal details of the field of the target are probed. The
independence on $x^+$ reflects the fact that the target wave
function is frozen during the interaction, an extreme time dilation.
At fixed rapidity all dominant contributions are determined by the
background field $b^+=\delta(x^-)\beta(\bm{x})$. Moreover, they only
enter in a very specific form, via the Wilson lines $U$.

For some generic operator made of these Wilson--line fields, $O[U]$,
the average of Eq.~\eqref{eq:dipole-cross} will be written as
\begin{equation}
  \label{eq:corrs}
  \left\langle O[U] \right\rangle_\y \,:=\, \int  \Hat{D}[U]\,  O[U]\, Z_\y[U]
\end{equation}
where $\Hat{D}[U]$ is a functional Haar-measure and $Z_\y[U]$
contains the detailed physics beyond the eikonal approximation
already incorporated by selecting $U$ as the relevant
degrees-of-freedom: $Z_\y[U]$ is the statistical weight for all
possible field configurations.

Most of our knowledge of $Z_\y[U]$ is perturbative: the JIMWLK
equation\footnote{For a first derivation of this equation in its
most compact form see Ref.~\cite{Weigert:2000gi}. The version
presented here is based on Ref.~\cite{Weigert:2005us}.} determines
the $\y$ dependence of this average:
\begin{equation}
  \label{eq:JIMWLK}
  \partial_\y \Hat Z_\y[U]\,=\,
  - {\cal H}[U]\,
  Z_\y[U],
\end{equation}
where the JIMWLK Hamiltonian ${\cal H}[U]$ is given by
\begin{align}
  \label{eq:JIMWLK-Hamiltonian-LO}
  {\cal H}[U] =-\frac{\alpha_s(\mu^2)}{2\pi^2}\, {\cal K}_{\bm{x z y}}\, \Big[
    U_{\bm z}^{a b}\left(i\Bar\nabla^a_{\bm x}i\nabla^b_{\bm y} +i\nabla^a_{\bm x}
    i\Bar\nabla^b_{\bm y}\right) + \left( i\nabla^a_{\bm x} i\nabla^a_{\bm
      y}+i\Bar\nabla^a_{\bm x} i\Bar\nabla^a_{\bm y}\right) \Big] \
      ,
\end{align}
with the LO kernel:
\begin{equation}
  \label{eq:JIMWLK-LO-kernel}
  {\cal K}_{\bm{x z y}} = \frac{(\bm{x}-\bm{z})\cdot(\bm{z}-\bm{y})}{%
    (\bm{x}-\bm{z})^2(\bm{z}-\bm{y})^2} \,=\,
     -\frac{ {\vdl}\cdot  {\vdr}}{\,{r_1}^2 \,{r_2}^2}\ ,
\end{equation}
where $\bm{x}$, $\bm{y}$ and $\bm{z}$ are transverse coordinates.
Here we also introduced a shorthand notation for the vectors
connecting the points in the transverse plain:
\begin{equation}
  \label{eq:distshort}
  {\vrr}=\bm{x}-\bm{y}\ ,\hspace{2cm}
  {\vdl}=\bm{x}-\bm{z}\ ,\hspace{2cm}
  {\vdr}=\bm{y}-\bm{z}\ ,
\end{equation}
and their lengths: $r_1:=|{\vdl}|$, etc. The notation in
(\ref{eq:JIMWLK-Hamiltonian-LO}) assumes an integration convention
over repeated coordinates appearing as an index in ${\cal K}_{\bm{x
z y}}$ and in the vector field operators $\nabla^a_{\bm x}$. The
Hamiltonian ${\cal H}[U]$~is~second order in left-- and
right--invariant vector fields $\nabla^a_{\bm x}$ and
$\Bar\nabla^a_{\bm x}$, which are Lie derivatives: they act on the
Wilson--line variables $U_{\bm x}$ according to\footnote{See
Ref.~\cite{Weigert:2000gi} for more details.}
\begin{equation}
\label{eq:nabla_acting} i\nabla^a_{\bm{x}} U_{\bm{y}} := -U_{\bm{x}}
t^a \delta^{(2)}_{\bm{x y}}\,; \quad \qquad i\Bar\nabla^a_{\bm{x}}
U_{\bm{y}} := t^a U_{\bm{x}} \delta^{(2)}_{\bm{x y}}.
\end{equation}
The terms in the square brackets on the r.h.s of
Eq.~(\ref{eq:JIMWLK-Hamiltonian-LO}) are grouped according to their
origin in real--emission and virtual corrections, respectively:
real--emission contributions involve an additional Wilson line
$U_{\bm z}^{a b}$ at transverse location $\bm z$.

The full derivation of Eq.~(\ref{eq:JIMWLK}) has been presented
exhaustively in Refs.~\cite{Weigert:2000gi,
  Balitsky:1996ub, Iancu:2000hn, Ferreiro:2001qy}. We nevertheless need to
  recall here how the JIMWLK Hamiltonian relates to Feynman diagrams, in
order to prepare its re-derivation with running--coupling
corrections. Rapidity dependence of $Z_\y[U]$ at LO is
driven by the lowest--order fluctuations $\delta A$ around the
background $\delta(x^-)\beta({\bm x})$ of Eq.~\eqref{eq:bgfield}. In
this sense, the LO JIMWLK Hamiltonian in
Eq.~\eqref{eq:JIMWLK-Hamiltonian-LO} is {\em constructed} such that
it adds the LO ``exchange'' and ``self-energy''
corrections to, say, an interacting $q\Bar q$ pair, represented by
its Wilson--line bilinear $U_{{\bm
    x}}\otimes U_{{\bm y}}^\dagger$:
\begin{align}
  \label{eq:JIMWLK-LO-diagram-cont}
  \ln(1/x)\,\, {\cal H}[U] \,\, U_{{\bm x}}\otimes U_{{\bm y}}^\dagger
  = &
  \parbox{2.7cm}{\includegraphics[width=2.7cm]{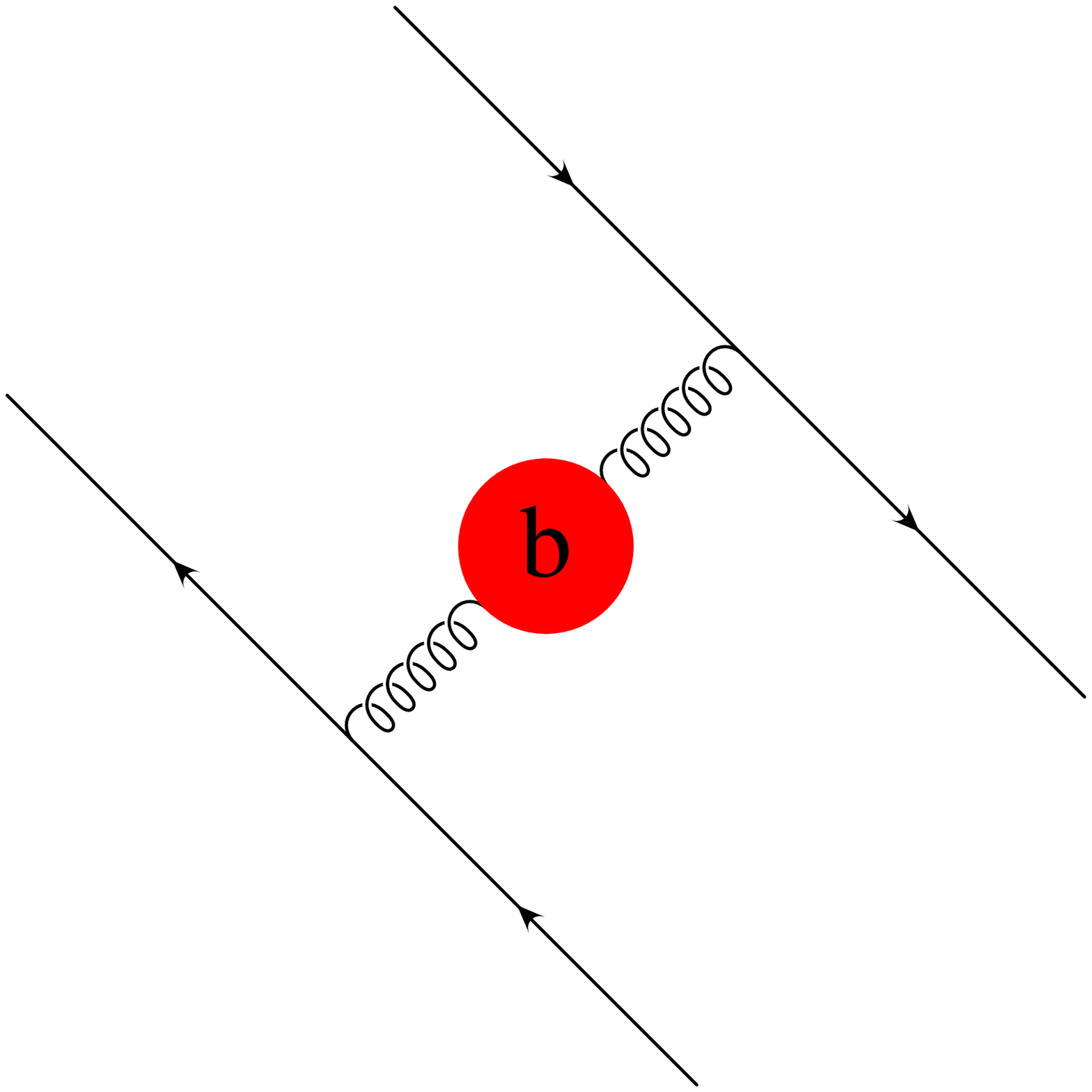}}
  +\parbox{2.7cm}{\includegraphics[width=2.7cm]{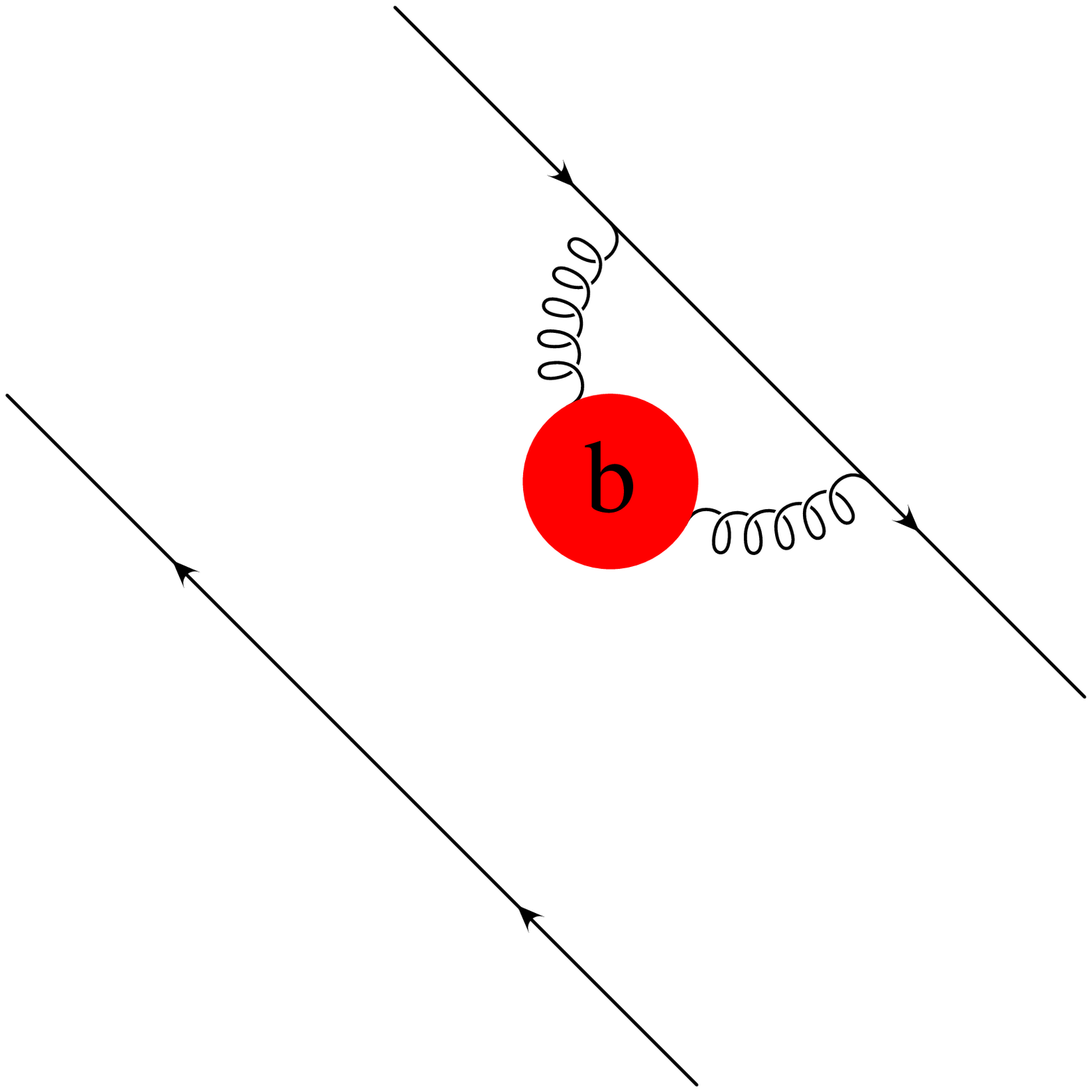}}
  +\parbox{2.7cm}{\includegraphics[width=2.7cm]{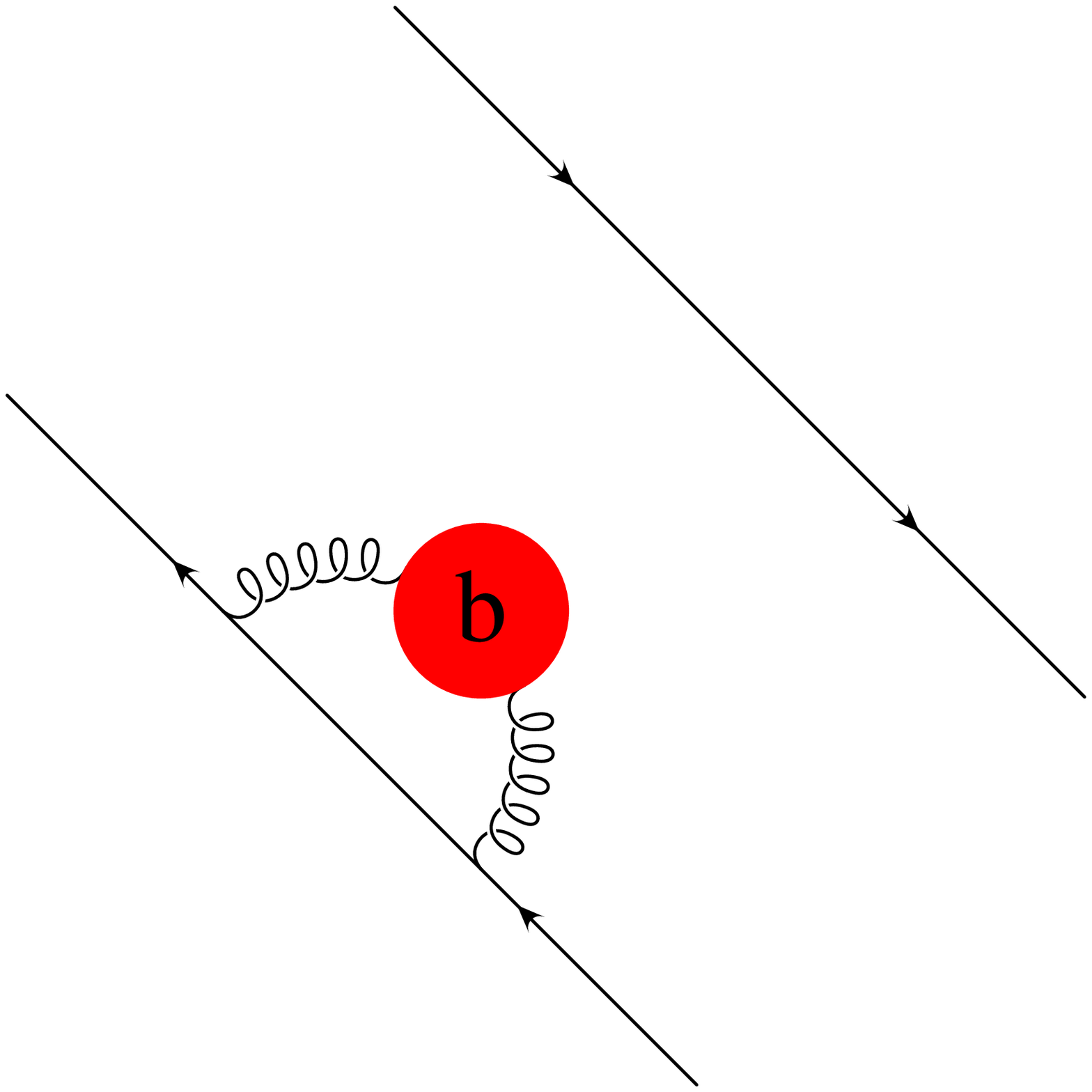}}
\end{align}
The diagrams shown in~\eqref{eq:JIMWLK-LO-diagram-cont} are Feynman
diagrams where the gluon propagator of the fluctuations
$\langle\delta A \delta A\rangle$ is taken in the background of the
strong target field $\delta(x^-)\beta({\bm x})$.

The correspondence to real and virtual diagrams becomes visible upon
resolving the Feynman diagrams into $x^-$ ordered diagrams of
light--cone perturbation theory. For instance,
\begin{align}
  \label{eq:interaction-diagrams}
  \parbox{2.7cm}{\includegraphics[width=2.7cm]{chiqqb}} = &
  \parbox{2.7cm}{\includegraphics[width=2.7cm]{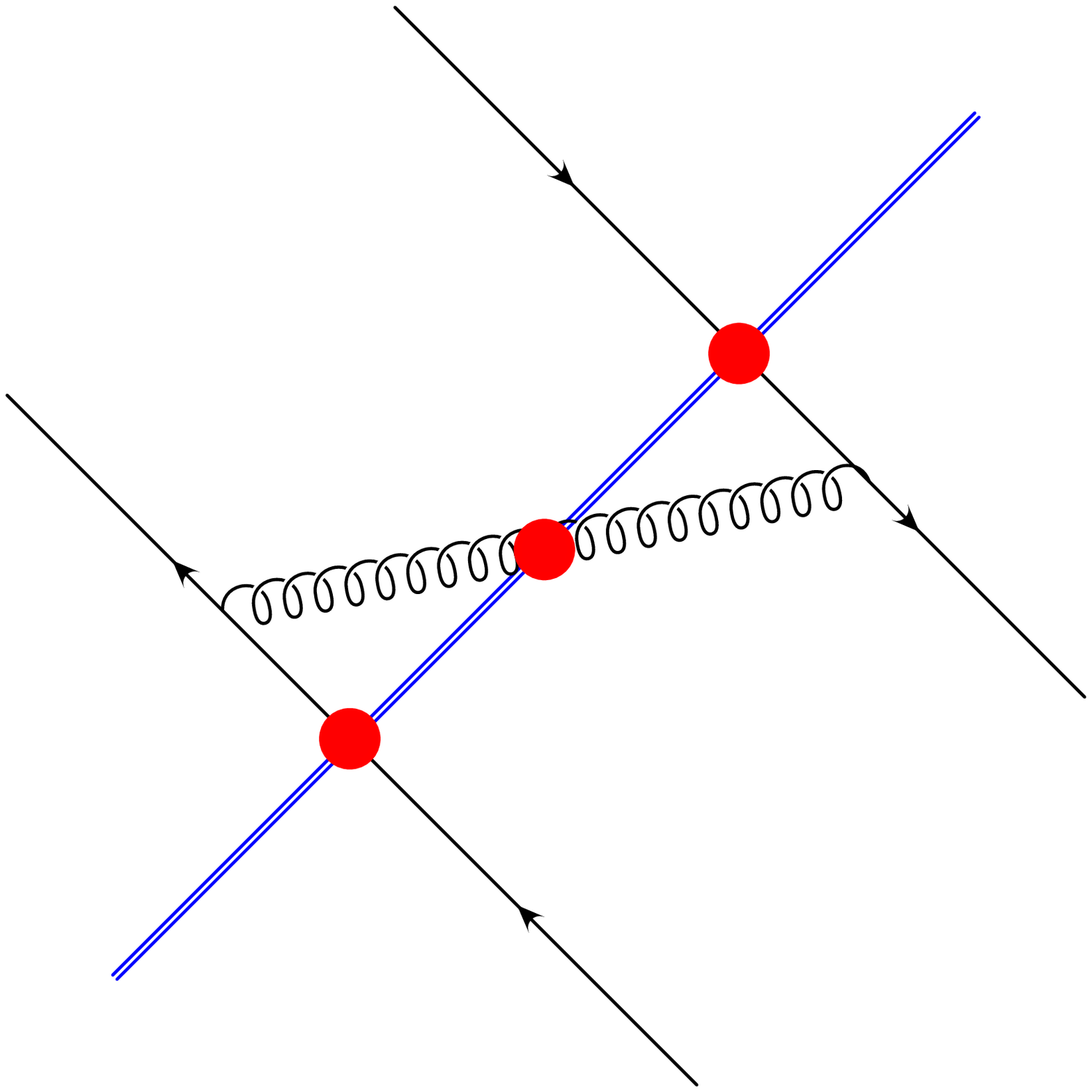}}
  +
  \parbox{2.7cm}{\includegraphics[width=2.7cm]{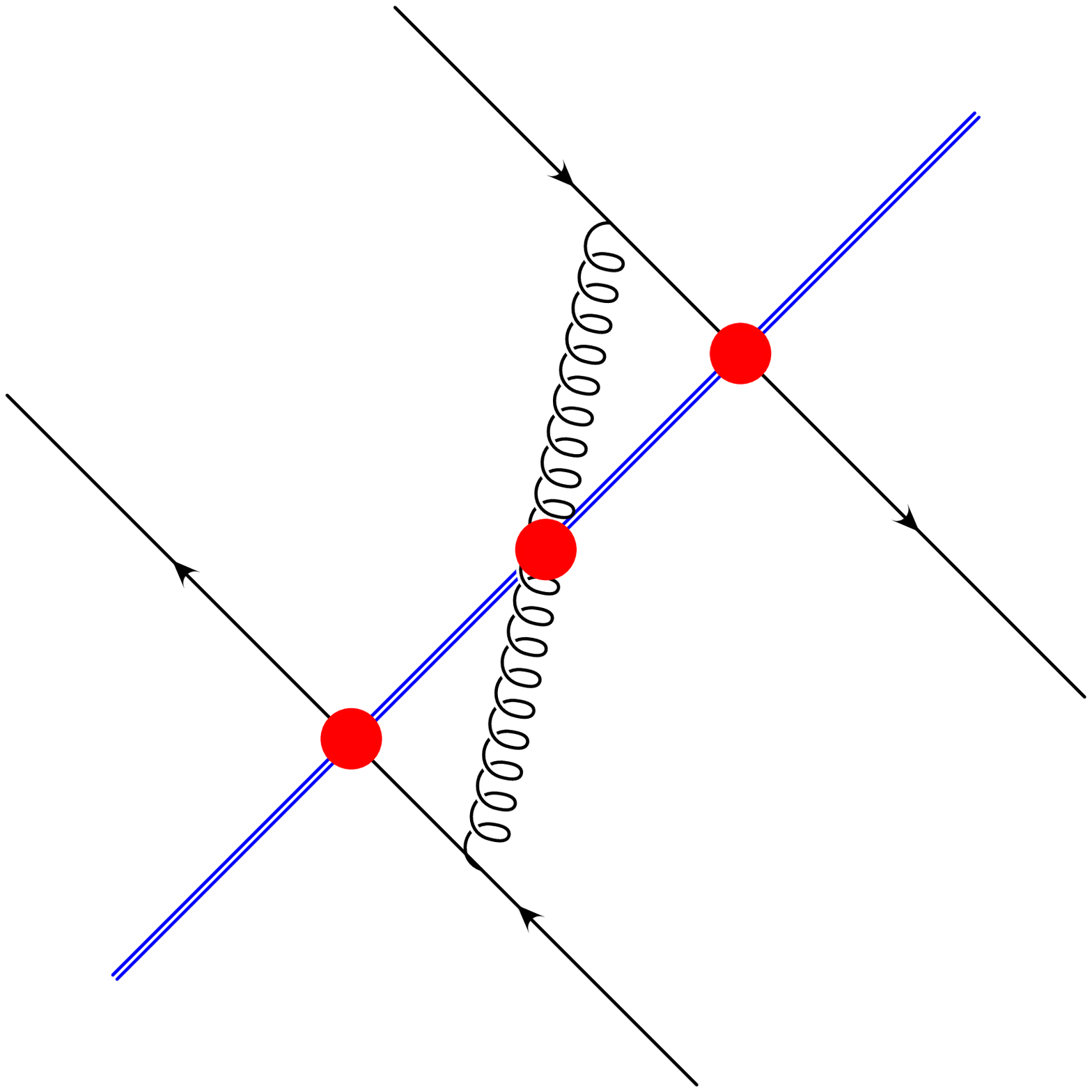}}
  +\parbox{2.7cm}{\includegraphics[width=2.7cm]{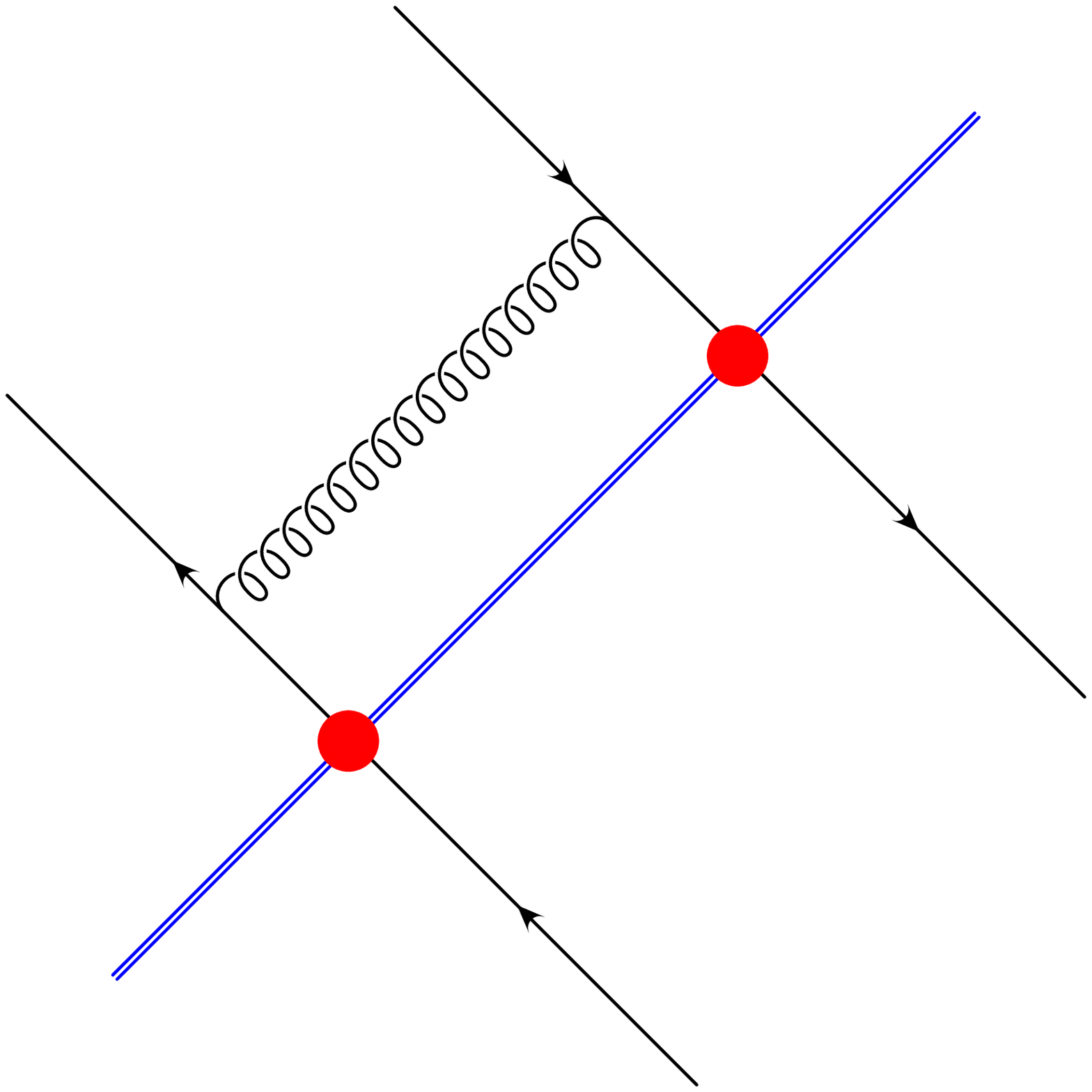}}
  +\parbox{2.7cm}{\includegraphics[width=2.7cm]{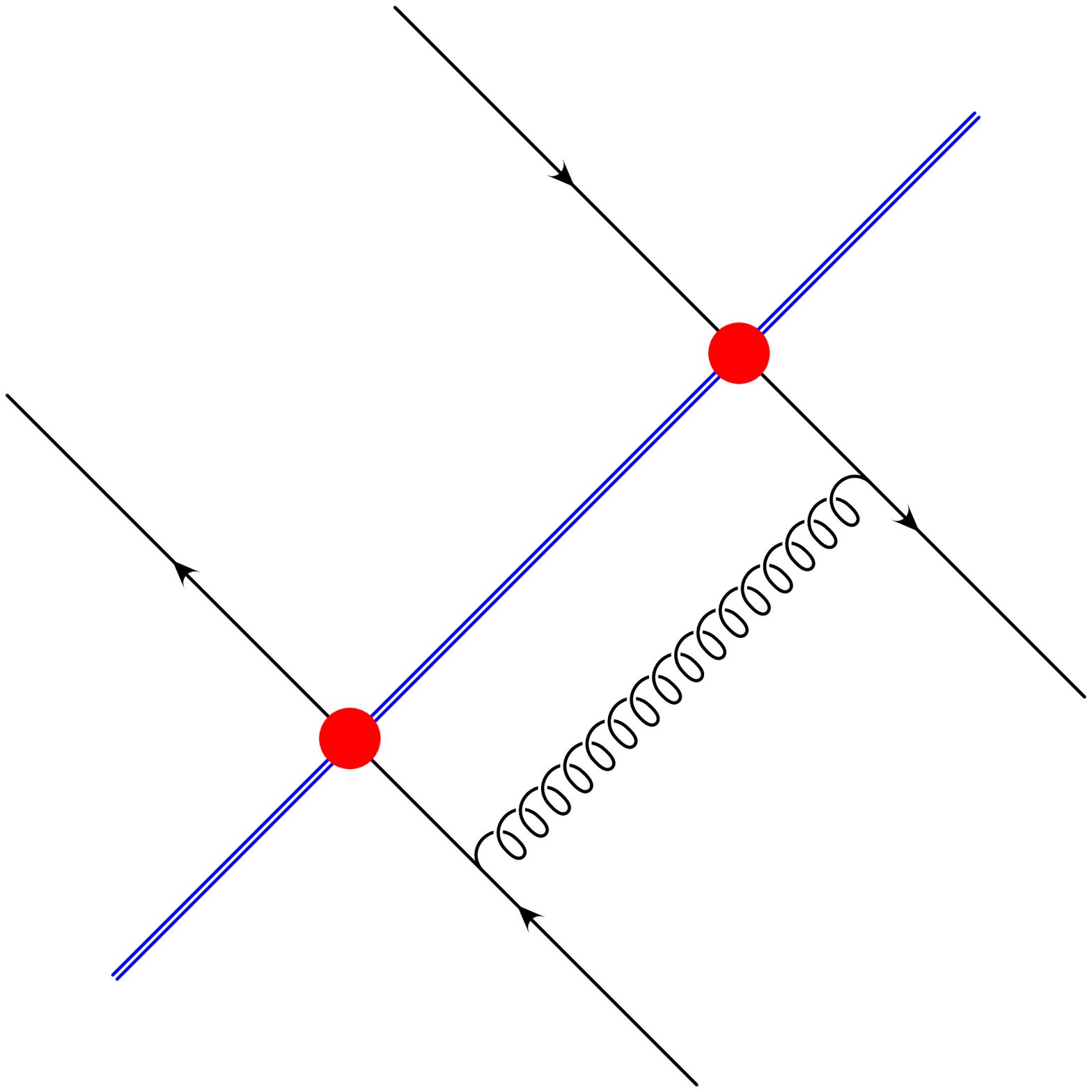}},
\end{align}
where the diagrams on the r.h.s. should be interpreted as diagrams
of light--cone perturbation theory. Light--cone time $x^-$ runs from
bottom right to top left, the two collinear\footnote{Note that these
two lines are actually separated only in the transverse direction.}
Wilson lines in this direction represent the dipole (the projectile)
and the target is shown as a perpendicular line at $x^-=0$, from
bottom left to top right. To be precise, the last two diagrams
in~(\ref{eq:interaction-diagrams}) are a shorthand notation for a
sum of two different $x^-$ orderings each:
\begin{equation}
  \label{eq:shorthand-diagrams}
  \parbox{2cm}{\includegraphics[width=2cm]{chiqqb-tup}}
  = \parbox{2cm}{\includegraphics[width=2cm]{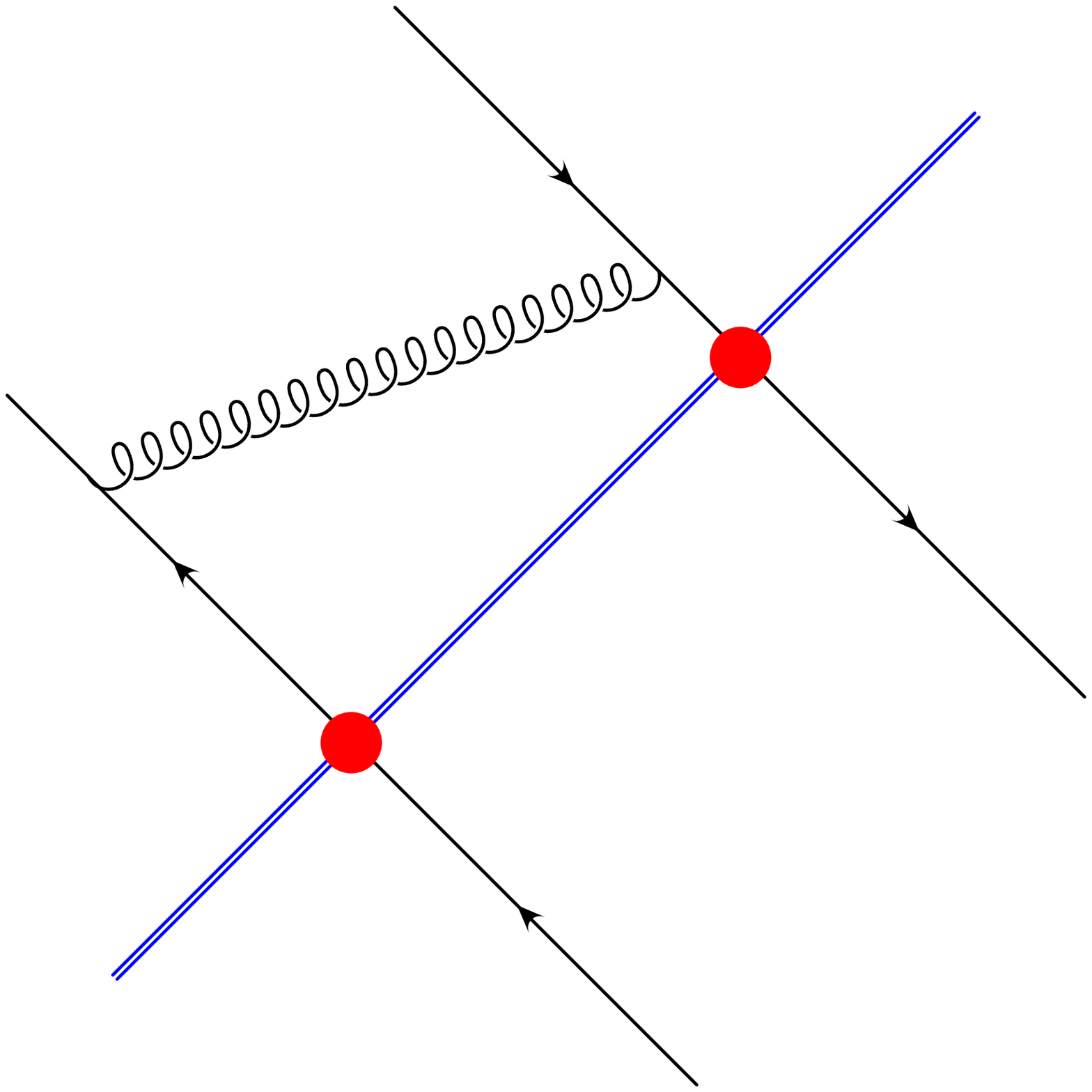}}
  +\parbox{2cm}{\includegraphics[width=2cm]{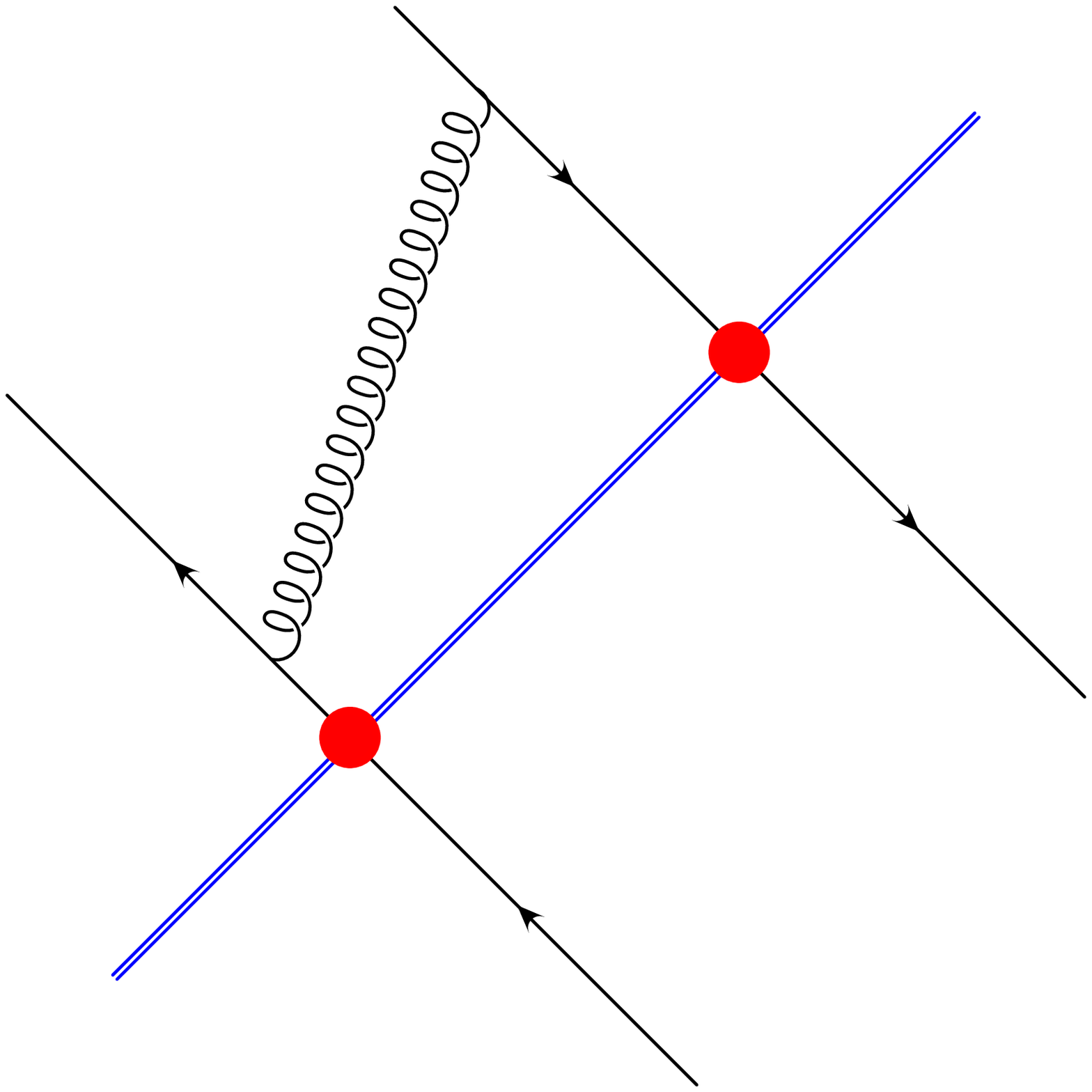}}
  \ ;\hspace{.5cm}
  \parbox{2cm}{\includegraphics[width=2cm]{chiqqb-tdown}}
  =
  \parbox{2cm}{\includegraphics[width=2cm]{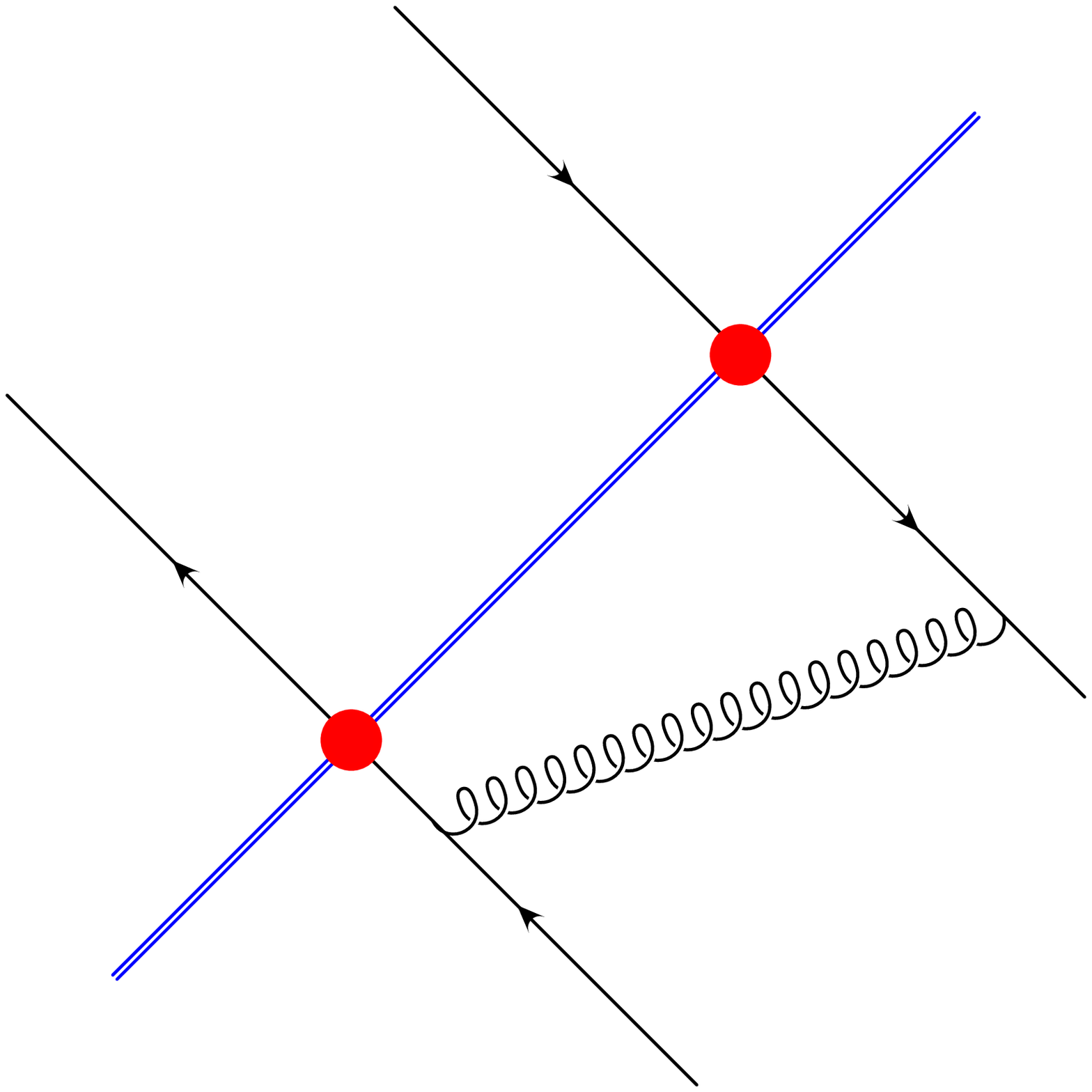}}
  +\parbox{2cm}{\includegraphics[width=2cm]{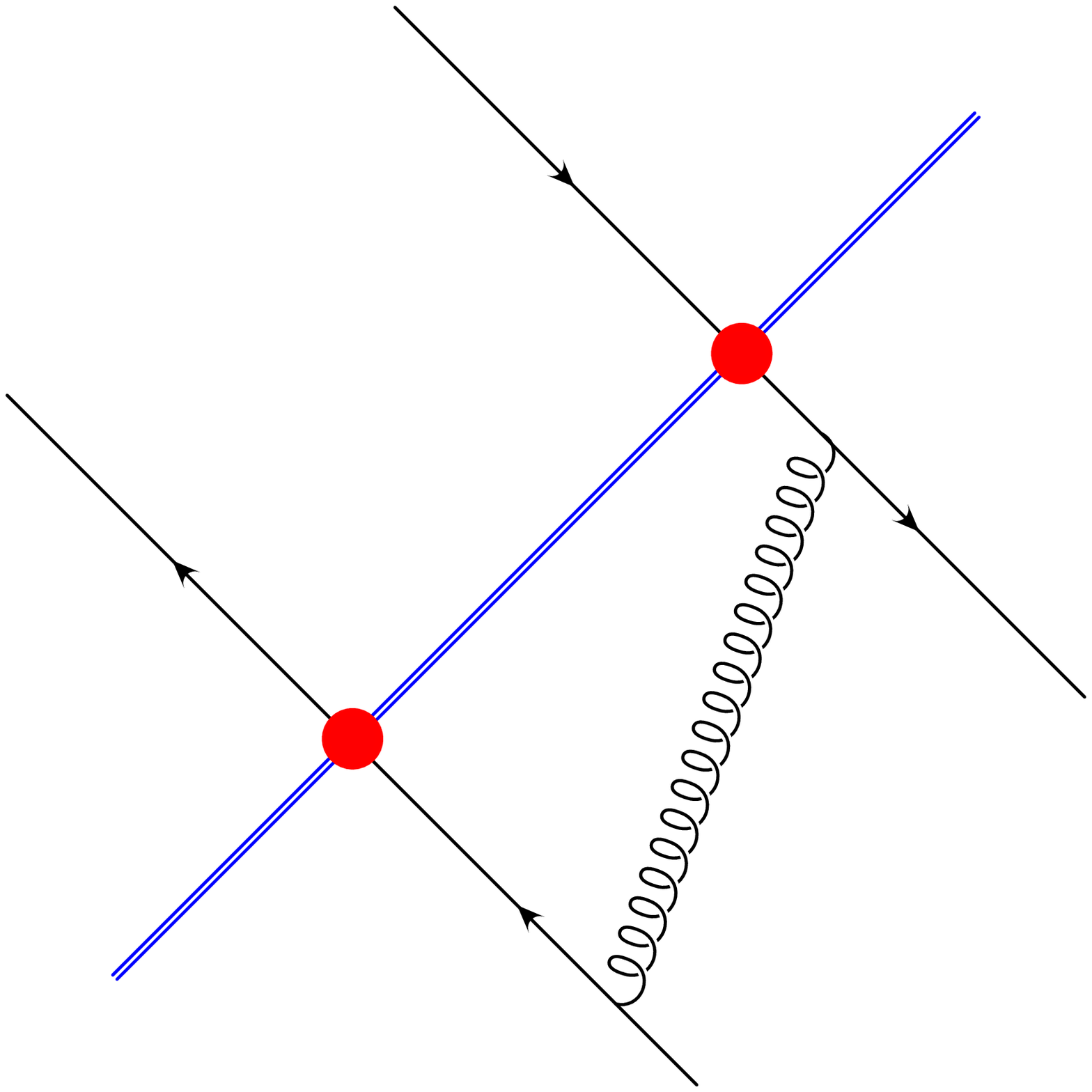}}.
\end{equation}
The factors of $U$, $U^\dagger$ and $U^{a b}$ representing the
interactions of a projectile quark, antiquark and gluon,
respectively, are indicated as large dots where these Wilson lines
cross the target line.  In a derivation of JIMWLK based on
projectile wave functions\footnote{See
Refs.~\cite{Mueller:1994rr,Mueller:1994jq,Mueller:1995gb,Chen:1995pa}
for a derivation of JIMWLK based on projectile wave functions in the
context of the dipole model,
Refs.~\cite{Kovchegov:1999yj,Kovchegov:1999ua} for the BK case, or
Ref.~\cite{Hentschinski:2005er} for a re-derivation of JIMWLK via
amplitudes.}, the contribution of the first two diagrams
of~\eqref{eq:interaction-diagrams} are associated with a situation
where the interacting gluon reaches the final state and in this
sense they correspond to {\em real--emission} diagrams.
Correspondingly, as indicated by the third dot, they contain an
additional Wilson line for this produced, eikonally interacting
gluon, which appears in (\ref{eq:JIMWLK-Hamiltonian-LO}) as $U_{\bm
z}^{a b}$. The remaining diagrams represent purely {\em virtual}
contributions in which the number of Wilson lines does not change.
Similarly, both real--emission and virtual corrections are present
in the self--energy--like diagrams in
(\ref{eq:JIMWLK-LO-diagram-cont}).

Note that in the absence of the target field~(\ref{eq:bgfield}), the
eikonal factors become trivial: $U\to 1$, and then there is strictly
no evolution. In this limit each individual diagram on the
r.h.s.~of~(\ref{eq:JIMWLK-LO-diagram-cont}) vanishes identically
owing to exact real--virtual cancellation:
in~(\ref{eq:interaction-diagrams}) the first two diagrams,
corresponding to real gluon emission, cancel against the last two
diagrams, which represent virtual corrections. Analogous
cancellations occur in this limit in the light--cone perturbation
theory decomposition of the self--energy--like diagrams in
~(\ref{eq:JIMWLK-LO-diagram-cont}).

As a functional equation, Eq.~(\ref{eq:JIMWLK}) is equivalent to an
infinite set of equations for $n$-point correlators of $U$ and
$U^\dagger$ fields in any representation of $SU(N_c)$, called the
Balitsky hierarchies~\cite{Balitsky:1996ub}. In order to obtain the
evolution equation for a correlator of a given composite operator
$O[U]$ made of $U$-fields, one first takes the $\y$ derivative of
(\ref{eq:corrs}) and then uses (\ref{eq:JIMWLK}) to replace the
$\partial_\y \Hat Z_\y[U]$ by $- {\cal H}[U]\,  Z_\y[U]$, obtaining:
\begin{equation}
  \label{eq:corrs_derivative}
   \partial_\y  \left\langle O[U] \right\rangle (\y) \,=\, - \int  \Hat{D}[U]\,
    O[U]\,  {\cal H}[U]\,  Z_\y[U],
\end{equation}
where the Hamiltonian still acts on $Z_\y[U]$. Using the self-adjoint nature
of ${\cal H}[U]$ with respect to the Haar measure (as ensured by the Lie
derivatives), one can then rewrite (\ref{eq:corrs_derivative}) as
\begin{equation}
  \label{eq:op-JIMWLK}
  \partial_\y \langle O[U] \rangle(\y) =
  -\left\langle \big({\cal H}[U] O[U]\big) \right\rangle(\y) \ .
\end{equation}
Here the JIMWLK Hamiltonian acts on $O[U]$. Observing that ${\cal H}[U]$
explicitly contains a $U$-operator, and that the number of $U$-operators
remains invariant when acted upon by the Lie derivatives $\nabla^a_{\bm{x}}$
(see Eq.~(\ref{eq:nabla_acting})) one understands that the evolution equation
for $\left< O[U]\right>$ must involve operators with more $U$ fields than the
original operator $O[U]$. Thus, the nonlinear nature of ${\cal H}[U]$ implies
that the r.h.s. of the equation depends on a new type of correlator of $U$
fields. To determine the rapidity dependence of $\langle O[U] \rangle(\y)$ one
will therefore need also the evolution equation of this new correlator, which
in turn will couple to yet higher composite operators containing more $U$
fields. Continuing the process one ends up with an infinite hierarchy of
equations, defined by the operator $O[U]$ used to start the process. The
derivation of the BK equation shown below provides a simple example for such a
hierarchy.

\subsubsection*{Truncation and the BK equation}

As the simplest correlator with immediate phenomenological relevance
we consider the two--point function of the dipole
operator\footnote{Below we will often use a shorthand notation where
the $\y$ dependence is not explicitly indicated.}:
\begin{equation}
  \label{eq:dipole}
  N_{\y,\bm{x}\, \bm{y}}:=  \langle \Hat N_{\bm{x y}} \rangle(\y) \hspace{2cm}
  \Hat N_{\bm{x}\, \bm{y}} := \frac{\tr(1 - U_{\bm{x}}^\dagger
  U_{\bm{y}})}{N_c}.
\end{equation}
Using Eq.~(\ref{eq:op-JIMWLK}) in the case of $\Hat N_{\bm{x y}}$
with the explicit expressions corresponding to the diagrams in
Eq.~\eqref{eq:JIMWLK-LO-diagram-cont}, one immediately obtains:
\begin{equation}
  \label{eq:pre-BK}
  \partial_\y \big\langle \Hat N_{\bm{x y}} \big\rangle(\y)
  = \frac{\alpha_s N_c}{2\pi^2}\int\!\! d^2 z
  \big(
  2{\cal K}_{\bm{x z y}}- {\cal K}_{\bm{x z x}}-{\cal K}_{\bm{y z y}}
  \big)
  \big\langle
  \Hat  N_{\bm{x z}}+
  \Hat  N_{\bm{z y}}- \Hat N_{\bm{x y}}-
  \Hat  N_{\bm{x z}}\, \Hat  N_{\bm{z y}}\big\rangle(\y).
\end{equation}
The linear combination of JIMWLK kernels ${\cal K}$ that appear in
this expression are in one-to-one correspondence with the three
diagrams in~\eqref{eq:JIMWLK-LO-diagram-cont}. They combine into the
very compact form of the BK kernel $\Tilde{\cal K}_{\bm{x z y}}$:
\begin{equation}
  \label{eq:BK-kernel}
  \Tilde{\cal K}_{\bm{x z y}} :=
  2{\cal K}_{\bm{x z y}}- {\cal K}_{\bm{x z x}}-{\cal K}_{\bm{y z y}}
  =\frac{(\bm{x}-\bm{y})^2}{%
    (\bm{x}-\bm{z})^2(\bm{z}-\bm{y})^2}
  \ .
\end{equation}

Clearly the r.h.s. of Eq.~\eqref{eq:pre-BK} depends on a 3-point
function containing operators with up to four $U^{(\dagger)}$
factors, that in general does not factorize into a product of two
2-point correlators:
\begin{equation}
  \label{eq:4-point-nonfac}
  \big\langle  \Hat  N_{\bm{x z}}\, \Hat  N_{\bm{z y}}\big\rangle(\y)
  = \big\langle  \Hat  N_{\bm{x z}} \big\rangle(\y) \, \times\,
  \big\langle \Hat   N_{\bm{z y}}\big\rangle(\y) \,+\,\text{corrections}\ .
\end{equation}
To completely specify the evolution of $\langle \Hat N_{\bm{x y}}
\rangle(\y)$ one therefore need to know $\big\langle \Hat N_{\bm{x
z}}\, \Hat N_{\bm{z
    y}}\big\rangle(\y) $. The latter, in its evolution equation, will couple
to yet higher $n$-point functions and thus one is faced with an
infinite hierarchy of evolution equations, as anticipated above. The
entire hierarchy (as well as others, corresponding to other
composite operators $O[U]$) is encoded in the single functional
equation~(\ref{eq:JIMWLK}). If one drops the corrections
in~\eqref{eq:pre-BK} and factorizes the correlators in the spirit of
a large--$N_c$ approximation, the hierarchy is truncated and reduces
to a single equation.  This truncation can be interpreted as an
independent scattering approximation and it leads to the
Balitsky-Kovchegov (BK)
equation~\cite{Kovchegov:1999ua,Balitsky:1997mk}:
\begin{equation}
  \label{eq:BK}
  \partial_\y N_{\y,\bm{x y}} = \frac{\alpha_s(\mu^2) N_c}{2\pi^2}\int\!\! d^2 z
  \ \Tilde {\cal K}_{\bm{x z y}}\
  \Big(
  N_{\y,\bm{x z}}+ N_{\y,\bm{z y}}-N_{\y,\bm{x y}}
  - N_{\y,\bm{x z}}\, N_{\y,\bm{z y}}\Big)
\ .
\end{equation}

\subsubsection*{Generic features and infrared stability}

\begin{figure}[htbp]
  \centering
  \begin{minipage}{7.1cm}
    \centering
    \includegraphics[height=5.1cm]{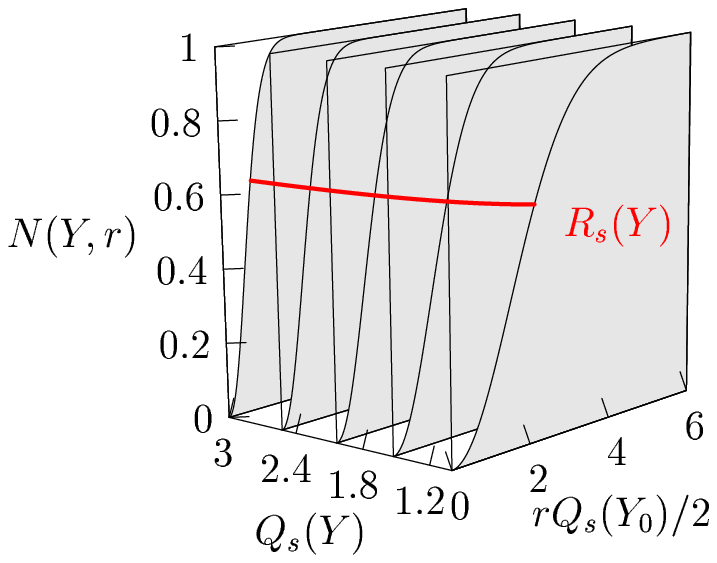}\\ (a)
  \end{minipage}\hfill
  \begin{minipage}{7.1cm}
    \centering
    \includegraphics[height=5.1cm]{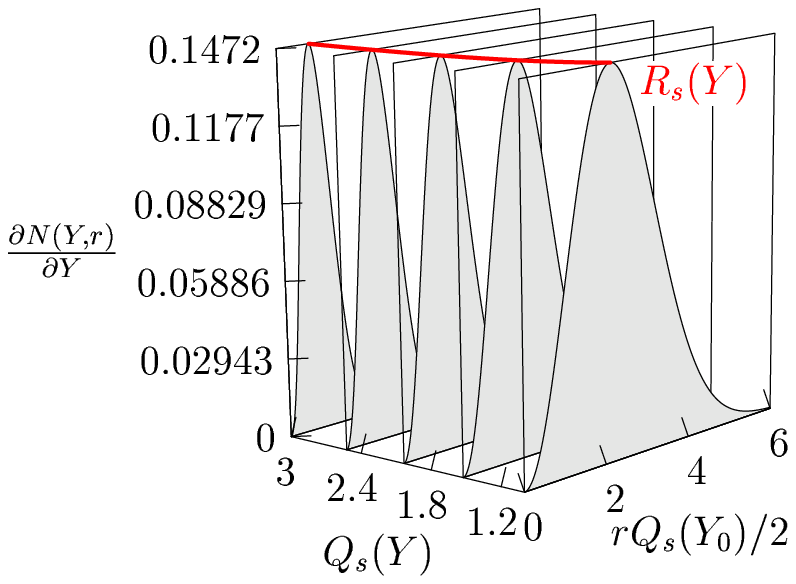} \\ (b)
  \end{minipage}

  \caption{\small \em Generic evolution trend for a single--scale
  dipole correlator.
    (a) shows $N(\y,r)$ as a function of the dipole size $r=\vert
    \bm{x}-\bm{y}\vert$ for several values of the saturation scale
    $Q_s(\y)$. $Q_s$ increases with $\y$; saturation then sets in at
    smaller distances.  (b) shows $\partial_Y N$ and thus the activity
    in a given evolution step as a function of the same variables.
    With increasing $Q_s(\y)$ contributions are centered at ever
    shorter distances. }
  \label{fig:generic-evol}
\end{figure}

Despite the complex nature of the evolution equation, it is possible
to gain insight into some generic features. It can be
proven~\cite{Weigert:2000gi} that the evolution equation possesses
an attractive fixed point at $\y\to\infty$ at which the system has
vanishing correlation length. For the evolution of a physical
correlator, such as $N_{\y,\,\bm{x y}}$, this implies a generic
trend as shown in Fig.~\ref{fig:generic-evol} (a): with
increasing~$\y$ saturation ($N_{\y,\,\bm{x y}}\to 1$) is reached at
ever shorter distances.
This leads to a further important property of the evolution
equation, namely its infrared stability: the evolution is not
affected by long--wavelength fluctuations beyond the characteristic
correlation length (the inverse of the saturation scale $Q_s$) where
$N_{\y,\,\bm{x y}}= 1$. In this way the saturation scale acts as an
effective infrared cutoff. Fig.~\ref{fig:generic-evol} (b)
demonstrates that the modes that contribute to the evolution have
momenta of order of $Q_s(\y)$. With increasing $\y$, the active
region moves towards the ultraviolet.

\subsection{What we mean by running coupling}
\label{sec:running-coupling-types}

Similarly to other evolution equations in QCD~\cite{Braun:2003rp},
the LO JIMWLK kernel is conformally
invariant\footnote{Note that while the LO \emph{kernel}
is conformally invariant, the solution of the equation is not: it is
characterized by a correlation length. This scale originates in the
initial condition, and it is preserved owing to the non-linearity of
the equation.}. As usual, breaking of this symmetry is expected to
appear through radiative corrections at ${\cal O}(\beta_0
\alpha_s^2)$ where $\beta_0$ is the leading coefficient of the
$\beta$ function,
\begin{equation}
\label{beta} \frac{d\alpha_s(\mu^2)/\pi}{d\ln \mu^2}=-\beta_0\,
\left(\alpha_s(\mu^2)/ \pi\right)^2-\beta_1\, \left(\alpha_s(\mu^2)/
\pi\right)^3\,+\,\cdots \,; \qquad\quad
\beta_0=\frac{11}{12}C_A-\frac16 N_f.
\end{equation}
At higher orders one expects corrections  ${\cal O}(\beta_0^{n-1}
\alpha_s^n)$, as well as ones associated with subleading
coefficients of the $\beta$ function, e.g. ${\cal
O}(\beta_1\beta_0^{n-3} \alpha_s^n)$. In
Sec.~\ref{sec:deriv-jimwlk-running} we compute these
running--coupling corrections using a dispersive representation of
the dressed gluon propagator in a background field. We will show
that while the general structure of the evolution equation, namely
Eq.~(\ref{eq:JIMWLK}), holds as at LO, the kernel itself
changes drastically with respect to Eq.~(\ref{eq:JIMWLK-LO-kernel})
--- see Eq.~(\ref{eq:new-JIMWLK}) or (\ref{eq:pert-exp}) below. A
similar generalization holds in the BK case, see Eqs.
(\ref{eq:BK-running-kernel}) and~(\ref{eq:BK-full}). JIMWLK and BK
evolution with running--coupling is therefore qualitatively
different from the fixed--coupling case in that at each step in the
evolution the coupling depends on the details of the evolving
configuration; moreover, different \emph{final states}, that are
characterized by different ``daughter dipole'' sizes, are weighted
differently in the r.h.s. of the evolution equation.

Although all previous derivations of JIMWLK and BK evolution
equations were restricted to the leading logarithmic approximation,
it has been clear for quite a while that running--coupling
corrections will be necessary to get quantitative
results\footnote{In Ref.~\cite{Rummukainen:2003ns} it has been
further emphasized that running--coupling corrections, where the
scale of the coupling depends on the scales involved in a {\em
single} evolution step, are needed to reduce active phase space in
the ultraviolet from about $6$ orders of magnitude to about one.
Despite this dramatic reduction of phase space, other qualitative
features of the solution of JIMWLK or BK equations were found to be
the same: for any set of initial conditions that interpolate between
color transparency at short distance and saturation at large
distance (as shown schematically in Fig.~\ref{fig:generic-evol}) the
system approaches an asymptotic line where the dipole correlator
reaches a near--scaling form.}. By making physically--motivated
scale choices for the renormalization scale of the coupling, several
authors found that the evolution rate $\lambda(\y)$ reduces by a
factor of two or more compared to
 the fixed--coupling case.

It is important to make a clear distinction between the actual
higher--order contributions to the kernel, which we compute in the
following, and ad hoc choices of scale for the coupling, which have
been often referred to as ``running coupling''. Such prescriptions
have been assumed in all numerical simulation of the BK
equation~\cite{Braun:2000wr, Kimber:2001nm,Armesto:2001fa,
Levin:2001et, Lublinsky:2001bc,
Golec-Biernat:2001if,Albacete:2004gw, Rummukainen:2003ns}. Similar
assumptions were used in the analytical estimates of
Ref.~\cite{Mueller:2002zm}, which agree well with numerical
simulations~\cite{Rummukainen:2003ns}. To put the results of the
present paper in context of previous work, we find it useful to
further distinguish between different categories of scale choices
that were made in the literature:
\begin{enumerate}
\item {\bf Fixed or essentially fixed coupling:}
  $\alpha_s(\mu^2)$ is treated as a constant or as a function of $\y$.
  Within this class, simulation results, starting from the same initial
  condition, can be related by re-scaling the evolution variable.
  For example, the scenario where the coupling depends on $Q_s(\y)$
can be related to the one where the scale of the coupling is fixed
as some $Q_0$ through the change of variables (see
also~\cite{Iancu:2002tr}):
\[
\y' = \frac{\alpha_s\left(Q_s^2(\y)\right) }{
\alpha_s\left(Q_0^2\right) } \,\y.
\]
Obviously, any ansatz of this sort would fail to capture the
essential physics of running coupling, and would inherit the
ultraviolate phase--space problem discussed in
Ref.~\cite{Rummukainen:2003ns}.
\end{enumerate}
A non-trivial change in the shape of the solutions with respect to
the fixed--coupling case occurs if the scale of the coupling is
determined by the size of the ``dipoles'' involved in the evolution.
In such cases the ultraviolate phase--space problem is generally
removed, since emission from very small objects is suppressed by
small coupling constants. The remaining two cases fall into this
category.
\begin{enumerate}\addtocounter{enumi}{1}
\item {\bf ``Parent--dipole'' running:} where the scale of the
coupling on r.h.s. of the BK equation~(\ref{eq:BK}) is assumed to
depend on the initial dipole size $\rr=|\bm x - \bm y|$, namely
\begin{equation}
\label{eq:parent_dipole_BK} \alpha_s\, (\mu^2) \Tilde{\cal K}_{\bm{x
z y}} \,\to\, \alpha_s(c^2/\rr^2) \,\Tilde{\cal K}_{\bm{x z y}},
\end{equation}
 where $c$ is a dimensionless ``scale
factor'' of order one\footnote{It was often taken as $c\approx 4$,
motivated by direct Fourier transform in the double logarithmic
limit.}. This has been the most common ansatz in numerical
simulations of the BK equation.

While this appears to be a natural ansatz in the context of the BK
equation, it can not be easily reconciled with the JIMWLK equation.
To see this, recall that (\ref{eq:JIMWLK}) is a functional equation.
Therefore, the Hamiltonian in Eq.~(\ref{eq:JIMWLK-Hamiltonian-LO})
cannot depend on any scales characterizing a particular operator
$O[U]$; all three transverse coordinates appearing in
(\ref{eq:JIMWLK-Hamiltonian-LO}) are integrated over internally.
When deriving the BK equation from JIMWLK as in (\ref{eq:pre-BK})
one obtains the BK kernel as a specific combination of the original
JIMWLK kernel:
\begin{align}
  \label{eq:BK_kernel_from_JIMWLK}
  {\alpha_s} \Tilde{\cal K}_{\bm{x z y}} \to
  \ 2 \, {\alpha_s(\mu^2_{\bm{x z y}})} {\cal K}_{\bm{x z y}}
  -{\alpha_s(\mu^2_{\bm{x z x}})}\,{\cal K}_{\bm{x z x}} -
  {\alpha_s(\mu^2_{\bm{y z y}})}\,{\cal K}_{\bm{y z y}},
\end{align}
where each term corresponds to one of the diagrams in
(\ref{eq:JIMWLK-LO-diagram-cont}).  Whatever one assumes of the
functional dependence of $\mu^2_{\bm{x z y}}$  on the coordinates,
since the self--energy--like diagrams are independent of the parent
dipole size $|\bm x - \bm y|$, Eq.~(\ref{eq:BK_kernel_from_JIMWLK})
does not lead to Eq.~(\ref{eq:parent_dipole_BK}).
\end{enumerate}
Even considering the BK case on its own, ``parent--dipole'' running may appear
artificial: there is no reason why the sizes of the dipoles \emph{produced} in
the same step in the evolution, which can of course be quite different from
the parent size, should not play an important r\^ole~\cite{Albacete:2004gw}.
This leads us to the final category:
\begin{enumerate}\addtocounter{enumi}{2}
\item {\bf Final--state--dependent evolution:} where the scale of the
coupling on r.h.s. of the BK (\ref{eq:BK}) or JIMWLK
(\ref{eq:JIMWLK-Hamiltonian-LO}) equations is assumed to depend on
all three distance scales present in a single evolution step, namely
the ``parent dipole'' $\rr=|\bm x - \bm y|$ and the two newly
produced ``daughter dipoles'', $\dl=|\bm{x}-\bm{z}|$ and
$\dr=|\bm{y}-\bm{z}|$. Obviously, in this case the coupling affects
the weight of different final states, depending on the transverse
location of the emitted gluon (the coordinate ${\bm z}$) that is
integrated over.

Several such models were proposed, e.g. $\sqrt{\alpha_s({c^2}/{\dl^2
})}\,\sqrt{\alpha_s({c^2}/{\dr^2 })}$ or
$\alpha_s\left(\text{max}\left\{{c^2}/{\dl^2 },{c^2}/{\dr^2
}\right\}\right)$, and found to be in fair agreement with
``parent--dipole'' running, see e.g.~\cite{Albacete:2004gw}. As
already mentioned, no deep justification of any of these models has
been provided.
\end{enumerate}

In the next section we compute running--coupling corrections to the
JIMWLK equation. In Sec. \ref{sec:BK-numerics} shall use this result
to derive the corrections for the BK case and study the consequences
numerically. As we will see, these corrections {\em do} depend
on the details of the final state at each evolution step and involve
all three scales. Moreover, in neither of the two equations do these
corrections naturally reduce to a single scale--dependent coupling
times the LO kernel.

\section{Derivation of JIMWLK evolution with running coupling}
\label{sec:deriv-jimwlk-running}

\subsection{Running coupling, Borel transform and the dispersive approach}
\label{sec:runn-coupl-disp}

We are interested in improving the leading--logarithmic result of
the JIMWLK and BK equations by including running--coupling effects,
namely corrections associated with the renormalization--group
equation (\ref{beta}). In general, running--coupling corrections are
important in QCD, for two reasons:
\begin{itemize}
\item{} They usually constitute a large part of the higher--order (notably NLO)
corrections~\cite{BLM,Brodsky:1997vq,Gardi:1999dq,Brodsky:2000cr,Neubert:1994vb,Beneke:1994qe,Ball:1995ni,Dokshitzer:1996qm}.
The main reasons for this are: (1) the average virtuality of gluons
is usually different (typically much lower) than the principal hard
scale, which is often used as the default renormalization point; and
(2) $\beta_0$ is sizeable. These effects are especially important
when the hard scale is low, since then the coupling is large and its
evolution (\ref{beta}) is fast.
\item{} They dictate the large--order asymptotic behavior of the perturbative
series, which is dominated by factorially increasing contributions,
${\cal O}\left(n! \beta_0^{n-1} (\alpha_s/\pi)^n\right)$, the
renormalons~\cite{Lautrup:1977hs,'tHooft:1977am}. In this way the resummation
provides some insight into the non-perturbative side of the problem:
the \emph{ambiguity} in summing the perturbative expansion, which is
expected to cancel in the full theory, indicates the parametric
dependence of non--perturbative corrections on the hard scales
(power--suppressed corrections) and provides some clue on the
potential size of these
corrections~\cite{Mueller:1984vh,Parisi:1978bj,David:1983gz,Zakharov:1992bx};
for a review see Refs.~\cite{Beneke:1998ui,Beneke:2000kc}.
\end{itemize}
Both these aspects are relevant in non-linear evolution. As
demonstrated in Sec.~\ref{sec:running-coupling-types}, the
multi-scale nature of the problem calls for a systematic study.

In this section we wish to briefly recall some basic ideas and
techniques for resummation of running--coupling corrections that
will be generalized and applied to the JIMWLK case in what follows.
To this end, consider some perturbatively calculable (infrared and
collinear safe) quantity, $R(Q^2/\Lambda^2)$, depending, for
simplicity, on a single external scale $Q^2$ and having an expansion
starting at order $\alpha_s(\mu^2)/\pi$ with the LO coefficient
normalized\footnote{Using this normalization $R(Q^2/\Lambda^2)$ can
also be interpreted as an ``effective charge''~\cite{
Grunberg:1980ja}.} to $c_{00}\equiv 1$:
\begin{equation}
\label{eq:R_series}
  R(Q^2/\Lambda^2)=\frac{\alpha_s(\mu^2)}{\pi}
  +\left[\left(c_{11}+\ln\frac{\mu^2}{Q^2}\right)\beta_0 +
    c_{10}\right]
  \left(\frac{\alpha_s(\mu^2)}{\pi}\right)^2
  +\ldots\ ,
\end{equation}
where $\beta_0$ is defined in (\ref{beta}) and the $c_{10}$ term is
a conformal coefficient, not associated with the running coupling.
We will work in the large--$\beta_0$ limit, where the $c_{10}$ term
is formally subleading. The well--known BLM prescription~\cite{BLM}
absorbs the NLO contribution that is leading in $\beta_0$ into the
LO by a scale choice:
$\mu^2_{\BLM}=Q^2\,\exp\left\{{-c_{11}}\right\}$. In this way also
higher--order corrections often become smaller. This becomes
intuitive upon looking at the momentum integral that is
approximated by $\alpha_s(\mu^2_{\BLM})/\pi$, where $\mu^2_{\BLM}$
acquires the interpretation of the average gluon virtuality. The
all--order resummation of running--coupling effects in the single
dressed gluon approximation,
\begin{equation}
\label{eq:R_series_large_beta_0}
  \left.R(Q^2/\Lambda^2)\right\vert_{\rm large \,\,\beta_0}\,=\,
  \left(\frac{\alpha_s(Q^2)}{\pi}\right) \left[1+\sum_{n=1}^{\infty}
  \,c_{nn}\,\times\,
  \,\left(\beta_0\frac{\alpha_s(Q^2)}{\pi}\right)^n\right],
\end{equation}
can be viewed as a generalization of this procedure, where instead
of a single optimal scale choice, the proper (observable--dependent)
weight is given to any specific gluon virtuality. In the context of
the dispersive approach presented below (see Eq. (\ref{R})), this
weight function is called the ``characteristic
function''~\cite{Dokshitzer:1995qm}.

Technically, the resummation of all ${\cal
O}(\beta_0^{n-1}\alpha_s^n)$  higher--order corrections becomes
feasible in QCD, owing to its simple relation with the resummation
of diagrams with an arbitrary number of fermion--loop insertions.
Because $\beta_0$ is linear in $N_f$, one can simply compute the
fermion--loop chain diagrams
and then replace $N_f$ by
the non--Abelian value of $-6\beta_0$, according to (\ref{beta}).
In more formal terms, one begins by considering the large--$N_f$
limit with fixed $N_f\alpha_s$, the leading term in the flavor expansion.
Clearly this limit
itself is not physically interesting, it just provides a tool
to identify running--coupling contributions in the approximation where the
non-Abelian $\beta$ function (\ref{beta}) is one loop,
the so--called large--$\beta_0$ limit.

The calculation of a gluon propagator, dressed by fermion--loop
insertions is simplified by the fact that the fermion loop itself is
transverse,
\[
\Pi_{\mu\nu}(k^2) = \left(g_{\mu\nu}-\frac{k_\mu k_\nu}{k^2}\right)
  \,k^2\, \Pi(k^2),
\]
and therefore, in any gauge such insertions affects only the
propagating particle pole $1/k^2$. The
all--order sum builds up a geometric series giving rise to a factor
$1/(1+\Pi(k^2))$. The resummed propagator takes the
form\footnote{Note that the $\xi$ term
in~\eqref{eq:resummed-propstructure-free-cov} is not affected as it
does not describe a physical mode. $\xi$ is the width of a Gaussian
approximation to a functional $\delta$-function that is meant to
implement the covariant gauge $\partial_\mu A^\mu=0$. Only for vanishing
width, $\xi=0$, do all gauge fields obey the gauge condition
strictly. The axial propagator is written for vanishing width, i.e.
the strict gauge condition. That is why the resummed terms multiply
the whole structure.}:
\begin{subequations}
  \begin{align}
    \label{eq:resummed-propstructure-free-cov}
    \text{covariant gauges:} & & \frac{1}{k^2}\frac{1}{1+\Pi(k^2)} &
    \left(g_{\mu\nu}-\frac{k_\mu k_\nu}{k^2}\right)+\xi
    \frac{1}{k^2}\frac{k_\mu k_\nu}{k^2}
    \\
    \label{eq:resummed-propstructure-free-ax}
    \text{strict axial gauge:} & & \frac{1}{k^2}\frac{1}{1+\Pi(k^2)} &
    \left(g_{\mu\nu}-\frac{k_\mu n_\nu+n_\mu k_\nu}{k\cdot n}+ \frac{k_\mu k_\nu
        n^2}{(k\cdot n)^2}\right)
\ ,
  \end{align}
\end{subequations}
with
\begin{equation}
\label{eq:Pi_ren}
  \left.\Pi(k^2)\right\vert_{\rm one-loop}
  =\frac{\alpha_s(\mu^2) \beta_0}{\pi}\,\ln\left(  -{k^2\,{\rm
  e}^{-\frac53}}/{\mu^2}\right),
  \end{equation}
  where the renormalization of the fermion--loop $\Pi(k^2)$ was
  done in the ${\overline {\rm MS}}$
  scheme\footnote{We denote the $\overline {\rm MS}$ coupling
  $\alpha_s^{\MSbar}(\mu^2)$ by $\alpha_s(\mu^2)$.}
  and where we already made the
  replacement: $N_f\,\to\, -6\beta_0$.

The next step would be to insert the dressed propagator into the relevant
Feynman diagrams and perform the momentum integration, $d^4k$.  Since the sum
over any number of $\Pi(k^2)$ insertions has already been done, by performing
the $k$-integration one would hope to get directly the \emph{resummed}
physical quantity $R(Q^2/\Lambda^2)$ in (\ref{eq:R_series_large_beta_0}).
Observing that the resummed propagator with (\ref{eq:Pi_ren}) has a Landau
singularity, one realizes that this direct all--order calculation cannot be
done.  As we explain below, a regularization of the sum is required even if
the coefficients $c_{nn}$ are all finite\footnote{This is indeed the case if
  $R$ is an observable, i.e. it requires no additional renormalization and has
  no infrared divergencies.}.  The simplest and most familiar way to see this
is to represent the effective running coupling\footnote{It is convenient to absorb the
  factor ${\rm e}^{-\frac53}$ from the renormalization of the fermion loop in
  (\ref{eq:Pi_ren}) into the definition of the coupling. We follow this
  convention and define $\alpha_s^V$ as the coupling in the $V$ scheme, which
  is related to the ${\overline {\rm MS}}$ scheme by $\Lambda_V^2={\rm
    e}^{\frac53}\Lambda^2$.} \emph{that includes the dressing}, as a Borel sum:
\begin{equation}
\label{eq:alpha_V_Borel}
 \frac{\alpha_s^V(-k^2-i0)}{\pi} := \frac{\alpha_s(\mu^2)}{\pi}
\,\frac{1}{1+\Pi(k^2)}\,=\,
\frac{1}{\beta_0}\, \int_0^{\infty}du\, T(u)\, \left(
-{k^2\,{\rm e}^{-\frac53}}/{\Lambda^2}\right)^{-u}
\ ,
\end{equation}
where for one--loop running coupling\footnote{Although we restrict the
  calculation to one--loop coupling where $T(u)\equiv 1$, we will keep writing
  $T(u)$ in any Borel representation: this allows for a straightforward
  generalization to two--loop running coupling
  following~\cite{Grunberg:1993hf,Gardi:2002xm}.},
namely upon using (\ref{eq:Pi_ren}), one simply gets
\[
\left.\frac{\alpha_s^V(-k^2-i0)}{\pi}\right\vert_{\rm one-loop}
= \frac{1}{\beta_0}\,\frac{1}{\ln\left(
-{k^2\,{\rm  e}^{-\frac53}}/{\Lambda^2}\right)},
\]
and therefore $T(u)\equiv 1$. It is now
possible to proceed with the calculation of the Feynman diagrams where the
only change is that the particle pole is modified into a cut:
\begin{equation}
\label{eq:Borel_replacement}
 \frac{1}{-k^2-i0} \,\to\,
\frac{1}{(-k^2-i0)^{1+u}},
\end{equation}
This modification of the propagator is known as Borel or analytic
regularization.

The all--order resummation of a given quantity $R$ in the large--$\beta_0$
limit can therefore be done using (\ref{eq:alpha_V_Borel}) by first performing
the momentum integration with the modified propagator
(\ref{eq:Borel_replacement}), which directly yields the Borel representation
of the sum in the single dressed gluon approximation, namely:
\begin{equation}
  \left.R(Q^2/\Lambda^2)\right\vert_{\rm large \,\,\beta_0}\,
  =\,\frac{1}{\beta_0}\,\int\limits_0^{\infty}du \,
  T(u) \left(Q^2/\Lambda^2\right)^{-u} B(u).
  \label{R_borel_}
\end{equation}

Upon performing the momentum integration for a typical observable in QCD one
would find that $B(u)$ has singularities along the positive real axis, which
is the integration axis. This obviously means that Eq.~(\ref{R_borel_}) is
ill-defined as it stands. The obstructing singularities are called infrared
renormalons, and they merely reflect the fact that the series for $R$ in
(\ref{eq:R_series_large_beta_0}) and thus also in (\ref{eq:R_series}), is non
summable. The problem only becomes manifest if one attempts to sum the series:
any finite order expansion --- here all the coefficients $c_{nn}$ in
(\ref{eq:R_series_large_beta_0})
--- are well--defined and finite. They can be
obtained by expanding $B(u)$ under the integral,
\begin{equation}
\label{eq:Borel_expand} B(u)\,=\, \sum_{n=0}^{\infty}
\,\frac{c_{nn}}{n!}\, \,u^n.
\end{equation}
The Borel function typically has a finite radius of convergence $u_0$, so
$c_{nn}$ grow as $n!/u_0^n$ at high orders\footnote{This asymptotic behavior
  can be modified by additional factors of the form $n^{\gamma/\beta_0}$,
  depending on the structure of the Borel singularity. For simplicity we
  assume here simple poles.} ; in the case of infrared renormalons in QCD, the
singularity is at $u=u_0>0$, so the coefficients all have the same
sign\footnote{ In contrast, for ultraviolate renormalons in QCD $u_0<0$, so
  there is sign oscillation at high order.}.  In terms of the momentum
integration these large perturbative coefficients are associated with
extremely small virtualities $k^2\to 0$, namely soft modes whose dynamics is
non-perturbative.  Thus, the non-existence of the sum is a reflection of the
fact that perturbation theory does not fully describe the dynamics, and
the observable is sensitive to some extent to non-perturbative contributions.  The
ambiguity can therefore be resolved by a proper separation prescription
between perturbative and non-perturbative corrections, either by an infrared
curoff, or by other means, for example by modifying the integration contour of
the Borel integral~(\ref{R_borel_}) in a particular way or by taking its
principal value. The same prescription must then be applied to regularize
the non-perturbative contribution.

An additional, general property is that the singularities in $B(u)$ occur at
integer --- and sometimes half integer --- values of $u$. This corresponds to the
fact that alternative definitions of the sum of the series (that arise e.g. by
modifying the integration contour in (\ref{R_borel_})) differ by integer  --- or
half integer --- powers of $\Lambda^2/Q^2$.  These ambiguities must be cancelled
by non-perturbative power corrections, and they can therefore serve as a
perturbative probe of such effects. In cases where an operator product
expansion applies, one can get a direct interpretation of the source of each
ambiguity in terms of local operators of higher dimension (or
twist)~\cite{David:1983gz,Mueller:1984vh}.  In such cases it is also possible
to trace the cancellation of
ambiguities~\cite{Beneke:1998ui,Gardi:2002bk,Braun:2004bu}. In the absence of
an operator product expansion, the renormalon technique often provides a
unique window into the non-perturbative regime: by identifying the ambiguities
in summing the perturbative series one learns about the parametric dependence
of power corrections on the hard scales and about their potential
size~\cite{Parisi:1978bj, David:1983gz, Zakharov:1992bx, Mueller:1984vh,
  Beneke:1998ui, Gardi:2002bk, Braun:2004bu, Bigi:1994em, Smith:1994id,
  Beneke:1994qe, Beneke:1995qe, Ball:1995ni, Dokshitzer:1995qm,
  Grunberg:1998ix, Gardi:1999dq, Gardi:2000yh, Cacciari:2002xb, Gardi:2003iv,
  Gardi:1998qr}.

Computing directly the Borel function $B(u)$ using
(\ref{eq:Borel_replacement}) may sometimes present technical difficulties. One
of the most effective techniques to deal with this problem is the dispersive
technique, which has been used in a variety of applications, see
e.g.~\cite{Bigi:1994em, Smith:1994id, Beneke:1994qe, Beneke:1995qe,
  Ball:1995ni, Dokshitzer:1995qm, Grunberg:1998ix, Gardi:1999dq, Gardi:2000yh,
  Cacciari:2002xb, Gardi:2003iv, Gardi:1998qr}.  In the next sections we shall
generalize this technique to the case of Weizs\"acker-Williams background
fields, in order to use it in the re-derivation of the JIMWLK equation. Let us
therefore review here the basic idea, and collect the necessary formulae.

The dispersive method, see e.g.~\cite{Ball:1995ni, Dokshitzer:1996qm,
  Beneke:1995qe}, recasts the dressed gluon propagator, and thus the running
coupling of (\ref{eq:alpha_V_Borel}) as a dispersive integral. As we shall
see, this allows one to compute a generic quantity $R(Q^2/\Lambda^2)$ to all
orders in the large--$\beta_0$ limit, by simply replacing the massless gluon
pole by a massive one, namely:
\begin{equation}
\label{eq:massive_gluon_replacement}
 \frac{1}{-k^2-i0} \,\to\,
\frac{1}{m^2-k^2-i0},
\end{equation}
instead of~\eqref{eq:Borel_replacement}.

The dispersive representation takes the form
\begin{equation}
  \label{eq:simple-dispersive}
  \frac{\alpha_s^V(-k^2-i0)}{\pi} \,=\,
  \frac{\alpha_s(\mu^2)}{\pi} \,\frac{1}{1+\Pi(k^2)} =\, \frac{1}{\beta_0}\int\limits_0^\infty dm^2
  \frac{\rho_V(m^2)}{m^2-k^2-i0}
  \ ,
\end{equation}
where $\rho_V$ is the discontinuity of the coupling on the time-like
axis,
\begin{equation}
\label{eq:discontinuity}
\rho_V(m^2):= -\frac{\beta_0}{\pi}\,{\rm
    Im}\left\{\alpha_s^V(-m^2-i0)/\pi\right\} =\frac{1}{\pi}
  \frac{\beta_0\alpha_s(\mu^2)}{\pi}\,\frac{ \text{Im}
    \left\{\Pi(m^2)\right\}}{\vert
    1+\Pi(m^2)\vert^2} \ .
\end{equation}
Alternatively, it is convenient to express
Eq.~(\ref{eq:simple-dispersive}) in terms of the
``effective\footnote{Let us emphasize that we do {\em not} make any
use of the dispersive representation to impose analyticity and
regularize the Landau singularity, as done for example in
Ref.~\cite{Shirkov:1997wi}. We merely use the dispersive method
to compute the Borel function, which serves as a generating function
for the perturbative coefficients according to
(\ref{eq:Borel_expand}) and as a tool to analyze infrared renormalon
ambiguities.} time--like coupling'',
\begin{equation}
  \label{eq:alpha-via-rho}
  \frac{\alpha_s^V(-k^2-i0)}{\pi}
  = \frac{1}{\beta_0}\int\limits_0^\infty dm^2 \frac{\rho_V(m^2)}{m^2-k^2-i0}
  = \,
  \frac{1}{\beta_0}
  \int_0^{\infty}\frac{dm^2}{m^2} A_{{\text{eff}}}^V(m^2)
  \left[-m^2\frac{d}{dm^2} \frac{1}{k^2-m^2}\right]
  \ ,
\end{equation}
where $A_{{\text{eff}}}^V(m^2)$ is defined by
\begin{equation}
  \label{eq:Aeff}
  \rho_V(m^2) =: \,
  m^2\frac{d}{d m^2}A_{\text{eff}}^V(m^2).
\end{equation}
The explicit expression for $A_{{\text{eff}}}^V(m^2)$
in the case of a one--loop running--coupling (namely
(\ref{eq:simple-dispersive}) with~(\ref{eq:Pi_ren}))
takes the form:
\[
\left.A_{{\text{eff}}}^V(m^2)\right\vert_{\rm one-loop}\,=\,\frac12 -\frac1\pi
\arctan\left(\frac1\pi\ln\frac
  {m^2}{\Lambda_V^2}\right).
\]
A two-loop expression for $A_{{\text{eff}}}^V(m^2)$ can be found in
Ref.~\cite{Gardi:1998qr}.

The all--order resummation of $R$ in the
large--$\beta_0$ limit can now be done using
(\ref{eq:alpha-via-rho}) by first performing the momentum
integration with a massive gluon (i.e. with the replacement of
Eq.~(\ref{eq:massive_gluon_replacement})) leaving the dispersive integral
over the mass undone. This leads to\footnote{Note that we ignore
pure
  power correction terms ($\Lambda^2/Q^2$) which distinguish this integral
  from the Borel sum we eventually compute. These will be totally irrelevant
  here. The interested reader is referred
  to~\cite{Ball:1995ni,Grunberg:1998ix,Gardi:1999dq}.}
\begin{align}
  \label{R}
  \left.R(Q^2/\Lambda^2)\right\vert_{\rm large \,\,\beta_0}\,
  = & \frac{1}{\beta_0} \int_0^{\infty}\frac{dm^2}{m^2}
  A_{{\text{eff}}}^V(m^2) \left[-m^2\frac{d}{dm^2} {\cal F} (m^2/Q^2)\right]
  \nonumber \\
  = & \frac{1}{\beta_0}\int_0^{\infty}\frac{dm^2}{m^2} \rho_V(m^2) \left[ {\cal
      F} (m^2/Q^2)-{\cal F} (0)\right],
\end{align}
where in the $m^2\to 0$ limit ${\cal F} (m^2/Q^2)$ yields the LO coefficient,
${\cal F} (0)=c_{00}=1$. By definition, ${\cal F} (m^2/Q^2)$ depends on the
specific quantity computed and it encodes perturbative as well as
power--correction information (see e.g.~\cite{Ball:1995ni,
  Dokshitzer:1996qm}). ${\cal F} (m^2/Q^2)$ is therefore called the
``characteristic function'' of the quantity or process under consideration.
Upon expanding the effective coupling $A_{{\text{eff}}}^V(m^2)$ in (\ref{R}) in powers
of $\ln (m^2/\Lambda^2)$ and integrating term by term (log moments) one
obtains the perturbative coefficients $c_{nn}$. On the other hand, expanding
${\cal F} (m^2/Q^2)$ at small $m^2$ and extracting the non--analytic
terms~\cite{Dokshitzer:1996qm} one obtains information on renormalon
ambiguities and thus on power corrections in $\Lambda^2/Q^2$.

As explained above, the perturbative and power--correction information encoded
in the characteristic function can be conveniently extracted from the Borel
representation (\ref{R_borel_}).  The Borel formulation also offers a natural
definition of the sum, using the principal value (PV) prescription. The PV Borel sum
can be related to a Euclidian--cutoff regularization~\cite{Gardi:1999dq},
and it has good analytic properties as a function of the hard
scales~\cite{Cacciari:2002xb,Andersen:2005bj}, e.g. it has no Landau singularities
and it is real--valued.
In order to relate the dispersive and the Borel formulations, (\ref{R})
and~(\ref{R_borel_}), respectively, one first computes the Borel
representation of the discontinuity of the coupling by using
(\ref{eq:alpha_V_Borel}) in (\ref{eq:discontinuity}); a straightforward
calculation yields:
\begin{equation}
\label{eq:discontinuity_Borel}
\rho_V(m^2)= - \int_0^{\infty} du\, T(u)\, \frac{\sin \pi u}{\pi} \,
e^{\frac53 u}\,\left({m^2}/{\Lambda^2}\right)^{-u}.
\end{equation}
Next, by substituting (\ref{eq:discontinuity_Borel}) in (\ref{R}) and changing
the order of integration one recovers the Borel representation
(\ref{R_borel_}) with
\begin{eqnarray}
  \label{eq:B-F-relation}
  B(u) &=&
  -e^{\frac53 u}\frac{\sin \pi u}{\pi}\,
  \int\limits_0^{\infty}d\zeta \,\zeta^{-1-u}
  \,\left[{\cal F} (\zeta)-{\cal F}(0)\right] \,=\,
-e^{\frac53 u}\frac{\sin \pi u}{\pi u}\,
  \int\limits_0^{\infty}d\zeta \,\zeta^{-u}
  \,\frac{d{\cal F} (\zeta)}{d\zeta} .
\end{eqnarray}
In the following we will use this formula to compute the Borel
function $B(u)$ from ${\cal F} (\zeta)$.

Finally, it should be emphasized that the resummed result for
$R(Q^2/\Lambda^2)$, computed with a single dressed gluon,
is exactly renormalization--scale invariant, despite using just a
subset of all perturbative corrections.
At power level, we shall define the sum using the principal value
prescription of~(\ref{R_borel_}) and use the
renormalon poles to analyze non-perturbative power corrections.

\subsection{The dressed propagator in the background field}
\label{sec:dress-prop-backgr}

At this point, we have described the method used to derive the
JIMWLK equation at fixed coupling
(Sec.~\ref{sec:evolution-equations}) and the generic tools used to
promote a one--loop calculation to one with a dressed gluon
(Sec.~\ref{sec:runn-coupl-disp}). These need now to be combined.
Since all the diagrams in~\eqref{eq:JIMWLK-LO-diagram-cont} involve
a single gluon, the dispersive technique should be applicable.
The crucial step is to construct a propagator that is both in the
background of Weizs\"acker-Williams field
\emph{and} is dressed by running--coupling corrections.
This is the main goal of this section.

\begin{figure}[ht]
  \centering
  \includegraphics[width=\textwidth]{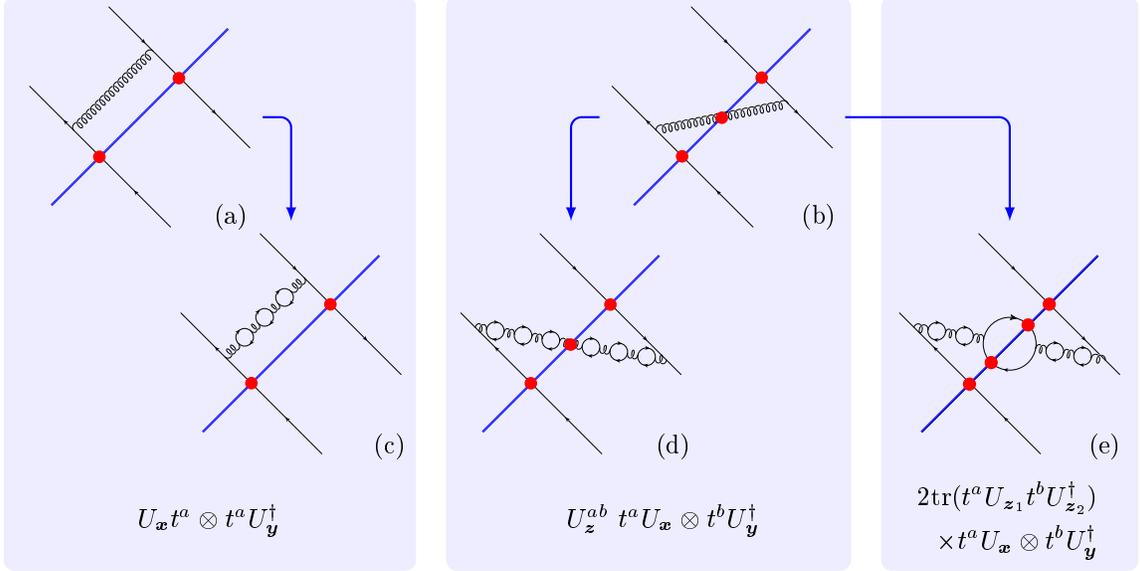}
  \caption{\small \em Virtual and real contributions at LO (a), (b) and
    NLO (c), (d) and (e). Vertically (as delineated by the background shading)
    they are grouped by the Wilson line structure induced by the target
    interaction.  The latter are indicated on the diagrams via large solid
    dots, their analytical form is listed at the bottom.  Diagram (e) induces
    a new structure via a $q\Bar q$ loop that interacts with the target.}
  \label{fig:NLO-diagrams}
\end{figure}

Independently of the particular technique adopted, the calculation of
running--coupling corrections to the kernel involves a subtlety which is
associated with the presence of new production channels, namely the production
of a well separated $q\bar{q}$ pair or a pair of gluons, starting at NLO.
These processes involve \emph{two} additional Wilson lines --- not one ---
on the r.h.s of the
evolution equation, namely an entirely new contribution to the JIMWLK
Hamiltonian (\ref{eq:JIMWLK-Hamiltonian-LO}).
Obviously, these contributions should not be considered a running--coupling
correction, and we do not aim to compute them here. As explained
in Sec.~\ref{sec:runn-coupl-disp} running--coupling corrections are usually
identified by considering diagrams with fermion--loop insertions, namely considering the
formal large--$N_f$ limit with fixed $N_f\alpha_s$ (the flavor expansion), and
then making the substitution $N_f\to-6\beta_0$. However, in the present case
this criterion is insufficient: similarly to
running--coupling corrections, $q\bar{q}$ pair production appears at LO in the
flavor expansion, as can be seen in Fig.~\ref{fig:NLO-diagrams}.
Moreover, the running--coupling corrections and $q\bar{q}$ pair production
are physically distinct only if the transverse separation of the pair is non-vanishing.
Therefore, the identification of running--coupling corrections to the kernel requires a
separation prescription. As we shall see below, the use of the dressed propagator in the
background field automatically implements such a prescription.

A related complication arises in the derivation of the JIMWLK Hamiltonian
with running coupling because of the possibility of interaction between
the background field and a produced $q\bar{q}$ pair.
Dressing the \emph{virtual} diagram, shown in Fig.~\ref{fig:NLO-diagrams}~(a),
poses no problems, since the gluon itself does not interact with the
target\footnote{At this point one might be worried that some of intermediate
  quark--gluon vertices (whose location is to be integrated over) might lie on
  different sides of the target line -- this is precluded by causality since
  the interaction is restricted to the light-like support of the target field
  $b^+=\delta(x^-)\beta({\bm x})$. Even the dressed gluon line cannot
  propagate back and forth across this hyperplane. This will be borne out by
  the explicit calculation.}.  For each LO virtual contribution
in~\eqref{eq:interaction-diagrams}, as well as their self--energy--like
analogues, there is just one type of dressed virtual counterpart; e.g.  in
Fig.~\ref{fig:NLO-diagrams}, diagram~(c) corresponds to dressing of
diagram~(a). The eikonal interactions with the target field remain unaffected
by the dressing. The dressed propagator in (c) reduces to the dressed
propagator in the absence of the target field, and the running--coupling
corrections may be directly obtained using the tools of
Sec.~\ref{sec:runn-coupl-disp}.

On the other hand, real--emission diagrams, such as the diagram in
Fig.~\ref{fig:NLO-diagrams}~(b), lead to two physically different
generalizations upon inserting fermion loops; these are shown in diagrams~(d)
and~(e), respectively. Diagram~(d) contains the same eikonal factors as the LO
diagram, while diagram~(e) contains a new channel: the production of a well
separated $q\Bar q$--pair, which interacts with the target field at two
\emph{distinct} points on the transverse plain ${\bm z}_1$ and ${\bm z}_2$.  Instead
of the single gluonic eikonal factor $U_{\bm z}$ as in diagrams~(b) and~(d),
in diagram (e) one encounters a factor $2\,\tr(t^a U_{{\bm z}_1} t^b U_{{\bm
    z}_2}^\dagger)$. The distinction between the two, however, is lost if the
transverse separation of the $q\Bar q$--pair (or the invariant mass of the
pair) becomes small: in this limit the $q\Bar q$--pair in the final state
becomes indistinguishable from a gluon.  This (by necessity) is mirrored in
the eikonal structure associated with the diagrams:
\begin{equation}
  \label{eq:eikonal-local-lim}
  \lim\limits_{{\bm z}_1,{\bm z}_2\to {\bm z}} 2\tr(t^a U_{{\bm z}_1} t^b
  U_{{\bm z}_2}^\dagger) = U_{\bm z}^{a b}
  \ .
\end{equation}
Therefore, while in the virtual diagrams running--coupling corrections can be
reconstructed solely from bubble sums as in Fig.~\ref{fig:NLO-diagrams} (c),
in the real--emission diagrams these corrections are associated with both
Fig.~\ref{fig:NLO-diagrams}~(d) and the local limit of~(e).  However, real and
virtual contributions are related by the requirement of probability
conservation: in the absence of interactions the dressed virtual terms must
cancel the dressed real contributions exactly. In the following we show that
it is possible to generalize the LO calculation
of~Sec.~\ref{sec:evolution-equations} in a way that manifestly implements all
these requirements. Specifically, the separation prescription between the
running--coupling corrections that we compute and
the $q\bar{q}$ production channel that we neglect, simply amounts to the replacement of
$2\,\tr(t^a U_{{\bm z}_1} t^b  U_{{\bm z}_2}^\dagger)$ of diagram (e)
by its local limit of (\ref{eq:eikonal-local-lim}), namely
$U_{\bm z}^{a b}$. As announced, this is automatically realized upon using
the dressed propagator in the background field to compute the diagrams of
Eq.~\eqref{eq:JIMWLK-LO-diagram-cont}.

The main ingredient in the LO calculation in~\cite{Weigert:2000gi} is the
propagator of the fluctuation $\delta A$ in the presence of the background
field $b^+$. The calculation sketched in Sec.~\ref{sec:evolution-equations}
was performed in the $A^-=0$ axial gauge in order to reduce the number of
diagrams that contribute. These diagrams are shown in
Eq.~\eqref{eq:JIMWLK-LO-diagram-cont}. The eikonal nature of the interaction
in these diagrams implies that one only needs the ``$++$'' component of the
fluctuation propagator, $\langle \delta A^+_x \delta A^+_y \rangle^{a b}$.
This remains true for the dressed propagator.

At LO, the gluon propagator can be expressed (see
Eq.~(\ref{eq:prop-int-leading}) below) in terms of an external tensor
structure and the propagator~$i\,G^{a b}_0(x',y')$ of a massless scalar field
in the adjoint color representation, propagating through the target
field~\cite{McLerran:1994vd}.  Our generalization will involve its massive
counterpart $i\,G^{a b}_m(x',y')$.  Since the structure of the propagator has
been discussed extensively in the
literature~\cite{McLerran:1994vd,Hebecker:1998kv}, we only describe here the
key ingredients that will be needed to apply the dispersive method and
re-derive the JIMWLK equation with running coupling.

\subsubsection*{Massive scalar propagator in the background field}

The starting point is a spectral representation of the massive scalar
propagator
\begin{equation}
  \label{eq:G-m-disp-def}
  G^{a b}_m(x,y) := \int\frac{d^4k}{(2\pi)^2} \frac{1}{k^2-m^2+i0}
  \
  [\phi_{k}(x)\phi^\dagger_{k}(y)]^{a b}\,,
\end{equation}
where we have combined a spectral integral over
the virtuality $\int d k^2$, and the sum over all states at fixed virtuality,
$\int {d^2{{k}_{\perp}}\, dk^-}/{ (2 k^+)}$,
into one 4-dimensional integral ${\int d^4k}$.

Here $\phi_{k}^{a c}(x)$ are the solutions of the Klein-Gordon equation at
virtuality $k^2=2k^+k^- -{\bm k}^2$ in the background field~$b^+$ of
Eq.~\eqref{eq:bgfield}:
\begin{equation}
  \label{eq:K-G-off-shell}
  (-D[b]^2-k^2)^{a b} \phi_{k}^{b c}(x) =0
  \ ,
\end{equation}
where $a$ and $b$ are color indices and $c$ is a basis label.
$D_\mu[b]= \partial_\mu -i g b_\mu$ is the (adjoint) covariant
derivative in the background field~\eqref{eq:bgfield}.
Explicitly,
\begin{align}
  \label{eq:phi-bg-sol}
  \phi^{a b}_k(x) = \frac{1}{(2\pi)^2} \int d^4p \,e^{-i p\cdot x} \,\delta\left(p^2-k^2\right)\,
  \delta\left(1-\frac{p^-}{k^-}\right) \ \int d^2z\ e^{-i(\bm{p}-\bm{k})\bm{z}}
  [U^{-1}_{x^-,-\infty}(\bm{z})]^{a b} \, ,
\end{align}
reflecting the fact that the interaction with the background field can
transfer \emph{transverse} momentum from the target, while it cannot modify
the conserved $k^-$ component nor the virtuality $k^2$.  The color structure
in (\ref{eq:phi-bg-sol}) is carried by the adjoint eikonal factor
$[U^{-1}_{x^-,y^-}(\bm{z})]^{a b}$ defined by
\begin{equation}
  \label{eq:U-def-one-side}
  [U^{-1}_{x^-,y^-}(\bm{z})]^{a b} :=
  \left[ {\sf P}\exp\left\{
    ig \int\limits_{y^-}^{x^-} dz^- \delta(z^-) \beta(\bm{z})
    \right\}\right]^{a b}
  \ .
\end{equation}

A straightforward calculation using~(\ref{eq:phi-bg-sol})
in~(\ref{eq:G-m-disp-def}) yields:
\begin{align}
  \label{eq:KGpropexpl}
  G^{ab}_m(x,y) = & \int\!\! \frac{ dk^-}{2k^- (2\pi)^3}
  \left[\theta(x^--y^-)\theta(k^-) - \theta(y^--x^-)\theta(-k^-)\right]
  \int\!\! {d^2p_\perp d^2q_\perp } \nonumber \\ & \hspace{1cm} \times \left[
    e^{-i p\cdot x} \int\! \frac{ d^2z_\perp}{(2\pi)^2}\
    e^{-i(\bm{p}-\bm{q})\bm{z}} [U^{-1}_{x^-,y^-}(\bm{z})]^{a b} e^{i
      q\cdot y} \right] \ ,
\end{align}
where the 4-momenta $p$ and $q$ obey the constraints
\[
{p}^+ 
= \frac{{\bm{p}}^2+m^2}{ 2k^-}\,;\qquad {q}^+= \frac{{\bm{q}}^2+m^2}{ 2k^-}\,;\qquad
{p}^-={q}^-=k^-.
\]
Eq.~\eqref{eq:KGpropexpl} differs from the free massive propagator by the
Wilson line (\ref{eq:U-def-one-side}) that represents the interaction with the target field.
Since the interaction in (\ref{eq:U-def-one-side}) is localized at $z^-=0$, this Wilson line
reduces to $U_{\bm{z}}$ (defined in (\ref{eq:U-via-b})) if
$x^->0>y^-$, to $U^\dagger_{\bm{z}}$ if $x^-<0<y^-$ and to~$1$ otherwise.
This implies~\cite{McLerran:1994vd,Hebecker:1998kv}
free propagation\footnote{With $U=1$ there is no obstacle to performing the $\bm z$
integration in~\eqref{eq:KGpropexpl}.} when the endpoints $x^-$ and $y^-$
are both on the same side of the $z^-=0$
hyperplane, while if they lie on opposite sides
one encounters a three--step process: free
propagation from the initial point onto the hyperplane, a current interaction
with the background field, and free propagation again from there to the
endpoint. Only in the latter case does the background field induce a change
of the transverse momentum in the propagator. It is important to note that
these features carry over to the fluctuations propagator $\langle \delta A^+_x
\delta A^+_y \rangle^{a b}$ in both the massless and the massive case.

\subsubsection*{Massive gauge--field propagator in the background field}

The relevant, ``++'' component of the fluctuation propagator at LO is
expressed in terms of the scalar--field propagator $i\,G^{a b}_0(x',y')$ as
\begin{equation}
  \label{eq:prop-int-leading}
  \langle \delta A^+_x \delta A^+_y\rangle^{a b} =
  - \left[\frac{1}{{\partial^-}}\right]_{x x'}  \left\{
    \minus
    D^2[B]_{x'}  i\, G^{a b}_0(x',y')
    \,+\,
    \overrightarrow{\partial^j_{x'}}
    i\, G^{a b}_0(x',y')
    \overleftarrow{\partial^j_{y'}}
  \right\}
  \left[\frac{1}{{\partial^-}}\right]_{y' y}
  \ .
\end{equation}
Here an integration convention over 4-vector coordinates $x'$ and $y'$ is
implied.
Note that one may simplify the first term in this expression using the defining
equation for $G_0$, namely
\[
D^2[b]_{x'} \, G^{a b}_0(x',y') = \delta^{(4)}(x'-y').
\]
We have chosen not to do this since $G^{a b}_0(x',y')$ represents the
propagating particle pole that will be affected by the dressing,
while the remainder of the expression will not change. To illustrate that
this is indeed the right procedure let us connect back to the free case,
or alternatively restrict ourselves to virtual diagrams such as
Fig.~\ref{fig:NLO-diagrams} (c), by replacing
\begin{align}
  \label{eq:to-int-repl}
  D^2[b]^{a b} \longrightarrow  \Box_{x'}
\hspace{2cm}
i G^{a b}_0(x,y) \longrightarrow
  \frac{i}{\Box}(x,y)\ \delta^{a b}
\end{align}
to obtain
\begin{equation}
  \label{eq:prop-free-leading}
  \left.\langle \delta A^+_x \delta A^+_y\rangle^{a b}\right\vert_{\rm free} =
  - \left[\frac{1}{{\partial^-}}\right]_{x x'}  \left\{
    \minus
    \Box_{x'} \Big[\frac{i}{\Box}\Big]_{x' y'}
    \,+\,
    \overrightarrow{\partial^j_{x'}}
    \Big[\frac{i}{\Box}\Big]_{x' y'}
    \overleftarrow{\partial^j_{y'}}
  \right\}
  \left[\frac{1}{{\partial^-}}\right]_{y' y} \delta^{a b}.
\end{equation}
This coordinate--space expression may be readily Fourier--transformed and
shown to coincide with the ``++'' component of the free axial--gauge
propagator\footnote{For our purposes the gauge vector $n$ projects out the
  minus component of a vector: $n\cdot a=a^-$.}
of~\eqref{eq:resummed-propstructure-free-ax}, which is
$({1}/{k^2})\,{2k^+}/{k^-}$.  One can in fact relate the two terms
in~\eqref{eq:prop-free-leading} respectively, to the contributions
of longitudinal and transverse polarizations in the conventions of
light--cone perturbation theory. To this end let us define
transverse polarizations by
\[
{\epsilon_T}^{\mu\,\lambda} := \left(\frac{{\bm k}\cdot{\bm
\epsilon}_T^\lambda}{k^-}, {\bm
    \epsilon}_T^\lambda,0\right) \qquad \text{based on}\,\,\,{\bm
  \epsilon}_T^\lambda:=\left(1,i(-1)^{\lambda+1}\right)/\sqrt{2}
\quad \text{with}\,\, \lambda=1,2
\ .
\]
Here we used the notation $(+,\perp,-)^\mu$ for the components of a
four--vector.  We may then recast
Eq.~\eqref{eq:resummed-propstructure-free-ax} (before dressing the propagator),
for our light--cone gauge (where $n^2=0$), as
\begin{align}
  \label{eq:transv-long-pol}
  \frac1{k^2}\left(g_{\mu\nu}-\frac{n^\mu k^\nu + k^\mu n^\nu}{k \cdot n}\right)  =
                   - \frac{n^\mu n^\nu}{(k\cdot n)^2}\ k^2 \frac1{k^2}
                   -\sum\limits_{\lambda=1,2} \frac{ {\epsilon_T^*}^{\mu\,\lambda}
                   {\epsilon_T}^{\nu\,\lambda} }{k^2}.
\end{align}
This uniquely identifies the second term in~\eqref{eq:prop-free-leading}
with transverse polarizations while
the first term is related to longitudinal contributions.

In~\eqref{eq:prop-free-leading} the factors $\frac{i}{\Box}(x,y)$ correspond
to the propagating particle pole $1/k^2$
in~\eqref{eq:resummed-propstructure-free-ax}, which get multiplied by
${1}/({1+\Pi(k^2)})$ upon dressing the gluon.  The point of separating the
simple tensor structure of $({1}/{k^2})\,{2k^+}/{k^-}$
by polarizations into two
terms as in~\eqref{eq:prop-int-leading} and~\eqref{eq:prop-free-leading} is that the
separation helps to extract the leading small--$x$ contribution: when used in
the diagrams of Eq.~\eqref{eq:JIMWLK-LO-diagram-cont}, only the second term
contributes a logarithm in $1/x$, while the first term is subleading.

Eq.~\eqref{eq:prop-free-leading} offers a straightforward route to
apply the dispersive method for {\em virtual} diagrams such as
Fig.~\ref{fig:NLO-diagrams}~(c).
As explained above, in this case the dressed propagator does not
interact with the target field, so the standard dispersive method
described in Sec.~\ref{sec:runn-coupl-disp} directly applies.
Following Eq.~(\ref{eq:massive_gluon_replacement}) one replaces
$1/k^2$ by $1/(k^2-m^2)$ or, in coordinate
language \[\Big[\frac{i}{\Box}\Big]_{x' y'} \to
\Big[\frac{i}{\Box+m^2}\Big]_{x' y'}.\]
After making this replacement in (\ref{eq:prop-free-leading}) the first term
may be further split according to
\begin{equation}
  \label{eq:firsttermsplit}
  \minus
  \Box_{x'}\Big[\frac{i}{\Box+m^2}\Big]_{x' y'}
  =\,\minus
  \,i\delta^{(4)}_{x' y'}
  \,\plus
  \, m^2 \Big[\frac{i}{\Box+m^2}\Big]_{x' y'}
  \ ,
\end{equation}
so the massive propagator finally takes the form
\begin{equation}
  \label{eq:prop-free-disp}
  \left.\langle \delta A^+_x \delta A^+_y\rangle_m^{a b}\right\vert_{\rm free} =
  - \left[\frac{1}{{\partial^-}}\right]_{x x'}  \left\{
    \,\minus
    \,i\delta^{(4)}_{x' y'}
    \,\plus
    \, m^2 \Big[\frac{i}{\Box+m^2}\Big]_{x' y'}
    \,+\,
    \overrightarrow{\partial^j_{x'}}
    \Big[\frac{i}{\Box+m^2}\Big]_{x' y'}
    \overleftarrow{\partial^j_{y'}}
  \right\}
  \left[\frac{1}{{\partial^-}}\right]_{y' y} \ \delta^{a b}.
\end{equation}
As we shall see below, the $\delta$-function term does not generate a
logarithm in $1/x$, while the other two terms do, and will therefore be
relevant for the JIMWLK equation.

The dispersive representation of the dressing was derived in the free case and
it therefore directly applies only to~\eqref{eq:prop-free-disp} and thus to
the resummation of running--coupling corrections in the virtual diagrams in
the JIMWLK Hamiltonian.  Nevertheless, a similar procedure can be applied to
real--emission diagrams, where it provides a natural \emph{definition} of the
running--coupling contribution, whose separation from the $q\bar{q}$-pair
production channel is a priori ambiguous. To this end we construct the massive
propagator in the background field by reversing the
replacements~\eqref{eq:to-int-repl} in~\eqref{eq:prop-free-disp}:
\begin{equation}
  \label{eq:prop-int-disp}
  \langle \delta A^+_x \delta A^+_y\rangle_m^{a b} =
  - \left[\frac{1}{{\partial^-}}\right]_{x x'}  \left\{
    \,\minus
    \,i\delta^{(4)}_{x' y'}
    \,\plus
    \,
    m^2\ i\,G^{a b}_m(x',y')
    +
    \overrightarrow{\partial^j_{x'}}
    i\, G^{a b}_m(x',y')
    \overleftarrow{\partial^j_{y'}}
  \right\}
  \left[\frac{1}{{\partial^-}}\right]_{y' y}.
\end{equation}
By using this propagator  in \emph{both}
real and virtual diagrams, the real--virtual cancellation mechanism is in
place, just as at LO. Thus, probability conservation is guaranteed.  It is by
this fundamental principal that the dressing of the \emph{virtual} diagrams by
fermion--loop insertions, which is uniquely determined by
(\ref{eq:prop-free-disp}) using the dispersive method, essentially dictates
the structure of running--coupling corrections in the JIMWLK kernel as a
whole.

\subsubsection*{Generalization of the dispersive approach}

To incorporate running--coupling corrections in the derivation of
the JIMWLK equation, we
will modify the fluctuation propagator in analogy with Eq.~(\ref{R}),
namely
\begin{align}
  \label{eq:deltaA-disp-replacement}
  \frac{\alpha_s}{\pi} \langle \delta A^+_x \delta A^+_y\rangle^{a b} \to
  \frac{1}{\beta_0}\int\limits_0^\infty \frac{dm^2}{m^2} \rho_V(m^2)
  \Big[\langle \delta A^+_x \delta A^+_y\rangle^{a b}_m -\langle \delta A^+_x
  \delta A^+_y\rangle^{a b}_0\Big] \ ,
\end{align}
where $\langle \delta A^+_x \delta A^+_y\rangle_m^{a b}$ is given in (\ref{eq:prop-int-disp}).
Note this modification can be traced back to a substitution in the scalar propagator
\begin{eqnarray}
  \label{eq:G-running-1}
  \frac{\alpha_s}{\pi} G^{a b}_0(x,y)
  =  \frac{\alpha_s}{\pi}\int\frac{d^4k}{(2\pi)^4} \frac{\,\,[\phi_{k}(x)\phi^\dagger_{k}(y)]^{a b}}{k^2+i0}
  \,
  & \to &
  \int\frac{d^4k}{(2\pi)^4}
\frac{\alpha_s^V(-k^2-i0)}{\pi}\,
  \frac{\,\,[\phi_{k}(x)\phi^\dagger_{k}(y)]^{a b}}{k^2+i0}
\\
  &=&
  \frac{1}{\beta_0}\int\limits_0^\infty \frac{dm^2}{m^2} \rho_V(m^2)
  \Big[G^{a b}_m(x,y) -G^{a b}_0(x,y)\Big]\nonumber
  \ ,
\end{eqnarray}
where in the final expression we used the dispersive representation of the
running coupling~\eqref{eq:alpha-via-rho}. Restoring the tensor structure then
yields~\eqref{eq:deltaA-disp-replacement}.

As explained above, our procedure, which uses the
propagator~(\ref{eq:prop-int-disp}) for both virtual and
real--emission diagrams guarantees probability conservation. Its
diagrammatic interpretation at the level of fermion loops can be
read off Fig.~\ref{fig:NLO-diagrams}: in the absence of interaction,
(the limit $U\to 1$), the sum of diagrams (d) and (e) reduces to
diagram (c) and their cancellation is complete; as the interaction
is turned on they become distinct and evolution takes place.  At the
same time a new channel, the production of a $q\bar{q}$ pair in
diagram (e), opens up.  Our procedure uses the leading--$N_f$
corrections in the virtual diagrams to identify (define)
running--coupling corrections.  This amounts to a specific
separation of the leading--$N_f$ corrections in the real--emission
diagrams into ones that are part of the running coupling and ones
that constitute the new production channel of diagram (e). By virtue
of Eq.  (\ref{eq:eikonal-local-lim}) the local limit of diagram (e)
is included in the running--coupling contribution.  As a
consequence, real--virtual cancellation holds {\em separately} for
the running--coupling corrections, and the remainder, namely
the new $q\Bar q$ production channel can be computed separately: it is
a well--defined NLO contribution to the kernel that is not associated with
the running coupling. This is confirmed by an
explicit calculation of the diagrams using light--cone perturbation
theory~\cite{HW-YK:2006}.

Finally, let us briefly comment on the separate terms in
Eq.~\eqref{eq:prop-int-disp}.  We begin by observing that, as in the
LO case, the first term in~\eqref{eq:prop-int-disp}, $\minus
\,i\delta^{(4)}_{x' y'}$, is suppressed at large $p^-$ and will
therefore not contribute to the evolution equations.  The two terms
containing $G_m$, however, will generate $\partial^-$ contributions
in the numerator, and will therefore contribute to
logarithmically--enhanced terms at small $x$ through $\int
{dp^-}/{p^-}=\ln(1/x)$.  Next, we note that the two surviving terms
are very different in nature: the last term is already present in
the LO fixed--coupling calculation, while the middle term does not
contribute there, as it vanishes in the $m\to 0$ limit. It starts
contributing at NLO.  This is in keeping with the different origin
of the terms: the transverse partial derivatives in the last term
reflect the Weizs\"acker-Williams field structure of the LO emission
kernel~\eqref{Kdef} that is entirely driven by transverse
polarizations. The middle term, in contrast, arises entirely from
longitudinal polarizations that are absent at leading order. As such
it has an entirely new dependence on the transverse coordinates not
present at LO.  The corresponding diagrammatic calculation
in~\cite{HW-YK:2006} confirms this structure and the association of
the terms with transverse and longitudinal polarizations; the latter
enter there (starting at NLO) via instantaneous contributions.

\subsection{Application to JIMWLK}
\label{sec:disp-runn-pres}

The dispersive method introduces running--coupling corrections by performing a
direct replacement of the gluon propagator in the {\em leading--order
  calculation} according to~\eqref{eq:deltaA-disp-replacement}.  Therefore,
the generalization of the JIMWLK equation to running coupling essentially
constitutes of recalculating the LO diagrams
of~\eqref{eq:JIMWLK-LO-diagram-cont} with the massive propagator of
(\ref{eq:prop-int-disp}) instead of that of (\ref{eq:prop-int-leading}).  In
the following we first present an explicit calculation of one of the diagrams,
and then generalize and derive the running--coupling corrections to the
Hamiltonian ${\cal H}[U]$ in (\ref{eq:JIMWLK-Hamiltonian-LO}) .

Let us first recapitulate the ingredients: in
(\ref{eq:JIMWLK-LO-diagram-cont}) we consider the calculation of the leading
logarithmic correction to the propagation of a fast moving dipole, a $q\Bar
q$-pair. As explained in Sec.~\ref{sec:jimwlk-equation}, this object can be
represented as a product of two Wilson lines along the classical trajectories
of the quarks.  We take the trajectories to lie along the minus light--cone
direction at $x^+=0$. The pair is then characterized by the transverse
locations $\bm{x}$ and $\bm{y}$ of the quark and antiquark, respectively.

The first diagram\footnote{This quantity was called $\Bar\chi^{q \Bar
    q}_{\bm{x y}}$ in~\cite{Weigert:2000gi}.}
in~\eqref{eq:JIMWLK-LO-diagram-cont} contains all the information we need: the
other two can be deduced from it.  It describes the leading $\ln(1/x)$
contribution to the exchange of a gluon between the two quarks in the presence
of Weizs\"acker-Williams background field.  The result
(see~\cite{Balitsky:1996ub,Weigert:2000gi}) involves the fluctuation
propagator and the Wilson lines that represent the external quark and
antiquark:
\begin{align}
  \label{eq:example-standard}
  \parbox{2cm}{\includegraphics[height=2.0cm]{chiqqb}}
  = \, & g^2 \int d x^-\, d y^- \big\langle \delta A^+_x \delta
  A^+_y\big\rangle_0^{a b} \notag \\[-.6cm] & \hspace{.5cm}\times \Big(
  \theta(-x^-) U_{\bm{x}} t^a +\theta(x^-) t^a U_{\bm{x}} \Big) \otimes \Big(
  \theta(y^-) U^\dagger_{\bm{y}} t^b + \theta(-y^-) t^b U^\dagger_{\bm{y}}
  \Big),
   \notag 
  \intertext{which, after some algebra involving the fluctuation
    propagator~\eqref{eq:prop-int-leading} (see below) may be recast as} = &
  -\frac{\alpha_s}{\pi^2} \ln(1/x) \int\!\! d^2\bm{z}\, {\cal K}_{\bm{x z y} }
  \Big( [U]^{a b}_{\bm{z}}\ t^a U_{\bm{x}} \otimes t^b
  U^\dagger_{\bm{y}} + [U^\dagger]^{a b}_{\bm{z}}\ U_{\bm{x}} t^a
  \otimes U^\dagger_{\bm{y}} t^b \notag \\ & \hspace{4cm} - \delta^{a b} \big[
  U_{\bm{x}} t^a \otimes t^b U^\dagger_{\bm{y}} +t^a U_{\bm{x}}\otimes
  U^\dagger_{\bm{y}} t^b \big] \Big).
\end{align}
The four terms on the r.h.s are in one--to--one correspondence with
the four diagrams in Eq.~\eqref{eq:interaction-diagrams}, and
${\cal K}$ is a purely two--dimensional scale--invariant kernel, given by
\begin{align}
  \label{Kdef}
  {\cal K}_{\bm{x z y}} = -\partial_x^j \partial_y^j \int
  \frac{d^2p\,d^2q}{(2\pi)^2} \frac{{\rm
      e}^{i\bm{p}(\bm{x}-\bm{z})}}{\bm{p}^2}\, \frac{{\rm
      e}^{i\bm{q}(\bm{z}-\bm{y})}}{\bm{q}^2} = \frac{(\bm{x}-\bm{z})\cdot
    (\bm{z}-\bm{y})}{%
    (\bm{x}-\bm{z})^2 (\bm{z}-\bm{y})^2}
  =  -\frac{ {\vdl}\cdot  {\vdr}}{\,{r_1}^2 \,{r_2}^2}
  = \frac12
  \frac{ \rr^2-(\dl^2+\dr^2)}{\,{r_1}^2 \,{r_2}^2}
\ ,
\end{align}
where we used the notation for ``parent--'' and ``daughter--dipoles''
introduced in~\eqref{eq:distshort}. The last formula shows explicitly that
this leading order kernel changes sign at a circle whose diameter is defined
by the ``parent.''  To arrive at the final expression in
(\ref{eq:example-standard}) with the kernel of (\ref{Kdef}) we have performed
the $x^-$ and $y^-$ integrals, using~\eqref{eq:prop-int-leading} with the
explicit formula for $G^{ab}_0(x,y)$, given by \eqref{eq:KGpropexpl} with
$m=0$.  These integrations give rise to factors $2p^-/\bm{p}^2$ and
$2q^-/\bm{q}^2$, respectively. The minus momentum components in the numerator
compensate the explicit  $\partial^-$ in the denominator
in~\eqref{eq:prop-int-leading}, while the additional $\int_0^\infty
{dk^-}/{k^-}$ in $G^{ab}_0(x,y)$ leads to the factor $\ln(1/x)$ upon imposing
cutoffs in~$x$. Note that the first term in the curly brackets
in~\eqref{eq:prop-int-leading} reduces to $\minus \,i\delta^{(4)}_{x' y'}$ and
it does not contribute to the equation since it is suppressed at large $p^-$.

Eqs. (\ref{eq:example-standard}) and (\ref{Kdef}) summarize the result for
one of the diagrams in~\eqref{eq:JIMWLK-LO-diagram-cont}
used in deriving the JIMWLK equation~\cite{Weigert:2000gi}.
As announced we will now use this diagram to
demonstrate how the JIMWLK equation can be promoted to running coupling by means of the
dispersive method. To this end we use the substitution~\eqref{eq:deltaA-disp-replacement}
obtaining
\begin{align}
  \label{eq:dispersive-replacement-direct}
  \frac{\alpha_s}{\pi}\begin{minipage}[m]{2.0cm}
    \includegraphics[height=2.0cm]{chiqqb}
  \end{minipage}
  \to \frac{1}{\beta_0}\int_0^{\infty}\frac{dm^2}{m^2} \rho_V(m^2) \left(
    \begin{minipage}[m]{2.0cm}
      \includegraphics[height=2.0cm]{chiqqb}
    \end{minipage}(m) -\begin{minipage}[m]{2.0cm}
      \includegraphics[height=2.0cm]{chiqqb}
    \end{minipage}(m=0) \right).
\end{align}
Recall that $\rho_V(m^2)$ is a universal object, the discontinuity of
the coupling on the time--like axis, defined in (\ref{eq:discontinuity}).
To compute running--coupling corrections using (\ref{eq:dispersive-replacement-direct})
we need to repeat the calculation of the diagram, but now with the
massive propagator of (\ref{eq:prop-int-disp}) instead of the massless
one~\eqref{eq:prop-int-leading}.
Doing this we find that Eq. (\ref{eq:example-standard}) is unmodified while
${\cal K}_{\bm{x z y}}$ is replaced by its massive counterpart:
\begin{align}
  \label{eq:Kmassive} {\cal K}_{\bm{x z y}} \to {\cal K}_{\bm{x z y}}^m =
-\left(\partial_x^j \partial_y^j \plus m^2\right)
\int
\frac{d^2p\,d^2q}{(2\pi)^2}
  \frac{{\rm e}^{i\bm{p}(\bm{x}-\bm{z})}}{\bm{p}^2+m^2}\, \frac{{\rm
      e}^{i\bm{q}(\bm{z}-\bm{y})}}{\bm{q}^2+m^2},
\end{align}
where the two terms $\partial_x^j \partial_y^j$ and $m^2$ originate in the
last and the middle terms in Eq. (\ref{eq:prop-int-disp}), respectively. As in
the LO case, the $\minus \,i\delta^{(4)}_{x' y'}$ term in Eq.
(\ref{eq:prop-int-disp}) does not contribute in the small--$x$ limit.

Let us now compute the integrals in (\ref{eq:Kmassive}) explicitly.
Denoting the ``daughter dipoles'' respective \emph{lengths} by
$ r_i :=\vert{\vrr}_{i} \vert$ ($i=1,2$) we obtain:
\begin{align}
\label{eq:Kmassive_result}
\begin{split}
{\cal K}_{\bm{x z y}}^m\,= &\,
{\cal K}_{\bm{x z y}}\, \dl m \,K_1(\dl m) \,\,\dr m \,K_1(\dr m)\,
\minus
\,
  m^2\, K_{0}(\dl\, m) K_{0}(\dr\, m)
\end{split}
\end{align}
where
${\cal K}_{\bm{x z y}}$ is the LO kernel of~(\ref{Kdef})
and $K_{0}(x)$ and $K_{1}(x)=-dK_0(x)/dx$ are $K$-Bessel functions. These functions
depend only on the {\rm lengths} of the vectors ${\vrr}_i$, not on the
angles --- a direct consequence of the angular integration in~(\ref{eq:Kmassive}).
Angular dependence appears here \emph{only} through the derivative $\partial_x^j \partial_y^j$,
giving rise to the factor ${\cal K}_{\bm{x z y}}$ in the first term in (\ref{eq:Kmassive_result}), just as at LO.

The next observation is that upon applying the dispersive method to the self--energy
diagrams in Eq.~\eqref{eq:JIMWLK-LO-diagram-cont}, in analogy with
(\ref{eq:dispersive-replacement-direct}), one recovers again the LO result with the kernel
of (\ref{eq:Kmassive}).
Thus, \emph{all} the diagrams share the same massive kernel ${\cal K}_{\bm{x z y}}^m$
of Eq.~\eqref{eq:Kmassive_result}. Since also the dispersive integral of
Eq.~\eqref{eq:dispersive-replacement-direct}
is common to all three diagrams, running--coupling corrections to the JIMWLK Hamiltonian
${\cal H}[U]$ of (\ref{eq:JIMWLK-Hamiltonian-LO}) appear through the replacement:
\begin{align}
  \label{eq:K-replacement-direct}
  \frac{\alpha_s}{\pi}{\cal K}_{\bm{x z y}}\,\to\,
  {\cal K}_{\bm{x z y}} \frac{1}{\beta_0}\int_0^{\infty}\frac{dm^2}{m^2}
  \rho_V(m^2) \Big[{\cal F} (\vdl m,\vdr m)-1\Big] \ ,
\end{align}
where we defined the ``characteristic function'' ${\cal F} (\vdl m ,\vdr m)$,
following the general discussion in Sec.~\ref{sec:runn-coupl-disp},
by extracting the LO kernel:
\begin{align}
\begin{split}
\label{Fdef}
{\cal K}_{\bm{x z y}}^m\,=:& \,{\cal K}_{\bm{x z y}}\ {\cal F} (\vdl m ,\vdr m)=
{\cal K}_{\bm{x z y}}\Big[ {\cal F}^T(\vdl m ,\vdr m)\,+\,{\cal F}^L(\vdl m ,\vdr m)\Big],\\
&{\cal F}^T (\vdl m ,\vdr m) \,= \,\dl m \,K_1(\dl m) \,\,\dr m \,K_1(\dr m)\, ,\\
&{\cal F}^L (\vdl m ,\vdr m) \,=\,
 m^2\,\frac{\,{r_1}^2 \,{r_2}^2}{ {\vdl}\cdot  {\vdr}} \, K_{0}(\dl\, m) K_{0}(\dr\, m)\,,
\end{split}
\end{align}
and used the fact that $\lim\limits_{m\to 0} {\cal F} (\vdl
m,\vdr m)=1$.
Here ${\cal F}^T$ and ${\cal F}^L$ correspond to the first and second terms in
(\ref{eq:Kmassive_result}), which in turn originate in the $\partial_x^j \partial_y^j$
and $m^2$  terms is (\ref{eq:Kmassive}), respectively.
As mentioned in Sec.~\ref{sec:dress-prop-backgr} (see Eq.~(\ref{eq:prop-int-disp}) there)
these terms are associated with transverse~($T$)
and longitudinal~($L$) gluons, respectively, which explains the notation in (\ref{Fdef}).
As we see, these two contributions to ${\cal K}_{\bm{x z y}}^m$
are rather different in nature: the first can be naturally
written as a product of the scale--invariant kernel
${\cal K}_{\bm{xzy}}$ times a function of the lengths of the
``daughter dipoles'', while the second brings in an entirely new structure that does not share the
angular dependence of the LO kernel.

We also introduce a notation for the ``effective charge''
\begin{equation}
  \label{eq:Rdef}
  R(\vdl\Lambda,\vdr\Lambda) :=
  \frac{1}{\beta_0}\int_0^{\infty}\frac{dm^2}{m^2} \rho_V(m^2)
  \Big[{\cal F} (\vdl m,\vdr m)-1\Big] \ ,
\end{equation}
which plays the r\^ole of a scale--dependent ${\alpha_s}/{\pi}$
in all our calculations: the substitution of
\begin{equation}
  \label{eq:new-JIMWLK}
  \frac{\alpha_s}{\pi}{\cal K}_{\bm{x z y}}\to {\cal M}_{\bm{x z y}} :=
  R(\vdl \Lambda,\vdr\Lambda)\ {\cal K}_{\bm{x z y}}
\end{equation}
in~\eqref{eq:JIMWLK-Hamiltonian-LO} promotes all the fixed coupling diagrams
entering the JIMWLK equation to running coupling. The running coupling
corrected JIMWLK Hamiltonian then reads
\begin{align}
  \label{eq:JIMWLK-Hamiltonian-NLO}
  {\cal H}[U] =-\frac{1}{2\pi}\, {\cal M}_{\bm{x z y}}\, \Big[
    U_{\bm z}^{a b}\left(i\Bar\nabla^a_{\bm x}i\nabla^b_{\bm y} +i\nabla^a_{\bm x}
    i\Bar\nabla^b_{\bm y}\right) + \left( i\nabla^a_{\bm x} i\nabla^a_{\bm
      y}+i\Bar\nabla^a_{\bm x} i\Bar\nabla^a_{\bm y}\right) \Big]
\  .
\end{align}
All other ingredients in the JIMWLK equation~\eqref{eq:JIMWLK} and the
definition of correlator averages~\eqref{eq:corrs} remain unchanged.
Eq.~\eqref{eq:new-JIMWLK}, and with it~\eqref{eq:JIMWLK-Hamiltonian-NLO}, is
our key result that fully implements running coupling in the JIMWLK equation.

\subsection{Borel representation of the resummed kernel}
\label{sec:borel-rep-of-resummed-kernel}

To systematically define the perturbative sum and analyze power
corrections in the JIMWLK kernel ${\cal M}_{\bm{x z y}}$, it is convenient to
write a Borel representation of the ``effective charge'' $R$,
in the form of Eq.~\eqref{R_borel_}:
\begin{align}
  \label{eq:K-replacement}
{\cal M}_{\bm{x z y}}={\cal K}_{\bm{x z y}}  \, R(\vdl \Lambda,\vdr \Lambda,)
 = \frac{1}{\beta_0} \int_0^{\infty}du & \ T(u)
  \left(\frac{\mu^2}{\Lambda^2}\right)^{-u} {\cal K}_{\bm{x z y}}\,B(u,\vdl\mu,\vdr\mu),
\end{align}
where the Borel function is related to the characteristic function ${\cal F}$
(see Eq.~\eqref{eq:B-F-relation}) by
\begin{align}
  \label{eq:Bdef}
    B(u,\vdl\mu,\vdr\mu)= &\, -{\rm e}^{\frac53 u} \frac{\sin \pi u}{\pi}\,
    \int_0^{\infty}\frac{dm^2}{m^2} \bigg(\frac{m^2}{\mu^2}\bigg)^{-u}
    \Big[{\cal F} (\vdl m,\vdr m)-1\Big] \ .
\end{align}
Owing to the different nature of the transverse and longitudinal contributions
to ${\cal F}$, it is convenient to deal with them separately also on the level
of the Borel function.  To this end we write
\begin{equation}
B(u,\vdl\mu,\vdr\mu) \,= \, B^T(u,\vdl\mu,\vdr\mu) +
B^L(u,\vdl\mu,\vdr\mu), \label{B_TL}
\end{equation}
corresponding to the two terms in (\ref{Fdef}).
Below we present explicit expressions for these functions in both momentum and
coordinate space.  Using (\ref{eq:Kmassive}) and (\ref{eq:Bdef}) we obtain
the following expressions as Fourier transformations from momentum space:
\begin{subequations}
  \label{eq:mom-space-Borels}
  \begin{align}
    \label{eq:mom-space-trans-res}
    {\cal K}_{\bm{x z y}}\, B^T(u,\vdl\mu,\vdr\mu) = &\, -e^{\frac53 u}
    \frac{\sin\pi u}{\pi} \int \frac{d^2p\,d^2q}{(2\pi)^2} {\text
      e}^{i\bm{p}\cdot\vdl}{\text e}^{\minus i\bm{q}\cdot\vdr} \notag \\
    &\times \int\limits_0^\infty \frac{dm^2}{m^2}
    \left(\frac{m^2}{\mu^2}\right)^{-u} (\minus 1)
    \left[ \frac{{\bm
          p}\cdot{\bm q}}{({\bm p}^2+m^2)({\bm q}^2+m^2)} -\frac{{\bm
          p}\cdot{\bm q}}{{\bm p}^2\, {\bm q}^2} \right] \notag \\ = & \,
    \minus
    e^{\frac53 u} \int \frac{d^2p\,d^2q}{(2\pi)^2}\
    e^{i\bm{p}\cdot\vdl}\ e^{\minus i\bm{q}\cdot\vdr} \ \frac{{\bm
        p}\cdot{\bm q}}{{\bm p}^2\, {\bm q}^2}\ \frac{{\bm
        q}^2\left(\frac{{\bm p}^2}{\mu^2}\right)^{-u} -{\bm
        p}^2\left(\frac{{\bm q}^2}{\mu^2}\right)^{-u}}{{\bm q}^2-{\bm p}^2}
\ ,
    \displaybreak[0]\\
    {\cal K}_{\bm{x z y}}\, B^L(u,\vdl\mu,\vdr\mu) = &\, -e^{\frac53 u}
    \frac{\sin\pi u}{\pi} \int \frac{d^2p\,d^2q}{(2\pi)^2} \
    e^{i\bm{p}\cdot\vdl}\ e^{\minus i\bm{q}\cdot\vdr} \notag \\ &\times
    \int\limits_0^\infty \frac{dm^2}{m^2} \left(\frac{m^2}{\mu^2}\right)^{-u}
    \frac{\minus m^2}{({\bm p}^2+m^2)({\bm q}^2+m^2)}
    \label{eq:mom-space-inst-res}
    \notag \displaybreak[0]\\ = & \minus
    e^{\frac53 u} \int
    \frac{d^2p\,d^2q}{(2\pi)^2}\ e^{i\bm{p}\cdot\vdl}{\text
      e}^{i\bm{q}\cdot\vdr} \ \frac{ \left(\frac{{\bm p}^2}{\mu^2}\right)^{-u}
      - \left(\frac{{\bm q}^2}{\mu^2}\right)^{-u} }{ {\bm p}^2-{\bm q}^2}
\ .
  \end{align}
\end{subequations}
These expression assume a perhaps more familiar form if we restrict ourselves
to one--loop running, where, with $T(u)=1$, expanding under the
Borel integral amounts to replacing the powers of the momenta ${\bm k}={\bm
  p}, {\bm q}$, i.e.  $\left({{\bm k}^2}/{\mu^2}\right)^{-u}$ by the
corresponding geometric series according to
\begin{equation}
  \label{eq:borel-to-geom}
  \left(\frac{{\bm k}^2}{\mu^2}\right)^{-u} \to
  \frac{\alpha_s(\mu^2)}{\pi}
  \frac1{1+\frac{\beta_0\alpha_s(\mu^2)}{\pi}
    \ln\left({{\bm k}^2}/{\mu^2}\right)}
\ .
\end{equation}
In fact, using~\eqref{eq:K-replacement} and~\eqref{eq:borel-to-geom},
the expressions in~\eqref{eq:mom-space-Borels} can be directly mapped
onto the expressions derived diagrammatically in~\cite{HW-YK:2006}.

We note that in the purely virtual case of
Fig.~\ref{fig:NLO-diagrams}~(c) the result simplifies in the expected
manner: there the ${\bm z}$ integral may be performed and, in the absence
of interaction with the target, it sets ${\bm q}={\bm p}$.
Then the {\em sum} of the integrands of transverse and longitudinal
contributions~\eqref{eq:mom-space-Borels} yields
\begin{equation}
  \label{eq:borel-virtual-simplification}
  \underbrace{\frac1{{\bm p}^2} \left(\frac{{\bm p}^2}{\mu^2}\right)^{-u} (1+u)}_{{\rm from}
  \,\, {\cal K}\, B^T}
\
\underbrace{-\frac1{{\bm p}^2} \left(\frac{{\bm p}^2}{\mu^2}\right)^{-u} u}_{{\rm from}
  \,\, {\cal K}\, B^L}
   =
  \frac1{{\bm p}^2} \left(\frac{{\bm p}^2}{\mu^2}\right)^{-u}.
\end{equation}
Via~\eqref{eq:borel-to-geom}, this corresponds to $(1/{\bm
  p^2})\,{\alpha_s({\bm p}^2\,e^{-\frac53})}/{\pi}$, which is the expected
  contribution of a gluon of transverse momentum ${\bm p}^2$
  that does not interact with the background field.
  Note that there is an important cancellation at NLO (and beyond) between the
separate contributions of transverse and longitudinal gluons.
On the level of the dispersive integrals used to define $B^T$ and $B^L$ this
cancellation is even more obvious: for ${\cal K}^m_{\bm{x z y}}$ (c.f.
Eq.~\eqref{eq:Kmassive}) it takes the form
\begin{align}
  \label{eq:mom-sum-trans-long-virt}
  \frac{{\bm p}\cdot{\bm q}+m^2}{({\bm p}^2+m^2)({\bm q}^2+m^2)}
  \xrightarrow{{\bm q}\,\to\,{\bm p}} \frac{1}{{\bm p}^2+m^2},
\end{align}
which has exactly the same interpretation.

Direct Fourier integration of the perturbative sum obtained through
(\ref{eq:borel-to-geom}) is not well defined owing to the Landau pole. In this
respect these all--order expressions are largely symbolic. Correspondingly,
the Borel integrals of both~\eqref{eq:mom-space-trans-res}
and~\eqref{eq:mom-space-inst-res} converge for $u\to\infty$ only when the
momenta are larger than the QCD scale $\Lambda$. Thus, when we exchange the
order of integrations and perform the Borel integration {\em after} the
Fourier transforms, the Borel integral will no longer be unambiguous for any
value of the coordinates.  As we shall see explicitly below, the
coordinate--space Borel function exhibits poles along the integration path, at
positive integer values of $u$.  This reflects ambiguities in summing the
series, which are indicative of power corrections (see the general discussion
in Sec.~\ref{sec:runn-coupl-disp}).  Knowing the location of the poles and the
parametric dependence of their residues on the hard scales involved, one can
estimate the effect of non-perturbative power corrections.  It is the
advantage of the Borel method to expose these power corrections in such a
clear and concise way.

With this in mind we now perform the Fourier transformation of the Borel
functions. The all--order expressions lead to infinite sums that can be
recast in terms of hypergeometric functions. The calculation is done using
integral representations of the Bessel functions. The integrals over the
dispersive mass intertwine the parameter integrations so that it becomes
necessary to use Mellin-Barnes representations to decouple them. The
Mellin-Barnes integrals are done last and lead to infinite sums of residues.
Details of the rather involved algebra are given separately for $B^T$ and
$B^L$ in Appendix~\ref{sec:FT-of-BT} and~\ref{sec:FT-of-BL}, respectively.

To express the results in a compact manner we define
\begin{equation}
  r_>  :=\max\left\{\dl,\dr \right\} \,;\qquad
  r_<  :=\min\left\{\dl,\dr \right\} \,;\qquad
  \xi:= \frac{{r}_<}{{r}_>}.
\end{equation}
We present two versions for each of the results for $B^T$ and $B^L$, as a
series in $\xi^2$ and in $1-\xi^2$, respectively. This is important for
numerical implementations near $\xi=0$ and $\xi=1$, respectively, and will be
used in Sec.~\ref{Sec:numerical} below. In addition, the first form
conveniently displays the singularity structure and the functional form of the
residues, while the second form is convenient for performing an expansion in
$u$, in order to extract the perturbative coefficients.

The coordinate--space result for $B^T$ is:
\begin{subequations}
  \label{eq:BT}
  \begin{align}
    \label{eq:BTa}
    {\cal K}_{\bm{x z y}} \, B^T(u,\vdl\mu,\vdr\mu)
    = & \frac{\vdl\cdot \vdr}{\dl^2\dr^2}
    \left(\frac{4\,e^{-\frac53 }}{r_>^2
        \mu^2}\right)^{-u}\frac{\sin(\pi u)}{\pi}\,
    \bigg\{\Gamma(1-u)\Gamma(-u)\, \nonumber\\ &\vspace*{-30pt}
    +\sum_{k=0}^{\infty} (\xi^2)^{1+k}\,
    \frac{\Gamma(2-u+k)\Gamma(1-u+k)}{\Gamma(1+k)\Gamma(2+k)}\,
    \,\notag\\&\vspace*{-30pt} \times
    \Big(\Psi(2-u+k)+\Psi(1-u+k)-\Psi(2+k)-\Psi(1+k)+\ln \xi^2\Big)\bigg\}
\ ,
    \intertext{which can be recast in terms of a hypergeometric series
      $\sideset{_2}{_1}\F$ using 15.3.11 of~\cite{AbramSteg}}
\begin{split}
    \label{eq:BTb}
    = & -\frac{\vdl\cdot \vdr}{\dl^2\dr^2}
    \left(\frac{4\,e^{-\frac53 }}{r_>^2 \mu^2}\right)^{-u}
    \frac{\sin(\pi u)}{\pi}\,
    \frac{u(1-u)\Gamma^2(-u)\Gamma^2(1-u)}{\Gamma(2-2u)}\\
    &\hspace*{50pt}\times \qFp{2}{1}{1-u,-u}{2-2u}{1-\xi^2}.
    \end{split}
  \end{align}
\end{subequations}
A useful alternative expression to (\ref{eq:BTb}) can be obtained using the
identity:
\begin{equation}
\label{eq:_hyper_identity}
\qFp{2}{1}{1-u,-u}{2-2u}{1-\xi^2}
\,=\,
\xi^2\,\bigg[1\,-\,\frac{1-\xi^2}{1-u}\frac{d}{d\xi^2}\,
\bigg]\,
\ \qFp{2}{1}{1-u,1-u}{2-2u}{1-\xi^2}.
\end{equation}
The result for $B^L$ is:
\begin{subequations}
  \label{eq:BL}
  \begin{align}
    \label{eq:BLa} {\cal K}_{\bm{x z y}} \, B^L(u,\vdl\mu,\vdr\mu) = &
    \minus
    \frac{\sin(\pi u)}{\pi} \frac1{{\bm r}_>^2} \left(\frac{4\,e^{-\frac53}}{
        {\bm r}_>^2 \mu^2 }\right)^{-u} \sum\limits_{k=0}^\infty
    \left(\frac{\Gamma(k+1-u)}{\Gamma(k+1)}\right)^2 \left(\xi^2\right)^k
    \notag \\ & \hspace*{50pt}
\times \Big( \ln(\xi^2) -2\Psi(k+1)+2\Psi(k+1-u)
    \Big)
\ ,
\intertext{which again can be expressed in terms of a hypergeometric
      function, this time using 15.3.10 of~\cite{AbramSteg},}
    \label{eq:BLb}
    = & \plus \frac{\sin(\pi u)}{\pi} \frac{1}{{\bm r}_>^2} \left(
      \frac{4\,e^{-\frac53}}{{\bm r}_>^2\mu^2} \right)^{-u}
    \frac{\Gamma^4(1-u)}{\Gamma(2-2u)}\ \
    \qFp{2}{1}{1-u,1-u}{2-2u}{1-\xi^2}
    \ .
  \end{align}
\end{subequations}
The result is fully symmetric in $\vdl$ and $\vdr$, in particular all
coefficients in an expansion in powers of $u$ (or $\alpha_s$) have
this property.

Our expressions for $B^T$ and $B^L$ can be put to use in several
ways. The first two pertain to obtaining information on the
perturbation theory:
\begin{itemize}
\item \emph{A generating function} for the perturbative
  expansion in powers of $\alpha_s(\mu^2)$ (at a fixed scale). To this end one
  expands the results for the Borel function around $u=0$ and
  then integrates over $u$ in (\ref{eq:K-replacement}) term by term. Assuming one-loop coupling
  ($T(u)=1$) this amounts to the replacement:
\begin{equation}
\label{eq:u_replacement}
  u^n \longrightarrow n!\left(\beta_0\,\frac{\alpha_s(\mu^2)}{\pi}\right)^{n+1}.
\end{equation}

\item Knowing the analytic structure of the Borel function, which has a leading
renormalon singularity at $u=1$, it is clear that the expansion: (1)
does not converge, and (2) it is not Borel summable; \emph{A
definition of the sum} in (\ref{eq:K-replacement}) is required.
Starting from the expansion, a natural prescription is to truncate
the sum at the minimal term. A more systematic regularization, which
has been found useful in several applications in QCD, see
e.g.~\cite{Cacciari:2002xb,Andersen:2005bj,Gardi:1999dq}, is the
Principal Value (PV) of the Borel integral in
(\ref{eq:K-replacement}).
\end{itemize}
In the next section (Sec.~\ref{sec:perturbative-expansions}) we will
perform the perturbative expansion of the JIMWLK kernel, study
numerically its apparent convergence in the first few orders, and
compare it to the PV Borel sum. We also examine the approximation of
the sum by scale setting. These issues are revisited in
Sec.~\ref{sec:BK-numerics} where we examine the effect of the newly
computed higher--order corrections on the evolution in the case of
the BK equation.

The third application goes beyond perturbation theory; we can
extract information on the power corrections from the integrand:
in~\eqref{eq:K-replacement}:
\begin{itemize}
\item The renormalon poles at positive integer $u$ in $B^T$ and $B^L$ can be
  used to infer the parametric form and significance of non-perturbative
  power corrections that are expected to affect the evolution kernel.
\end{itemize}
This is discussed in Sec.~\ref{sec:power-corrections} for a single
evolution step and re-visited in Sec.~\ref{sec:BK-numerics} when
solving the evolution equations numerically.

\section{Perturbation theory}
\label{sec:perturbative-expansions}

Having computed the Borel function $B(u)$ entering
Eq.~(\ref{eq:K-replacement}), we have essentially determined the
expansion coefficients for $R(\vdl\Lambda,\vdr\Lambda)$, and thus
for the JIMWLK evolution kernel, to all orders in the
large--$\beta_0$ limit. To get the coefficients one expands the
expressions for $B^T(u,\vdl\mu,\vdr\mu)$ and
$B^L(u,\vdl\mu,\vdr\mu)$ in
Sec.~\ref{sec:borel-rep-of-resummed-kernel} under the integral in
(\ref{eq:K-replacement}),
\begin{equation}
B^T(u,\vdl\mu,\vdr\mu)= 1+\sum\limits_{n=1}^\infty
    b^T_n(\vdl\mu,\vdr\mu)u^n; \qquad
    B^L(u,\vdl\mu,\vdr\mu)=\sum\limits_{n=1}^\infty
    b^L_n(\vdl\mu,\vdr\mu)u^n
\end{equation}
and integrates over the Borel variable term by term\footnote{Since
we are using one--loop running coupling, this amounts to making the
replacement of Eq.~(\ref{eq:u_replacement}).}, to obtain:
\begin{align}
  \label{eq:pert-exp}
 {\cal M}_{\bm{x z y}} = R(\vdl\Lambda,\vdr\Lambda){\cal K}_{\bm{x z y}} =
 &\, {\cal   K}_{\bm{x z y}}\,\frac{\alpha_s(\mu^2)}{\pi}\,
   \left(1 +\sum\limits_{n=1}^\infty n!
    \left(\beta_0\frac{\alpha_s(\mu^2)}{\pi}\right)^n
    b_n(\vdl\mu,\vdr\mu)\right) \ ,
  \intertext{with } b_n(\vdl\mu,\vdr\mu) := &\,
  b^T_n(\vdl\mu,\vdr\mu)+ b^L_n(\vdl\mu,\vdr\mu) \ .
\end{align}
The purpose of this section is to study the effect of these
perturbative corrections to the JIMWLK evolution kernel.

Explicit expressions for the expansion coefficients will be
presented in Eqs. (\ref{eq:BTexpansion}) and (\ref{eq:BLexpansion})
below. Before looking into the details, let us briefly recall what
one expects on general grounds following the discussion in
Sec.~\ref{sec:runn-coupl-disp}:
\begin{itemize}
\item{} {\bf The all--order sum is renormalization scale invariant:}
  Eq.~(\ref{eq:K-replacement}) is renormalization--scale inavriant.  This
  means that also (\ref{eq:pert-exp}) shares this property, if it is formally
  considered to all orders. However, the choice of the expansion parameter
  would make any finite--order partial sum scale dependent.
\item{} {\bf The series in non-summble owing to infrared renormalons, which
    amount to power--suppressed ambiguities:} Going beyond the level of a
  finite--order partial sum, in an attempt to compute
  $R(\vdl\Lambda,\vdr\Lambda)$ \emph{to all orders}, one finds infrared
  renormalon ambiguities. Let us explain how they arise here: since $B(u)$ in
  (\ref{eq:K-replacement}) has a finite radius of convergence ($u=1$), $b_n$
  in (\ref{eq:pert-exp}) are characterized at high orders by geometrical
  progression with no sign oscillation. With the explicit $n!$ growth in
  (\ref{eq:pert-exp}), it is obvious that the series would not converge, but
  instead reach a minimal term, and then start diverging. An optimal
  perturbative approximation may be defined by truncation at the minimal term.
  This, however, is inconvenient since the truncation order may (and in our
  case does) depend on all the scales in the problem (in our case $\rr^2$,
  $\dl^2$ and $\dr^2$). A more systematic definition of the sum can be made
  using the Borel sum~(\ref{eq:BTexpansion}).  This integral does not exist as
  it stands, since the poles appear at real positive values of $u$, on
  the integration path. One can {\em define} the sum by shifting the contour
  above or below the real axis, or by choosing the Principal Value (PV)
  prescription. The differences between these choices are
  \emph{power--suppressed} in the hard scales: they scale as integer powers of
  $\Lambda^2\rr^2$, $\Lambda^2\dl^2$ and $\Lambda^2\dr^2$ (up to logarithms,
  see Sec.~\ref{sec:power-corrections}). The size of this ambiguity is
  obviously controlled by the residues of $B(u)$ in (\ref{eq:BTexpansion}) and
  can therefore be explicitly computed. It is also similar to the magnitude of
  the minimal term in the series.  Since these ambiguities are expected to
  cancel against non-perturbative power corrections, they provide important
  information on such corrections to the kernel, that are otherwise unknown.
  We will return to discuss these corrections in detail in
  Sec.~\ref{sec:power-corrections}.
\end{itemize}

\begin{figure}[t!]
  \centering
 \begin{minipage}[t]{7.1cm}
    \centering
  \includegraphics[width=7cm]{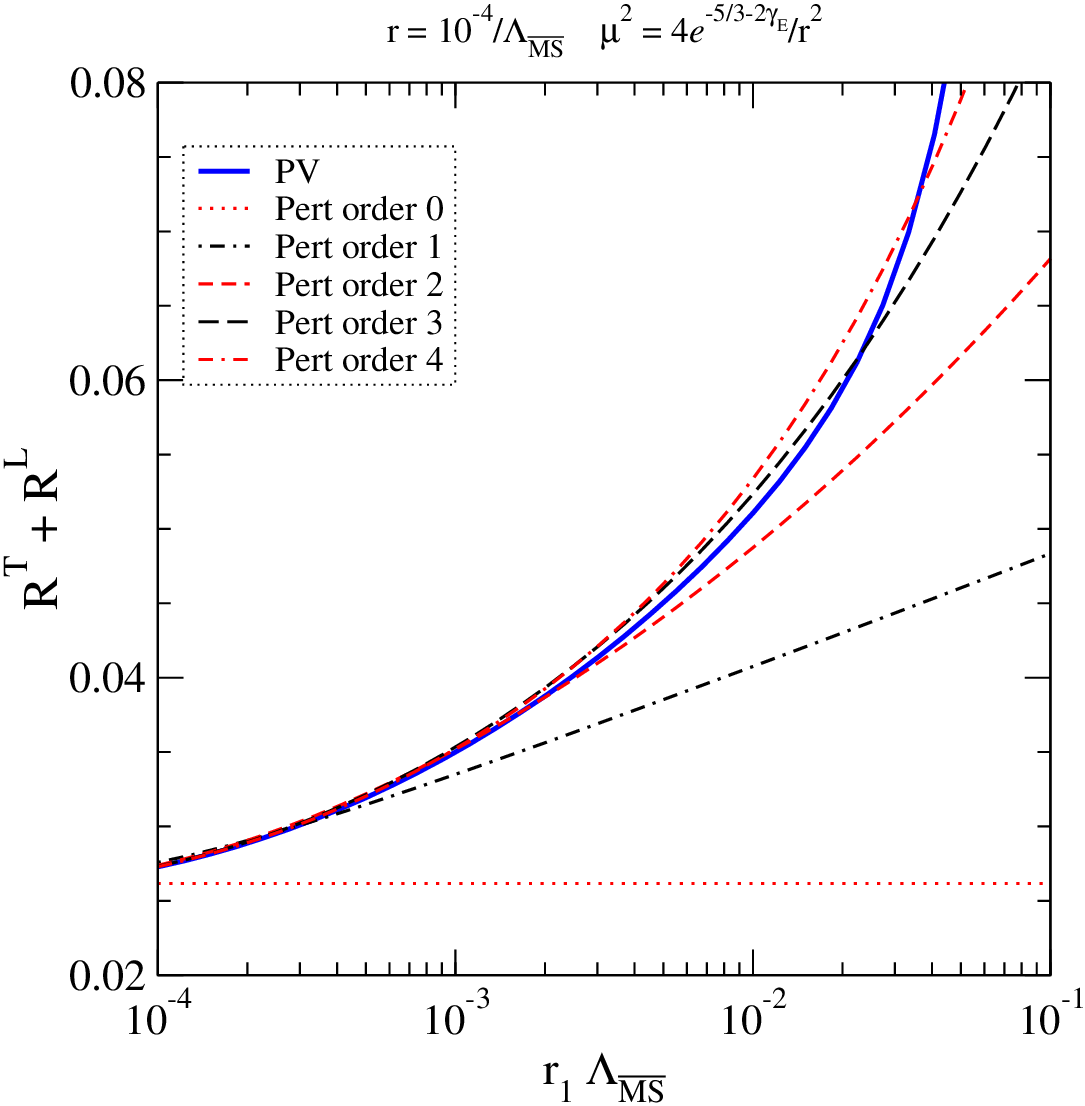}\\ (a)
\end{minipage}
\hfill
 \begin{minipage}[t]{7.1cm}
    \centering
  \includegraphics[width=7cm]{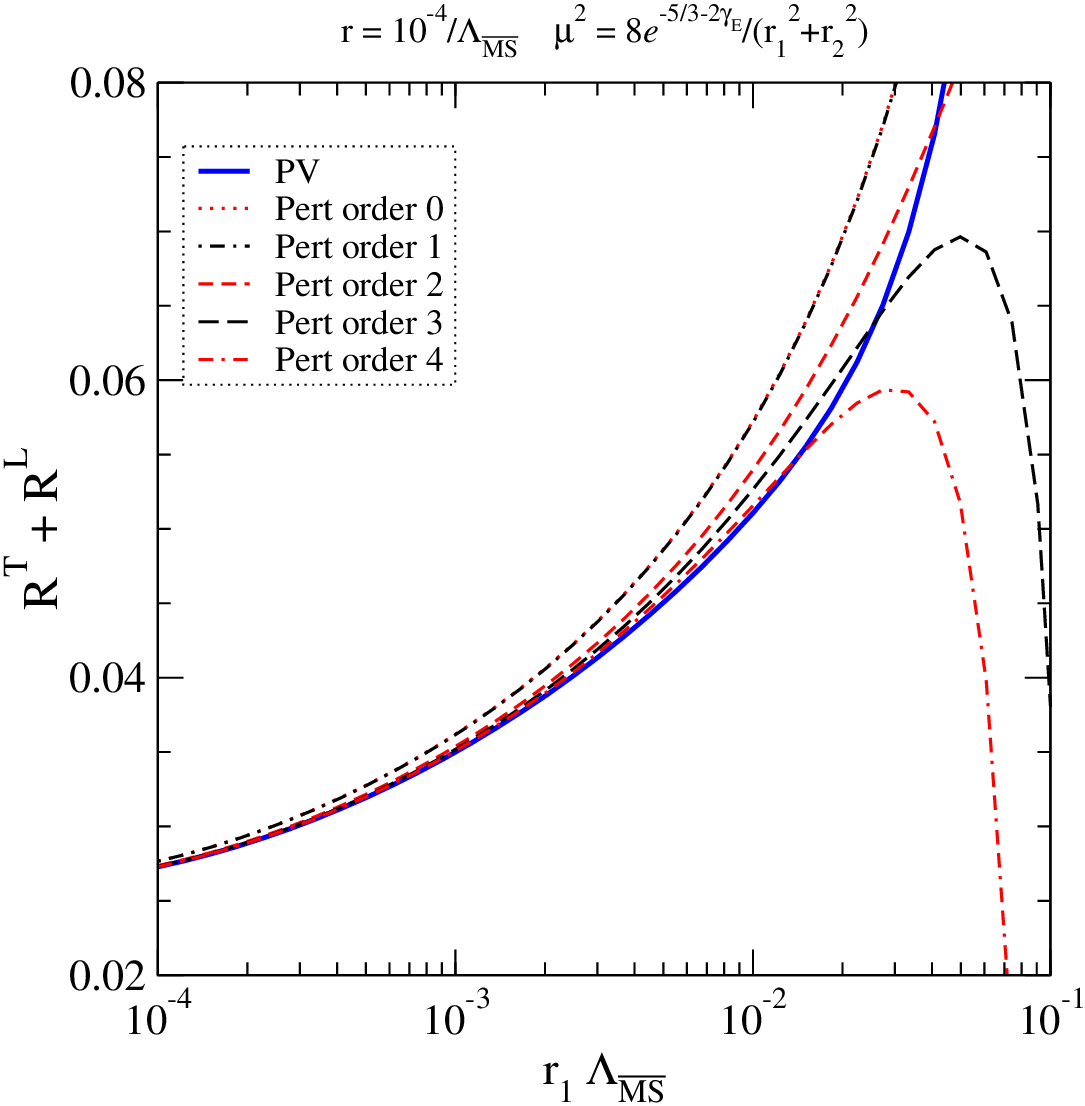}\\ (b)
  \end{minipage}
  \caption{\small \em $R(\vdl\Lambda,\vdr\Lambda)$ as
    defined with a PV regularization of the Borel sum (\ref{eq:K-replacement}),
    and the convergence of perturbation theory at
    the first few orders, according to~\eqref{eq:pert-exp},
    with $\mu^2=4\,\exp{\left(-\frac53-2\gamma_E\right)}/\rr^2$ (in (a)) and
    $\mu^2=8\, \exp{\left(-\frac53-2\gamma_E\right)}/(\dl^2+\dr^2)$ (in (b)).
   $R$ is shown as a function of $\dl$, with $\rr\Lambda = 10^{-4}$,
   $\vrr \parallel \vdl \parallel \vdr$ and $\dr=\dl+\rr$.
}
  \label{fig:conv-pert-theory}
\end{figure}

Our default choice for defining the all--order sum of the series is
the PV prescription. Having made this choice, we will examine the
convergence of the perturbative expansion in the first few orders,
and demonstrate (Fig.~\ref{fig:conv-pert-theory}) the dependence of
this (apparent) convergence on the region of phase space as well as
on the choice of the expansion parameter.

Technically, the implementation of the PV regularization in the
present case is complicated by the fact that Eqs.~\eqref{eq:BT}
and~(\ref{eq:BL}) contain terms with double poles\footnote{The
current example is by no means unique in this
  regard. The same occurs also in the absence of large target fields, e.g. for
  the well--studied example of the Adler function $D(Q^2)=4\pi^2
  d\Pi(Q^2)/dQ^2$,
  see~\cite{Broadhurst:1992si,Lovett-Turner:1995ti,Beneke:1998ui}.} in
  $u$.
We cope with this in the standard technique by isolating the double
pole contributions and using integration by parts. The explicit
expressions used in defining $R$ are given in
Appendix~\ref{sec:PV-def-pert-sum}; they are in turn based on the
explicit expression for the first-- and second--order poles
extracted in Appendices~\ref{sec:poles-BT} and~\ref{sec:poles-BL}.

The numerical result for the PV Borel sum is shown in
Fig.~\ref{fig:conv-pert-theory}. To appreciate the qualities of this
prescription as means of defining the all--order sum, it is useful
to compare it with increasing--order partial sums. The expansion of
the Borel functions can be conveniently obtained starting with
expressions (\ref{eq:BLb}) for $B^L$ and (\ref{eq:BTb}) with
(\ref{eq:_hyper_identity}) for $B^T$, which involve
$\qFp{2}{1}{1-u,1-u}{2-2u}{1-\xi^2}$, a special case of the type~E
function expanded in~\cite{Kalmykov:2006pu} (see Eq. (4.29) there).
It is convenient to define the
 dimensionless variable
\begin{equation}
\label{eq:Omega} \Omega:= \frac{4\,e^{-\frac53-2\gamma_E}}{r_>^2
\mu^2}
\ ,
\end{equation}
which appears as the argument of the logarithms in the coefficients.
The expansion for $B^T$ takes the form
\begin{subequations}
  \label{eq:BTexpansion}
  \begin{align}
    {\cal K}_{\bm{x z y}}\, & B^T(u,\vdl\mu,\vdr\mu) = \,
\frac12\frac{\rr^2-(\dl^2+\dr^2)}{\dl^2\dr^2}\left[
    1+\sum\limits_{n=1}^\infty
    b^T_n(\vdl\mu,\vdr\mu)u^n \right]
\ ,
    \intertext{where}
    \label{eq:BTexpansion-b}
    b^T_1(\vdl\mu,\vdr\mu) = &\,
    -\ln (\Omega)
    -\frac{\xi^2 \ln(\xi^2)}{1-\xi^2} \ ,
    \displaybreak[0]\\
    \label{eq:BTexpansion-c}
    b^T_2(\vdl\mu,\vdr\mu) = &\, \frac{1}{2}\,\Big(b^T_1(\vdl\mu,\vdr\mu)\Big)^2
    -\frac{\pi^2}{6}-\frac12\frac{\Big(\xi^2 \ln(\xi^2)\Big)^2}{(1-\xi^2)^2}
    +\frac{1+\xi^2}{1-\xi^2} \text{Li}_2(1-\xi^2)
\ ,
\displaybreak[0]\\
    \label{eq:BTexpansion-d}
    b^T_3(\vdl\mu,\vdr\mu) = &\,\frac{1}{1-\xi^2}
    \Bigg\{-{\displaystyle \frac {1}{6}}(1-\xi ^{2} )\,
    \ln^{3}(\Omega) - {\displaystyle \frac {1}{2}} \,\ln^{2}(\Omega )\,\xi
^{2}
\,\ln(\xi ^{2}) \notag \displaybreak[0]\\
& + \left[(\xi ^{2} + 1)\,\left(\ln(1 - \xi ^{2})\, \ln(\xi ^{2}) +
{\rm Li}_2(\xi ^{2}) \right)-
{\displaystyle \frac {\xi ^{2}\,\pi ^{2}}{3}} \right]\,\ln(\Omega )
\notag \\
& + 2\,\xi ^{2}\,\ln^{2}(\xi ^{2})\,\ln(1
 - \xi ^{2}) + 3\,\xi ^{2}\,\ln(\xi ^{2})\,
 {\rm Li}_2(\xi ^{2}) + (2\,\xi ^{2} + 2)\,{\rm Li}_3(1 - \xi ^{2})
\notag\\
& - 2\,\xi ^{2}\,{\rm Li}_3(\xi ^{2}) + \left( - {\displaystyle
\frac {4}{3}}  + {\displaystyle \frac {10\,\xi ^{2 }}{3}}
\right)\,\zeta_3 - 4\,\xi ^{2}(1-\xi ^{2})\,\left.
\frac{dS_{2,2}(z)}{dz} \right\vert_{z=1 - \xi ^{2}}\Bigg\}
\ .
  \end{align}
\end{subequations}
${dS_{2,2}(z)}/{dz}$ in (\ref{eq:BTexpansion-d}) is the first
occurrence of a Nielsen polylogarithm,
\begin{equation}
S_{a,b}(z):= \frac{(-1)^{a+b-1}}{(a-1)!b!}
\int_0^1\frac{d\xi}{\xi}\ln^{a-1}(\xi)\ln^b(1-\xi z)
\ .
\end{equation}
At higher orders in the expansion one
encounters~\cite{Kalmykov:2006pu} higher Nielsen polylogarithms, as
well as other harmonic polylogarithms~\cite{Remiddi:1999ew}.

The longitudinal part, $B^L$ starts at order $u$, corresponding to
${\cal O}(\beta_0 \alpha_s^2)$. The expansion takes the form:
\begin{subequations}
  \label{eq:BLexpansion}
  \begin{align}
    {\cal K}_{\bm{x z y}}\, B^L(u,\vdl\mu,\vdr\mu) = &\,
    \sum\limits_{n=1}^\infty {\cal K}_{\bm{x z y}}b^L_n(\vdl\mu,\vdr\mu)u^n \ , \intertext{where,
      from~\eqref{eq:BLb}, we find} {\cal K}_{\bm{x z y}}\,
    b^L_1(\vdl\mu,\vdr\mu) =
     &\,\minus \frac{1}{{\bm r}_>^2}\frac{\ln(\xi^2)}{1-\xi^2}\label{eq:BLexpansion1}
    \ ,
    \\
    {\cal K}_{\bm{x z y}}\, b^L_2(\vdl\mu,\vdr\mu) = &\, \plus \frac{1}{{\bm r}_>^2}
    \frac{\ln(\xi^2)\ln(\Omega) +2\text{Li}_2(1-\xi^2)}{1-\xi^2}
\ ,
 \\
    \begin{split}
{\cal K}_{\bm{x z y}}\, b^L_3(\vdl\mu,\vdr\mu) = &\, -\frac{1}{{\bm
r}_>^2}\, \frac{1}{1-\xi^2}\Bigg[\ln(\xi ^{2})\,{\rm Li}_2(\xi ^{2})
- 2\,{\rm Li}_3(\xi ^{2}) +
2\,\zeta_3 - 4\,{\rm Li}_3(1 - \xi ^{2}) \\
& + {\displaystyle \frac {1}{2}} \,\ln(\xi ^{2})\, \ln^{2}(\Omega )
+ 2\,\ln(\Omega )\,{\rm Li}_2(1 - \xi ^{2})\Bigg].
\end{split}
  \end{align}
\end{subequations}
It is interesting to note that there is a qualitative difference
between the leading behavior of the transverse and the longitudinal
contributions, respectively. The sign of the LO, transverse contribution to
the kernel ${\cal M}_{\bm{x z y}}$ in (\ref{eq:pert-exp})
can be directly read of Eq.~(\ref{Kdef}); it is
\begin{align}
\label{eq:circle}
\begin{split}
\begin{array}{lll}
{\cal   M}_{\bm{x z y}}^T &> 0 \quad \text{inside the circle:}\,\,&\dl^2+\dr^2<\rr^2\\
{\cal   M}_{\bm{x z y}}^T &< 0 \quad \text{outside the circle:}\,\,&\dl^2+\dr^2>\rr^2,
\end{array}
\end{split}
\end{align}
while the NLO longitudinal contribution to ${\cal M}_{\bm{x z y}}$,
in (\ref{eq:BLexpansion1}), is \emph{always} positive.

With the explicit expressions of~(\ref{eq:BTexpansion}) and
(\ref{eq:BLexpansion}) we can now study the convergence of the
perturbative series in~\eqref{eq:pert-exp} and compare the
finite--order results to the PV definition of the all--order sum.
Of course, at this point we must\footnote{To avoid the ultraviolet
problems of a conventional
  fixed--coupling calculation one in fact needs to choose $\mu$ as \emph{some}
  function of $\dl$, $\dr$ and $\rr$ instead of a single homogeneous scale
  independent of the configuration encountered~\cite{Rummukainen:2003ns}.}
make a choice of scale $\mu$.
Since ${\cal M}_{\bm{x z y}} = R(\vdl\Lambda,\vdr\Lambda)\,{\cal
K}_{\bm{x z
    y}}$ is renormalization invariant (i.e. $\mu$--independent) one may choose
{\em any} function of $\dl$, $\dr$ and $\rr$ as long as the convergence of the
perturbative series is good enough. As we shall see, this choice makes a
significant difference for the apparent convergence at the first few orders.

A priori, a natural choice (see
Sec.~\ref{sec:running-coupling-types}) may be the ``parent dipole''
size: $\mu^2=c/\rr^2$. A posteriori, knowing the coefficients in
(\ref{eq:BTexpansion}) and (\ref{eq:BLexpansion}),
one can optimize the scale to reduce the
size of subleading corrections. One possibility is to introduce BLM scale
setting~\cite{BLM} by requiring that the entire NLO coefficient
$b_1(\vdl\mu,\vdr\mu)$ would vanish identically\footnote{To express
the ${\cal   K}_{\bm{x z y}}$ in the transverse part in terms of
dipole lengths only we substitute: $-{\vdl\cdot
\vdr}=(\rr^2-\dl^2-\dr^2)/2$, as in (\ref{Kdef}) above.}:
\begin{eqnarray}
\label{eq:mu_BLM} &&{\cal   K}_{\bm{x z y}}b_1^T(\vdl\mu,\vdr\mu)+
{\cal K}_{\bm{x z y}}
b_1^L(\vdl\mu,\vdr\mu)=0\nonumber \\
&&\qquad \Longrightarrow \qquad \mu^2_{\BLM} =
\frac{4}{r_>^2}\,\exp\left\{
-\frac{5}{3}-2\gamma_E+\frac{\xi^2\,\ln(\xi^2)}{1-\xi^2}\,
\left(1+\frac{2r_>^2}{\rr^2-r_>^2-r_<^2}\right)\right\}.
\end{eqnarray}
In our case, however, this approach is too simplistic: since the
leading--order kernel vanishes on the circle $\dl^2+\dr^2=\rr^2$, as implied
by (\ref{eq:circle}), while the NLO contribution associated with $B^L$ does
not, (\ref{eq:mu_BLM}) develops an artificial singularity \emph{within the
  parturbative region}. Consequently, with this particular choice of scale,
higher--order terms would not be bounded on the circle, while no special
difficulty would arise there otherwise.  While better possibilities for
optimizing the scale exist --- for example, setting the BLM scale such that
the NLO \emph{transverse} contribution would vanish $b_1^T(\vdl\mu,\vdr\mu)=0$
--- we will not take this avenue here.  Let us just note in passing that the
multi--scale nature of the problem, which is reflected in the dependence on
the parent and daughter dipoles --- or in momentum space by the separate Borel
powers of the transverse momentum before and after the interaction with the
target (Eq.~(\ref{eq:mom-space-Borels})) --- renders any scale--setting
procedure in terms of a \emph{single} coupling unnatural.  Further progress in
this direction will be reported in Ref.~\cite{HW-YK:2006}.  Here we consider a
couple of simple examples for the scale:
\begin{equation}
  \mu^2 = \frac{4e^{-5/3-2\gamma_E}}{\rr^2}
  \mbox{~~~~ and ~~~~}
  \mu^2 = \frac{8e^{-5/3-2\gamma_E}}{\dl^2+\dr^2}.
  \label{eq:mu-setting}
\end{equation}
The choice of the constant $4e^{-5/3-2\gamma_E} \approx 0.24$ is
motivated by Eq.~(\ref{eq:mu_BLM}).

In Fig.~\ref{fig:conv-pert-theory} we show a numerical comparison of the
order--by--order expansion with the PV Borel sum.
As expected, at short distance scales perturbation theory is
well--behaved\footnote{If we allow the constant in
(\ref{eq:mu-setting}) to vary significantly compared the choice indicated, the
convergence in both Figs.~\ref{fig:conv-pert-theory}~(a) and~(b) becomes
appreciably worse, an obvious consequence of the logarithmic terms in the
coefficients, see (\ref{eq:BTexpansion}) and~(\ref{eq:BLexpansion}).}:
the first few orders in~\eqref{eq:K-replacement} gradually approach
the PV Borel sum. For larger distances, the corrections become large,
and the minimal term in the series
is reached sooner (see Fig.~\ref{fig:conv-pert-theory}~(b)).
Eventually, for $r_1\Lambda$ of order of a few times $10^{-1}$, the series
diverges right from the start. Note that even in that region the PV Borel sum
is uniquely defined, but as we shall see in the next section, power corrections
are large as well.

It is interesting to further observe that there is a significant difference in
the apparent convergence in the first few orders between
Figs.~\ref{fig:conv-pert-theory}~(a) and~(b): the ``parent dipole'' scale
setting in Fig.~\ref{fig:conv-pert-theory}~(a) has significant corrections
even at short distances, while with the scale choice of
Fig.~\ref{fig:conv-pert-theory}~(b) the first few orders provide a better
approximation. It should be emphasized though that
Fig.~\ref{fig:conv-pert-theory} shows a particular situation where $\dl,\dr\gg
\rr$.  If the dipole sizes are of comparable magnitude, there is no
significant difference between these two choices of scale\footnote{%
  For clarity, let us comment that we have chosen to show the function $R$
  with $\bm\dl$ parallel to $\bm\dr$.  This avoids the apparent divergence
  $(\bm\dl\cdot\bm\dr)^{-1}$ in the definition of $R^L$; this divergence is
  exactly canceled in the BK kernel when $R$ is multiplied by $K_{\bm x \bm z
    \bm y}$.}. The fact that ``parent dipole'' running fails to approximate
the kernel for $\dl,\dr\gg \rr$, makes a difference for evolution when $R_s\gg
\dl,\dr\gg \rr$.  We observe that parent dipole running in this region
generically underestimates the perturbative sum.  Accordingly, in
Sec.~\ref{sec:simulation-results}, we will see that evolution is slower with
the ``parent dipole'' ansatz than with the Borel sum.

\begin{figure}[htb]
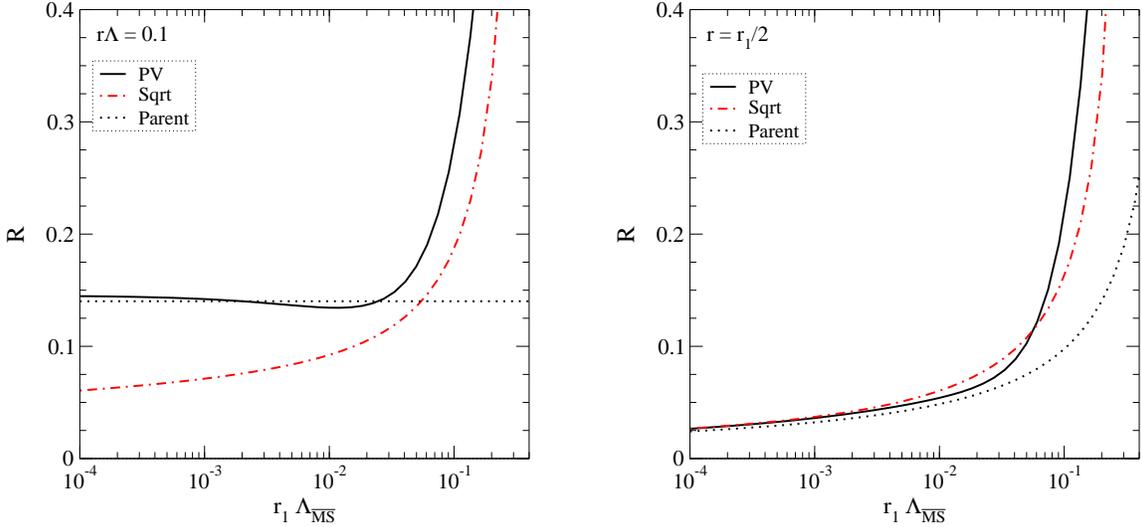

  \centering
  \includegraphics[width=7cm]{comp_Rall2}\hfill
  \includegraphics[width=7cm]{comp_Rall3}
  \caption{\small \em Comparison of ad hoc implementations for effective
    couplings $R$ calculated with the principal value (PV) prescription, the
    ``square root'' prescription
    $[\alpha_s(c^2/\dl^2)\,\alpha_s(c^2/\dr^2)]^{1/2}/\pi$, and the parent dipole
    running $\alpha_s(c^2/\rr^2)/\pi$.  On the left panel the size of the parent
    dipole is fixed, $\rr\Lambda = 0.1$ (and $\dr = \dl+\rr$); on the right
    panel the ratio of the dipole sizes is fixed, $\dl = 2\rr$, $\dr = 3\rr$.}
  \label{fig:compare-all}
\end{figure}

In Fig.~\ref{fig:compare-all} we compare the PV Borel sum with
two ad hoc running--coupling formulae, the ``parent dipole'' running and
the ``square--root daughter--dipole'' running:
\begin{equation}
  R_{\rm parent} =  \alpha_s^{\MSbar}\left({c^2}/{\rr^2}\right)/\pi
  =\frac{1}{\beta_0\ln\left(c^2/(r^2\Lambda^2)\right)};
  \qquad \quad
  R_{\rm sqrt} =
  \Big[\alpha_s^{\MSbar}\left({c^2}/{\dl^2}\right)\,
  \alpha_s^{\MSbar}\left({c^2}/{\dr^2}\right)\Big]^{1/2}/\pi.
  \label{eq:parent-sqrt}
\end{equation}
In these expressions we fix the scale
$c^2=4\,e^{-\frac53-2\gamma_E}$, as motivated by Eq.~\eqref{eq:mu-setting}.
This ensures that these couplings would match the PV
Borel sum soon enough when both $\dr, \dl \ll 1/\Lambda$. This is shown on
the right panel of Fig.~\ref{fig:compare-all}.  We note here that
$c^2$ here is much smaller than the value $c^2=4$ arising from a
Fourier transform in the double logarithmic limit.
From Fig.~\ref{fig:compare-all} it is clear that different definitions of
running coupling can be made to match reasonably well at small enough dipole
sizes, at least in parts of phase space, if the scales are chosen
appropriately. On the other hand, at small dipole
sizes, already at $0.1$--$0.3 \times 1/\Lambda$ the coupling diverge strongly.
This related of course to the smallness of the effective scale at which the
coupling is evaluated,
$\mu^2_{\rm eff} \sim 4\,e^{-\frac53-2\gamma_E}/r^2$.
As we shall see below, the effect of this divergence on the evolution rate
(in the BK approximation) is somewhat less important than for $R$. This is to be
expected, as evolution rate is an inclusive observable, and as such has
reduced sensitivity to the phase--space region where the effective coupling
is large.

\section{Non-perturbative power corrections}
\label{sec:power-corrections}

After discussing the perturbative features in some detail, let us
now turn to estimates the non-perturbative corrections to the JIMWLK
kernel ${\cal M}_{\bm{x z y}}$. The renormalon poles in $R$ allow us
to determine the type of power corrections, and their parametric
form, based on the singularities of $B^T$ and $B^L$. The original
expressions in (\ref{eq:BT}) and (\ref{eq:BL}) have simple and
double poles at integer values of $u$. As explained in the
Appendices~\ref{sec:poles-BT} and~\ref{sec:poles-BL}, we perform
integration by parts over the double pole terms to arrive at a
representation of the Borel integral where the integrand is composed
of simple poles only; in this way the double poles have been
converted into a simple ones accompanied by a
logarithmic--enhancement factor, $\ln(\Omega)$ where $\Omega$ is
given in Eq.~(\ref{eq:Omega}). In this representation, the explicit
expressions for the renormalon poles are the following:
\begin{subequations}
\label{eq:B-single-pole}
\begin{align}
  \label{eq:BT-single-pole}
  {\cal K}_{\bm{x z y}}B^T(u,\vdl\mu,\vdr\mu)
 = &
 \,{\cal K}_{\bm{x z y}}\,
  \left\{
  \left(\frac{4\,e^{-\frac53}}{r_>^2
      \mu^2}\right)^{-1}\frac{-2+(1-\xi^2)}{u-1}
  \right.  \notag \displaybreak[0]\\ &
  +\sum\limits_{m=2}^\infty \sum\limits_{n=2(m-1)}^\infty
  \frac{(-1)^m}{u-m}\frac{(1-\xi^2)^n}{\Gamma(n+1)}
  \left(\frac{4\,e^{-\frac53}}{r_>^2 \mu^2}\right)^{-m}
  \frac{\Gamma(1-m+n)}{\Gamma(n-2(m-1))\Gamma(m-1)} \notag \\ & +
  \sum\limits_{m=2}^\infty \sum\limits_{n=2}^{m-1}
  \frac{(1-\xi^2)^n}{\Gamma(n+1)} \frac{ 2(-1)^{m-n}}{(u-m)} \notag \\ &
  \left.
  \times \frac{d}{d m}\left[ \left(\frac{4\,e^{-\frac53}}{r_>^2
        \mu^2}\right)^{-m} \frac{\Gamma(2m-(n+1))}{
      \Gamma(m)\Gamma(m-1)\Gamma(m-n)\Gamma(m+1-n)} \right]\right\}
  \notag \\ & \, +\text{regular contributions}
\end{align}
and
\begin{align}
  \label{BL-single-pole-part}
  {\cal K}_{\bm{x z y}} & B^L(u,\vdl\mu,\vdr\mu)
 = \,\minus\, \sum\limits_{m=1}^\infty
  \sum_{n=m}^\infty \frac{-(-1)^m}{u-m}\frac{1}{r_>^2} \left(
    \frac{4\,e^{-\frac53}}{{\bm r}_>^2\mu^2} \right)^{-m}
  \frac{(1-\xi^2)^n}{\Gamma(n+1)}
  \frac{(\Gamma(n+1-m))^2}{(\Gamma(m))^2\Gamma(n+2-2m)} \notag \\ & \minus
  \sum\limits_{m=1}^\infty \sum\limits_{n=0}^{m-1} \frac{2 (-1)^{m-n}}{u-m}
  \frac{(1-\xi^2)^n}{\Gamma(n+1)} \frac{1}{r_>^2}
  \frac{d}{d m}\left[\left(
      \frac{4\,e^{-\frac53}}{{\bm r}_>^2\mu^2} \right)^{-m}
    \frac{\Gamma(2m-(n+1))}{(\Gamma(m))^2(\Gamma(m-n))^2} \right]
  \notag \\ & \, +\text{regular contributions}
\ .
\end{align}
\end{subequations}

\begin{figure}[htb]
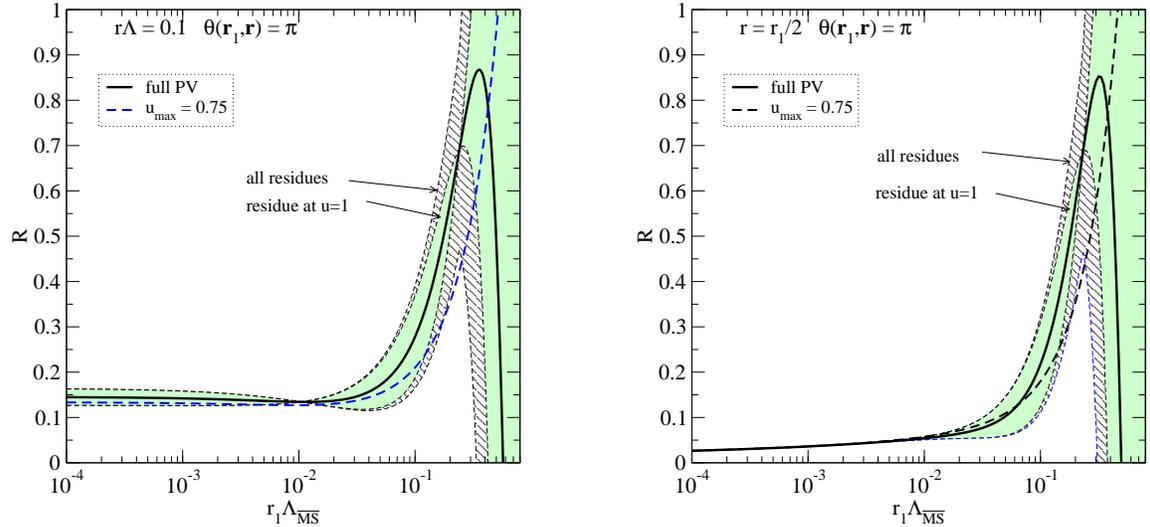

  \centering
  \parbox{.45\textwidth}{\includegraphics[width=.45\textwidth]{R_pv_tl_res_pi_2}}
  \hfill
  \parbox{.45\textwidth}{\includegraphics[width=.45\textwidth]{R_pv_tl_res_pi_3}}
  \caption{\small \em $R_{\rm PV}$ and the expected range of power
    corrections for fixed $\rr\Lambda=0.1$ (left panel), and fixed
    ratio $\dl=2\rr$, $\dr=3\rr$ (right panel) with $\dl$ parallel to $\dr$.
    The central solid line
    corresponds to the principal value result. The shaded regions estimate the
    relevance of power corrections by adding and subtracting $\pi
    |\text{residue}|(u=m)$. The inner band takes the first residue at $u=1$,
    and the outer hashed band is the sum of absolute values of
    contributions from all residues.  The dashed line shows the
    result when the $u$-integral has been cut at
    $u_{\rm max}=0.75$,
    before the first pole is encountered.}
  \label{fig:R-non-pert-uncert}
\end{figure}
To determine the ambiguity associated with each renormalon we should
first sum transverse and longitudinal contributions, and then
isolate the residue at fixed $u=m$.

Power corrections are expected to follow this ambiguity structure.  Let us
therefore introduce a non-perturvative parameter $C_m$, of order 1, relating
each power correction to the corresponding renormalon residue. The underlying
assumption is that genuine non-perturbative effects would be of the same order
as the ambiguities. The parametric dependence of the corrections on the hard
scales then follow directly from the residues in Eq.~\eqref{eq:B-single-pole}:
they are written as powers of $r_>^2 \Lambda^2$ times some function of
$\xi^2$, with additional logarithmic terms in $\Omega$. The dependence soft
scales is subsumed in the coefficients $C_m$. In doing so we write the full
kernel ${\cal M}_{\bm{x z y}}$ as a sum of perturbative contributions ${\cal
  M}_{\bm{x z y}}^{\text{PV}}$ and power corrections $\delta{\cal
  M}^{(m)}_{\bm{x z y}}$:
\begin{align}
  \label{eq:calM+power}
  {\cal M}_{\bm{x z y}} =
  {\cal M}_{\bm{x z y}}^{\text{PV}} + \sum\limits_{m=1}^\infty
  \delta{\cal M}^{(m)}_{\bm{x z y}}
\ .
\end{align}
As an example we explicitly present the power correction $\delta{\cal
M}^{(1)}_{\bm{x z y}}$ corresponding to the leading $u=1$ renormalon:
\begin{align}
  \label{eq:leading-residue}
  \delta{\cal M}^{(1)}_{\bm{x z y}}
  =C_1\frac{\pi}{ \beta_0}
  \left\{{\cal K}_{\bm{x z y}}
    (1+\xi^2)  \minus
    \frac{1}{r_<^2} \left[\ln\left(\xi^2\right)
      -2\ln\left(\frac{4\,e^{-\frac53-2\gamma_E}}{{\bm r}_>^2\Lambda^2}
      \right)\right] \right\}
      \frac{1}{4}\,r_>^2\Lambda^2\,e^{\frac53}
  \ .
\end{align}
Fig.~\ref{fig:R-non-pert-uncert} shows $R$ and the expected size of power
corrections as functions of $\dl$, for fixed parent dipole $\rr =
0.1/\Lambda$, and for fixed ratio of dipole sizes $\dr=\dl/2$ (here $\dr =
\dl+\rr$ and $\dr$ is parallel to $\dl$). The power corrections are assumed
here to be of the order of the ambiguity in choosing an integration contour in
the Borel plane, $\pi$ times the absolute
value of the residues at $u=m=1,2,3\ldots$.  The relative importance of
renormalon poles can be seen from the width of the error band in
Fig.~\ref{fig:R-non-pert-uncert}. The contribution from the pole at $u=1$,
Eq.~\eqref{eq:leading-residue}, is shown separately. It is evident that this
first pole strongly dominates, except very close to $1/\Lambda$, as expected
from the hierarchy of powers in the analytical expressions.

The key feature of the power corrections in
Fig.~\ref{fig:R-non-pert-uncert} is the fact that they quickly die
away at small scales\footnote{Note that on the left panel of
Fig.~\ref{fig:R-non-pert-uncert} $\dr$ approaches $\rr =
0.1/\Lambda$ as $\dl\rightarrow 0$.}. This feature is very robust
and it does {\em not} depend on the detailed prescriptions used to
estimate the size of the power corrections.  Just to give an
example, one might estimate the uncertainty by cutting off the $u$
integral at some value $u_{\text{max}}<1$, before one encounters the
first renormalon pole. This is shown in
Fig.~\ref{fig:R-non-pert-uncert} for $u_{\rm max} = 0.75$ with
dashed lines. Clearly this prescription leads to an alternative
estimate for the perturbative sum for $R$ which is roughly within
the error band of the full PV integral.

From Fig.~\ref{fig:R-non-pert-uncert} it becomes strikingly clear
that the poles limit the precision of our knowledge of the
(non-perturbative) evolution kernel at large distances. We will see
in Sec.~\ref{sec:BK-numerics}, when we discuss evolution in the
context of the BK equation, that for $Q_s$ near $1-2$GeV, where
large dipoles contributes significantly to the evolution, power
corrections {\em must} be taken into account if one wants to
quantitatively determine the evolution. If we wish to apply JIMWLK
or BK evolution starting from $Q_s$--values in this range, both the
evolution rate and the generic functional form of the initial
condition for the dipole cross section $N_{\y_0,\bm{x y}}$ receive
sizeable non-perturbative contributions.

\section{Evolution in the BK approximation}
\label{sec:BK-numerics}

Our discussion so far focused on the \emph{kernel} of the JIMWLK
equation. We computed and resummed running--coupling corrections to
the kernel and extracted an estimate for non-perturbative power
corrections. All these affect only the ``effective charge'' in front
of the leading--order kernel, although in a way that strongly
depends on the size of the evolving dipole, and that varies
significantly over the transverse phase space of the newly produced dipoles.
In order to explore
the consequences of these results on the \emph{evolution}, one
should clearly consider the solution of the equation over a
sufficiently large range in rapidity. Only in this way would it be
possible to examine the sensitivity of observable quantities, such
as the rate of evolution, to the corrections computed.

The purpose of the present section is to perform a first exploration
of this kind, by considering the evolution in the case of the BK
equation. To this end, let us repeat the derivation of the BK
equation, as described in Sec.~\ref{sec:evolution-equations},
starting with the JIMWLK equation with the Hamiltonian of
Eq.~(\ref{eq:JIMWLK-Hamiltonian-NLO}), which includes
running--coupling corrections. A simple way to see the way the
corrections enter is to first write the leading--order BK equation
with the kernel separated according
Eq.~\eqref{eq:JIMWLK-LO-diagram-cont} into exchange and
self--energy--like diagrams:
\begin{align}
  \label{eq:BK-sep}
  \partial_\y N_{\y,\bm{x y}} = \frac{ N_c}{2\pi}\int\!\! d^2 z &
  \left[2\frac{\alpha_s(\mu^2)}{\pi}{\cal K}_{\bm{x z y}}
  -\frac{\alpha_s(\mu^2)}{\pi} {\cal K}_{\bm{x z x}}
  -\frac{\alpha_s(\mu^2)}{\pi}{\cal K}_{\bm{y z y}} \right]
 \\ & \notag
  \hspace*{40pt}\times\,\Big(
  N_{\y,\bm{x z}}+ N_{\y,\bm{z y}}-N_{\y,\bm{x y}}
  - N_{\y,\bm{x z}}\, N_{\y,\bm{z y}}\Big)\ .
\end{align}
The result obtained using (\ref{eq:JIMWLK-Hamiltonian-NLO}) by
repeating the steps of Sec.~\ref{sec:evolution-equations} simply
amounts the substitution~\eqref{eq:new-JIMWLK}, separately for each
of the three terms in (\ref{eq:BK-sep}):
\begin{align}
  \label{eq:BK-running-kernel}
  \frac{\alpha_s}{\pi} \Tilde{\cal K}_{\bm{x z y}} &\,\to \Tilde{\cal
    M}_{\bm{x z y}}= \ 2 \, R(\vdl\Lambda,\vdr\Lambda)\, {\cal K}_{\bm{x z y}}
  -R(\vdl\Lambda,\vdl\Lambda)\,{\cal K}_{\bm{x z x}} -
  R(\vdr\Lambda,\vdr\Lambda)\,{\cal K}_{\bm{y z y}}
\end{align}
where the vectors $\vrr_i$ are defined in (\ref{eq:distshort}).
Our final result for the resummed BK equation is then written as
\begin{equation}
  \label{eq:BK-full}
  \partial_\y N_{\y,\bm{x y}} = \frac{ N_c}{2\pi}\int\!\! d^2 z
  \
  \Tilde{\cal M}_{\bm{x z y}}\,\,
  \Big(
  N_{\y,\bm{x z}}+ N_{\y,\bm{z y}}-N_{\y,\bm{x y}}
  - N_{\y,\bm{x z}}\, N_{\y,\bm{z y}}\Big)
  \ .
\end{equation}

\subsection{Numerical implementation}
\label{Sec:numerical}

The numerical solution of the BK evolution equation~\eqref{eq:BK-full}
requires care. In the following we briefly describe the choices we made
in our implementation.
Firstly, we consider only translationally--invariant and
spherically--symmetric solutions; we set $\bm{x} \rightarrow 0$ and
$N_{\bm{xy}} \rightarrow N_{|\bm{y}|}$. Despite this, the
$\bm{z}$-integral in
Eq.~\eqref{eq:BK-full} renders the evolution equation in essence
two-dimensional.  The simplest way to perform this integral is to discretize the
two-dimensional space using a finite regular square lattice.  This was used
in~\cite{Rummukainen:2003ns} in order to compare BK with JIMWLK evolution.
This approach, however, restricts the ratio of ultraviolet and infrared
cutoffs (the ratio of the lattice size to the lattice spacing) to, at most,
$\sim 10^4$, and hence strongly limits the $\y$ range over which the shrinking
correlation length $R_s(\y)$ can be resolved on such a lattice. A solution to
this problem is to use discretization with higher resolution power at small
distances, as was done in~\cite{Albacete:2004gw}.

To achieve this we discretize $N_r$ on an even logarithmic scale,
$r_n = r_0 \exp(n\Delta)$, using $\sim 250$ points with $r_{\text{
min}} = e^{-22}/\Lambda$ and $r_{\text{max}} = 1/\Lambda$. The
two-dimensional integral in Eq.~(\ref{eq:BK-full}) is evaluated using
nested Simpson integrations in the
($\log |\bm z|$, $\arg\bm z$) coordinates.
While the values of $|\bm{z}|$ and $|\bm{y}|$ are
restricted to discrete $r_n$-values, $|\bm{z-y}|$ is not.  Thus, it
is necessary to interpolate the value of $N_{|\bm{z-y}|}$ in
Eq.~\eqref{eq:BK-full} using the known points $N_{r_n}$.  We have
checked the stability of the simulation by changing the number of
discretization points, and verified that this leads to a
negligible change in the results.

In order to evaluate the $R$-functions in the kernel $\Tilde{\cal M}$ of
(\ref{eq:BK-running-kernel}),
we need to integrate over the Borel function $B(u)$.  However, as
described in Sec.~\ref{sec:perturbative-expansions}, one cannot
directly use the expressions in Eqs.~(\ref{eq:BTa}) to~(\ref{eq:BLb})
when integrating over the positive real axis,
because of the presence of single and double infrared
renormalon poles there. As
described in App.~\ref{sec:PV-def-pert-sum},
the double poles were converted analytically
into simple ones by means of integration by parts.  The integration over
the simple poles was performed by a numerical implementation of the
Principal Value prescription. This implementation is based on different
analytic formulae, depending on the value of $\xi=r_< / r_>$:
for $\xi^2 < 0.8$ we use
\eqref{eq:RT-single-pole-version} and
\eqref{eq:RL-single-pole-version} for $R^T$ and $R^L$, switching
over to \eqref{eq:RT-single-pole-version-2} and
\eqref{eq:RL-single-pole-version-2} when $\xi^2 > 0.8$.  The
integrals are completely stable when the threshold value of $\xi^2$
is changed within the interval $(0.01,0.99)$, and our default value of $0.8$
is chosen to optimize the evaluation speed. The numerical evaluation of the
kernel is further stabilized by pulling the integrands of the three
$R$-functions appearing in \eqref{eq:BK-running-kernel} under a
single integral.

If the parent dipole is much smaller than the daughter dipoles, $\rr
\ll \dl \sim \dr$, the kernel $\tilde{\cal M}$ contains large terms
that cancel each other almost completely, leading to numerical
difficulties.  This can be dealt with by recognizing that the large
contributions arise in fact from the leading terms in the series expressions
of Borel functions Eqs. \eqref{eq:RT-single-pole-version-2} and
\eqref{eq:RL-single-pole-version-2} appearing in \eqref{eq:BK-full}
(in this parameter region $1-\xi^2 \approx 0$). Combining the first
terms analytically stabilizes the evaluation of the series.

The evolution in rapidity $\y$ is performed with the second-order
Runge-Kutta method, with a ``time'' step of $\delta\y = 0.05 \ldots 0.1$.
Again, we observed no significant difference in the results upon varying the
step size.

\subsection{Simulation results}
\label{sec:simulation-results}

\begin{figure}[tb]
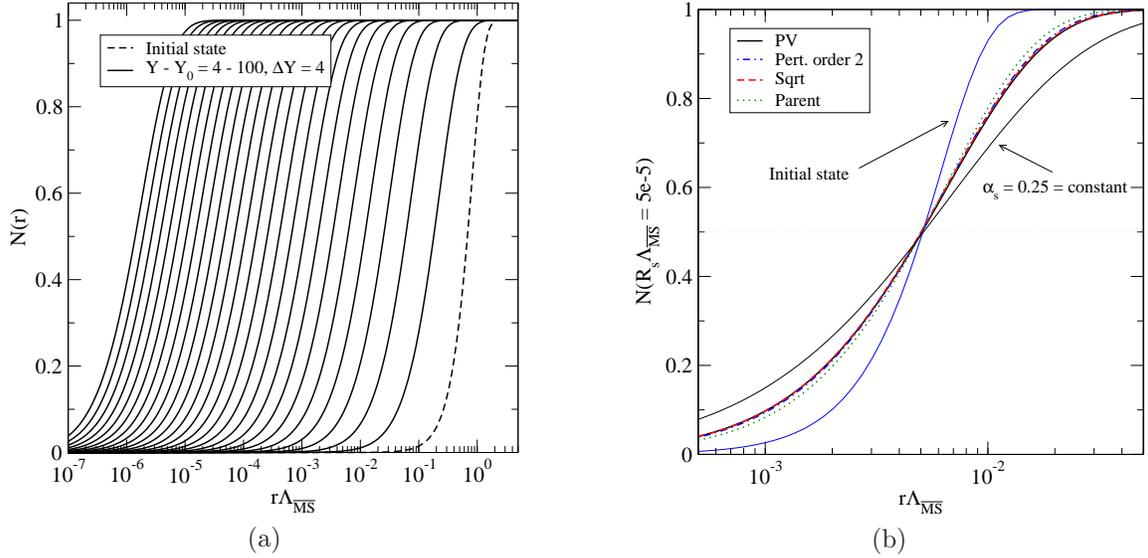

  \centering
  \begin{minipage}{.45\textwidth}
    \centering
    \includegraphics[width=\textwidth]{N_tau_evo_T_L.eps}
    \\ (a)
  \end{minipage}
  \hfill
  \begin{minipage}{.45\textwidth}
    \centering
    \includegraphics[width=\textwidth]{N_tau_mod.eps}
    \\ (b)
  \end{minipage}

  \caption{\small \em (a): The evolution of the function $N(r)$
  as a function of $\y
    = \ln 1/x$, shown in intervals of $\Delta\y=4$. After initial
    settling down the function evolves towards smaller $r$ while
    approximately preserving its shape.  The main effect of the
    running coupling is to slow down the evolution at small $r$.
    (b): The shape of $N(\y,r)$ using the computed running--coupling
    corrections (PV) as well as some running--coupling models,
    taken at $\y$ where the saturation scale
    {\hbox{$R_s(\y) =
    0.005/\Lambda$}},
    where we define $R_s$
    through $N(\y,R_s)=1/2$.  Different
    running--coupling forms display similar shapes for $N(r)$, albeit slightly
    steeper for the parent--dipole running; much shallower than the
    initial state $N_{\rm ini} = 1-e^{-r^2/r_0^2}$,
    but much steeper than the fixed coupling shape shown for
    comparison.}
  \label{fig:N-scaling}
\end{figure}

The key feature of JIMWLK and BK evolution is scaling of the solutions with
$Q_s(\y)$ after the initial--state effects have died out.
This scaling is exact in the fixed--coupling limit and has been
shown to be retained to a very good approximation in simulations with ad hoc
implementations of running coupling.  Also our all--order resummation result
shows this feature, as is demonstrated in Fig.~\ref{fig:N-scaling}.

The initial state in our simulations is $N_{\rm ini}(r) = 1-\exp[-r^2/r_0^2]$,
where $r_0$ sets the initial scale.  The evolution tends to flatten $N(\y,r)$
until the scaling shape has been reached.  However, the shape remains
considerably steeper than with fixed coupling, see
Fig.~\ref{fig:N-scaling}~(b). From this plot we also see that the
scaling shape is quite similar for the running coupling
corrections computed in this paper and the ad hoc prescriptions shown. Thus,
the asymptotic shape of $N(\y,r)$ is largely insensitive to the functional dependence
of the scale of the coupling on the phase space.

\begin{figure}[tb]
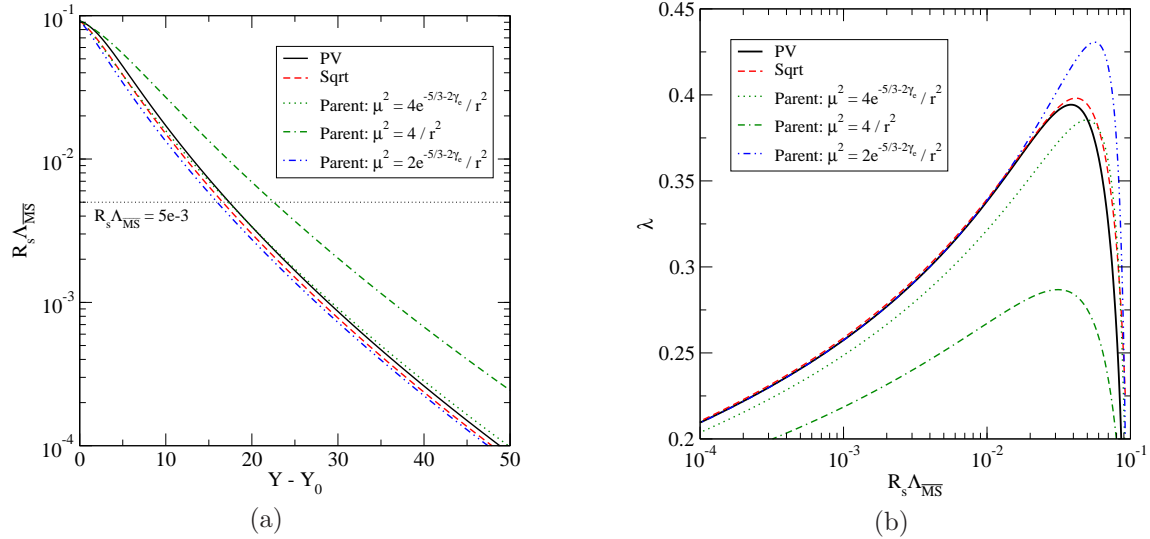

  \centering
  \begin{minipage}{.45\textwidth}
    \centering
    \includegraphics[width=\textwidth]{tau_rs_all_mod2.eps}
    \\ (a)
  \end{minipage}
  \hfill
  \begin{minipage}{.45\textwidth}
    \centering
    \includegraphics[width=\textwidth]{rs_lambda_all_mod2.eps}
    \\ (b)
  \end{minipage}

  \caption{\small \em (a): the evolution of the saturation scale
    $R_s(\y)$ with the computed running--coupling corrections (PV), in comparison
with a few different running--coupling models.  The evolution
    starts at initial \hbox{$R_s(\y=\y_0) \approx 0.1/\Lambda$}.  The
    horizontal dotted line shows the value of $R_s$ where $N(r)$
    was shown in Fig.~\ref{fig:N-scaling}.
    (b): the evolution speed $\lambda(\y)$ plotted against the saturation
    scale $R_s(\y)$.  The evolution starts at large $R_s$ values; the
    rapid increase at $R_s\Lambda \sim 0.1$ is an initial--state
    effect. Several choices of scale are presented for the
    ``parent dipole'', demonstrating the sensitivity to the scale and
    the fact that upon tuning the scale, this model too can be made
    to agree with the resummed result for $\lambda(\y)$ at small $R_s$.
  }
  \label{fig:R-Y-all}
\end{figure}

In Fig.~\ref{fig:R-Y-all} (a) we show the evolution of the
saturation scale $R_s \Lambda$ as a function $\y$; here the
full PV Borel sum result is compared with
the parent--dipole running and the ``sqrt'' ansatz of~\eqref{eq:parent-sqrt}.
The correlation length $R_s$ characterizes the scale where $N(\y,r)$ is
changing and saturation sets in; here we define $R_s(\y)$ through the condition $N(\y,R_s(\y)) =
1/2$.  The initial $R_s(\y=\y_0)$ is $ \approx 0.1/\Lambda$.  We see that
$R_s(\y)$ rapidly evolves towards smaller values.

We note that a different definition of $R_s(\y)$ would naturally give
somewhat different curves.  However, when the system has evolved
sufficiently long so that it reaches the scaling form, all reasonable
definitions should yield $R_s(\y)$ values that differ only by a
constant factor.  Above all, the evolution speed
\begin{equation}
\label{eq:ideal-lambda}
  \lambda(\y) = \frac{1}{Q_s^2(\y)}\frac{\partial Q_s^2(\y)}{\partial \y}
= - \frac{1}{R_s^2(\y)}\frac{\partial R_s^2(\y)}{\partial \y}\,; \qquad\qquad Q_s
:= 1/R_s
\end{equation}
should be the same.

As with the ad hoc prescriptions, the most interesting consequence of running
coupling on JIMWLK and BK evolution is the fact that it drastically slows down
the evolution compared to the purely fixed--coupling case
by restricting the active phase
space to within one order of magnitude of the characteristic scale
$Q_s(\y)$~\cite{Rummukainen:2003ns}.  This qualitative feature
does not depend on
the details of the implementation of the running--coupling effects.
The precise value of the evolution speed, as expressed by the evolution rate
$\lambda(\y) $ on the other hand, does depend on these details.  For fixed
coupling, scaling with $Q_s$ is exact, and $\lambda$ becomes a constant
proportional to $\alpha_s(\mu^2)$~\cite{Iancu:2002tr}. With running coupling,
scaling is approximate, and $\lambda$ becomes a function of $\y$ that will
receive both perturbative and non-perturbative corrections via
$R(\vdl\Lambda,\vdr\Lambda)$, which we already examined numerically in
Fig.~\ref{fig:R-non-pert-uncert}.

While using Eq.~\eqref{eq:ideal-lambda} with $N(\y,R_s(\y)) = C$
(where $C=1/2$ was used above) is certainly possible,
during the initial stages of
the evolution it is sensitive to the particular value of the
constant $C$ chosen.  A more robust definition is the one used
in~\cite{Rummukainen:2003ns}, using a $1/\rr^2$ moment of the
evolution equation as an operational definition of $\lambda$:
\begin{equation}
    \label{eq:running-lambda-integral-def}
    \lambda(\y) = \frac{N_c}{2\pi} \int
    \frac{d^2r}{\rr^2} \int\,d^2z\ {\cal M}_{\bm{x z y}}\
    \big( N(\y,\dl^2)
      + N(\y,\dr^2) - N(\y,\rr^2)
      - N(\y,\dl^2) N(\y,\dr^2)
      \big)
\end{equation}
We will follow this convention here.

\begin{figure}[htb]
  \centering
  \includegraphics[width=0.45\textwidth]{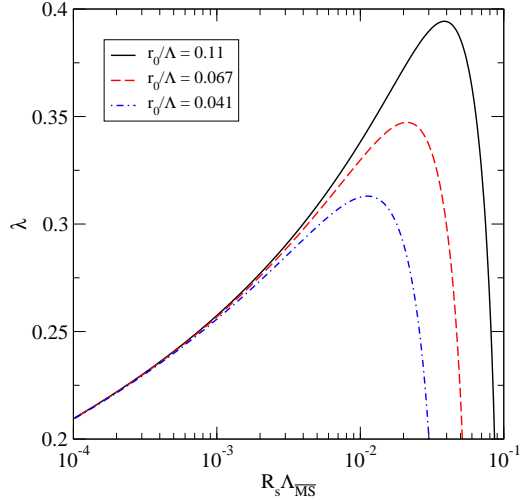}
  \caption{\small \em The evolution speed
    $\lambda(R_s)$ for different initial conditions $N_{\rm ini}(r) =
    1-\exp(-r^2/r^2_0)$, using the PV Borel sum.
    As the initial--state effects vanish, $\lambda(R_s)$ approaches
    a universal evolution trajectory.}
  \label{fig:lambda-initial}
\end{figure}

Our results for $\lambda(\y)$ are shown in Fig.~\ref{fig:R-Y-all} (b).  Here
we plot $\lambda$ against $R_s(\y)$; in this way the result is independent of
the initial value for $\y$.
The initial--state effects are visible as a very rapid increase of
$\lambda$ near the initial (large) $R_s$.
As soon as these effects die out, the evolution approaches an
initial--state--independent curve (see Fig.~\ref{fig:lambda-initial}).
On this curve $N(r)$ has reached the scaling shape shown
in Fig.~\ref{fig:N-scaling}, and $\lambda$ decreases slowly as $R_s$
decreases at large $\y$. This is entirely driven by the logarithmic
decrease of the coupling at short distances.

In this plot one can also observe the sensitivity of the evolution rate
to running--coupling corrections. We show the resummed result computed by the
PV prescription (full line) in comparison with LO evolution where the scale is
set in variety of ways. LO parent--dipole running with $\mu^2=4/r^2$ strongly
under--estimates the evolution rate \emph{even} at very small correlation lengths
$R_s$. This clearly demonstrates the significance of higher--order
running--coupling corrections, which are included in the PV Borel sum.

Interestingly, as seen in Fig.~\ref{fig:R-Y-all} (b),
one can approximate the PV Borel sum result
(at sufficiently small $R_s$) with a variety of different functional forms,
\emph{provided} that an appropriate choice of scale is made.
Thus, despite the large differences in the
``effective charge'' between the resummed result and the
models considered --- which are particularly significant for large dipoles ---
the differences in~$\lambda$, at sufficiently
small $R_s$, reduce to an overall phase--space independent
multiplicative factor in the scale of the coupling.
With the specific scales given in Eq.~\eqref{eq:mu-setting} the ``square root''
ansatz approximates the PV Borel sum well, whereas parent--dipole running
slightly underestimates it. In the latter case, choosing
$\mu^2=2{\rm e}^{\frac53-2\gamma_E}/r^2$ yields a very good approximation to
$\lambda$ for $R_s\Lambda\lsim 10^{-2}$, but somewhat over--estimates it
at larger $R_s$.  The possibility to approximate $\lambda$ in
(\ref{eq:running-lambda-integral-def}) well \emph{as a function of
the correlation length} (provided it is sufficiently small)
in terms of a phase--space independent coupling, such as the parent--dipole one,
is probably related to the scaling property of $N(\y,\rr)$, namely the fact that
its shape, which determines the weight given to different final states in
(\ref{eq:running-lambda-integral-def}) is invariant.

In Fig.~\ref{fig:lambda-initial} we show the evolution rate
$\lambda(\y)$ as a function of $R_s$ using different initial conditions.
The trajectories approach a universal curve, which is independent
of the initial condition.

\begin{figure}[tb]
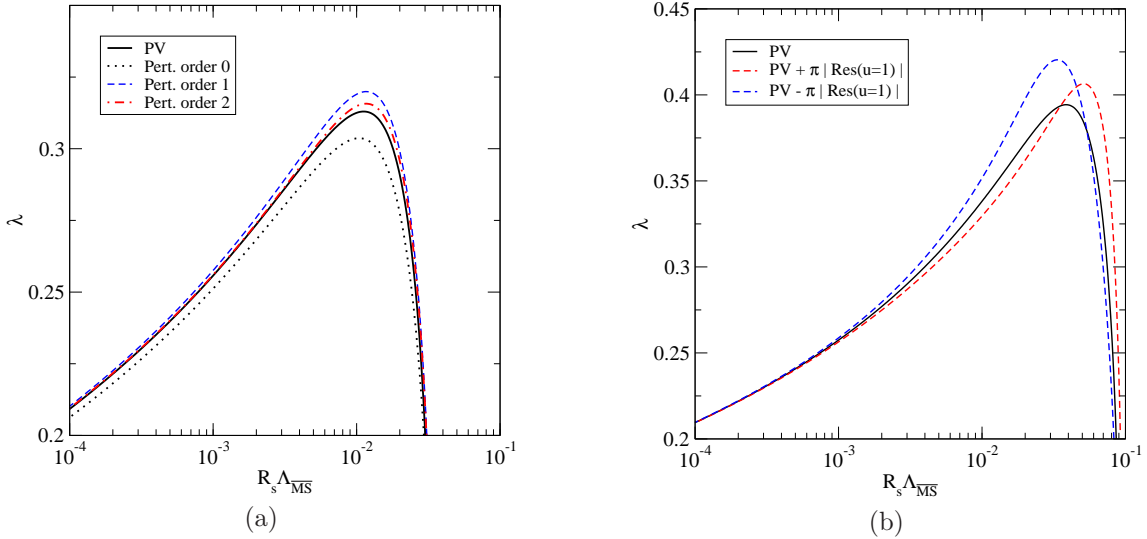

  \centering
  \begin{minipage}{.45\textwidth}
    \centering
    \includegraphics[width=\textwidth]{rs_lam_pv_pt_comp.eps}
    \\ (a)
  \end{minipage}
  \hfill
  \begin{minipage}{.45\textwidth}
    \centering
    \includegraphics[width=\textwidth]{rs_lambda_pv.eps}
    \\ (b)
  \end{minipage}

  \caption{\small \em (a):~Evolution using perturbatively
    expanded kernel, Eq.~\eqref{eq:pert-exp} with
    $\mu^2={8e^{-5/3-2\gamma_E}}/({\dl^2+\dr^2})$.
    (b):~Estimating the non-perturbative influence on $\lambda(R_s)$
    by adding and subtracting $\pi\times (\mbox{residue at $u=1$})$ to
    the PV evolution kernel.  The effect of the residue dies out
    quickly as $R_s$ becomes smaller. }
  \label{fig:lambda-nonpert-uncertainties}
\end{figure}

In Fig.~\ref{fig:lambda-nonpert-uncertainties}~(a) we show the
evolution rate as a function of $R_s$ using the perturbatively expanded kernel,
Eq.~\eqref{eq:pert-exp} at different truncation orders. Here the
scale of the coupling is chosen to be
$\mu^2={8e^{-5/3-2\gamma_E}}/({\dl^2+\dr^2})$ as in
(\ref{eq:mu-setting}).
We note that in contrast with the Borel sum,
the perturbative expansion suffers from a Landau pole when the
daughter dipoles become too large. These Landau singularities
obstruct the phase--space integration, so the large $\rr_i$ region
must be cut out. To avoid any visible effect by this cut we
performed the simulation of
Fig.~\ref{fig:lambda-nonpert-uncertainties}~(a) with relatively
small initial\footnote{To study the fixed--order perturbative result
at larger $R_s$ with this or a similar phase--space dependent scale
one would have to consider the effect of the cut or
introduce another regularization of the Landau pole.} $R_s$. In the
small $R_s$ regime shown, one observe good convergence towards the
PV Borel sum. At sufficiently small $R_s$ only the leading--order
result differs significantly from the PV one.
This is of course specific to the (almost optimal) choice of scale
made here. Much larger corrections appear for a generic scale choice
as we have already seen in Fig.~\ref{fig:R-Y-all}.

Fig.~\ref{fig:lambda-nonpert-uncertainties}~(b) shows the effect of the
estimated power corrections on evolution.
The relative importance of the
power corrections decrease with $Q_s(\y)$: the active phase
follows $Q_s(\y)$  towards the ultraviolet.
With increasing $Q_s(\y)$ the uncertainty induced by power
corrections dies away very rapidly. Despite this
general trend, it is apparent that a quantitative determination of evolution
speed for current experiments where $Q_s=1-2$ GeV,
requires to take power
corrections into account.
To test the predictive power of the approach, it is important to
compare several distinct observables, and address
the universality of the relevant power corrections.

\section{Conclusions}
\label{sec:conclusions}

We have presented a first calculation of running--coupling
corrections to both JIMWLK and BK equations. We have shown that both
these equations, which were originally derived at leading
logarithmic accuracy with strictly fixed coupling, can indeed be
promoted to the running--coupling case: the general structure of the
equations remains the same, while the kernel receives corrections.
Running--coupling corrections are singled out from other radiative
corrections in the following way: with or without running coupling,
the r.h.s of the evolution equation of any correlation involves
\emph{just
  one} more Wilson line (the produced gluon) than the l.h.s (the evolving
object); other radiative corrections entering at the NLO involve up
to \emph{two} more Wilson lines. The number of additional Wilson
lines grows further at higher orders. It remains for future work to
fully generalize the non-linear small--$x$ evolution equations to
NLO.  Some work in this direction, which is complementary to ours,
has already been done~\cite{Lublinsky:2001bc, Balitsky:2001mr,
  Gotsman:2004xb}.  A full NLO generalization exists of course for the linear
case of the BFKL equation, and our expectation is that it exists in
the non-linear case as well.  Since the BFKL equation is known in
full at NLO, one can make a useful comparison. This goes beyond the
scope of the present paper.  One important fact, however, that we do
learn from the BFKL limit, is that running--coupling corrections
constitute a significant part of the total NLO correction. This, we
expect holds also in the JIMWLK and BK cases.

The significance of the running coupling in the context of JIMWLK and BK
evolution has been acknowledged long ago. For one thing, in fixed--coupling
evolution the active phase space extends way into the ultraviolet,
while practically any dependence of the coupling on the scales
present at a single evolution step reduces the active phase
space to within one order of magnitude around the correlation length
$R_s$~\cite{Rummukainen:2003ns}. For this reason {\em all}
simulations of JIMWLK or BK evolution have been performed with running
coupling, implemented using some educated guess as to its scale dependence.
The simulations presented in this paper are the first where the scale
dependence is computed from QCD perturbation theory. While
\emph{qualitative} features of the effects of running coupling on the evolution
--- e.g. the decrease of the evolution rate $\lambda(Y)$ with
decreasing correlation length $R_s(Y)$ at large $Y$ --- are similar
to what has been observed before, the scale factor itself has been
determined here for the first time. As shown in
Fig.~\ref{fig:R-Y-all}, using the correct scale is crucial in
obtaining a quantitative estimate of the evolution rate. With this
predictive power, we can hope that non-linear evolution equations
with running coupling would become directly applicable to small--$x$
phenomenology at LHC energies.

Technically, running--coupling corrections are computed, as usual,
by focusing on a specific set of diagrams where a single gluon is
dressed by fermion--loop insertions, making use of the linearity of
$\beta_0$ in the number of flavors $N_f$. To perform this
calculation we utilized in this paper the dispersive approach, where
the all--order sum of vacuum--polarization insertions is traded for
a dispersive integral in the ``gluon mass''. While this technique
has been used extensively in the past, using it to compute the
diagrams entering the JIMWLK Hamiltonian is non-trivial for two
reasons: first, the large--$N_f$ limit by itself is not sufficient
to disentangle running--coupling effects from the new production
channel of a $q\bar{q}$ pair; second, the presence of the strong
Weizs\"acker-Williams background field may interfere with the
``dressing'' by interacting with the fermions. Guided by the
correspondence between real and virtual corrections owing to
conservation of probability, we could nevertheless disentangle
running--coupling corrections from other contributions and
generalize the dispersive technique to the case of the background
field. To this end we derived a dispersive representation of the
dressed gluon propagator in the background field. This formal
development may well have other applications.

By computing the JIMWLK Hamiltonian using the dispersive technique
we could go beyond the NLO ${\cal
  O}(\beta_0\alpha_s^2)$ running--coupling corrections to the kernel,
  and resum ${\cal   O}(\beta_0^{n-1}\alpha_s^n)$  corrections \emph{to all
  orders} in perturbation theory. Besides the obvious advantage of obtaining
an exactly renormalization--scale invariant kernel, the all--order
calculation offers a unique window into the non-perturbative side of
the problem. The fact that infrared--finite evolution equations can be
established in the high--energy limit, does not mean of course that
the dynamics governing the evolution is purely perturbative;
non-pertubative dynamics affects the evolution though
power--suppressed corrections.  By performing an all--order resummation
one can get access to this infrared sensitivity by looking at the
ambiguity in separating perturbative and non-perturbative
contributions; these are the infrared renormalons.  Using Borel
summation we identified explicitly the ambiguities in defining the
perturbative sum, and in this way established an estimate for the
parametric dependence the of non-perturbative effects on the hard
scales involved and the potential magnitude of these effects.

We find that both perturbative and non-perturbative corrections
modify the evolution kernel in a non-trivial way. In particular,
these corrections depend on all the different scales present: the
``parent dipole'' $\rr$ as well as the two ``daughter dipoles'',
$\dl$ and $\dr$. In this way different final states are weighted
differently. An interesting feature at NLO and beyond is the
appearance of two classes of contributions: one that is proportional
to the LO kernel, which is associated uniquely with transversely
polarized gluons, and one that is not, which is associated with
longitudinal polarizations. Both perturbative and non-perturbative
corrections to the kernel, propagate through the evolution and
affect the rate at which the saturation scale $Q_s(\y)$ flows
towards the ultraviolet with increasing energies. Conversely, the
saturation scale itself determines the active phase space, leading
to the decoupling of soft modes at large $\y$.

We studied here the effect of the newly computed corrections on two
levels: first by looking at the ``effective charge'' controlling the
contribution to the r.h.s. of the evolution equation, and then by
solving the BK equation numerically and studying the effects on the
evolution itself. We found, on both levels, that the
running--coupling corrections are significant.
Our simulations have shown that non-perturbative corrections are
strongly suppressed at high energies, reflecting the fact that the
increase of $Q_s(\y)$ with energy moves the active phase space along
with it towards shorter distances. At presently accessible energies,
where the saturation scale is estimated to be $Q_s(\y)\sim
1-2$ GeV, power corrections are definitely relevant. Going
from high to low energies, one may view the breakdown of the
perturbative evolution as the onset of the Soft Pomeron. If we
accept this premise, our calculation opens a new way to think about
the Soft Pomeron: as the energy is lowered the number of
power--suppressed terms that need to be included in the evolution
kernel increases. A possible determination of the power terms from
data can thus be attempted at intermediate energies, before the
power expansion breaks down. A detailed examination of this idea
must address the distinction between the initial condition to the
evolution and the corrections to the kernel itself.

In conclusion, we made an impotent step in extending the framework
of non-linear evolution equations at small $x$ to include
running--coupling and power corrections. Nevertheless, this endeavor
is by no means complete: small--$x$ evolution at high densities
presents many challenges, some of which we touched upon in this
paper. This includes for example the generalization of the
non-linear evolution equations to the NLO; the detailed comparison
with the NLO BFKL kernel; understanding the relation with DGLAP
evolution and higher--twist corrections to the twist expansion,
gaining better understanding of the initial condition for the
evolution; the development of (experimentally accessible!)
observables of different degree of inclusiveness, which are
sensitive to the dynamics underlying the evolution; and the
determination to non-perturbative corrections affecting the
small--$x$~evolution.

\section*{Acknowledgments}

E.G. wishes to thank Francesco Hautmann for illuminating
discussions. H.W. is very grateful to Ian Balitsky for sharing his insights on
computing running--coupling corrections to small-$x$ evolution. H.W.
also acknowledges extensive discussions with Yuri Kovchegov that
lead to an independent investigation of this topic in terms of an
explicit diagrammatic calculation presented in~\cite{HW-YK:2006}.

The work of H.W. is supported in part by the U.S. Department of
Energy under Grant No. DE-FG02-05ER41377. JK and KR were partly
supported by the Academy of Finland contract 104382.

\appendix

\section{$B^T$ in coordinate space}
\label{sec:FT-of-BT}

We begin with ${\cal F}^T$ in $d$ dimensional coordinate space.  The regulator
is only needed to have separately finite terms in~\eqref{eq:borel-via-F-resc}
and may be removed once the cancellation of infinite contributions is manifest
by setting $d\to 2$.  The simplest way to obtain a dimensionally regulated
expression for ${\cal F}^T$ is to start from Eq.~\eqref{eq:Kmassive} and to
promote the momentum integrals from $2$ to $d$ dimensions. One then
exponentiates the denominators using two Schwinger parameters $t_1$ and $t_2$.
This leads to
\begin{align}
  \label{eq:reg-calFT} {\cal F}^T(\vdl m,\vdr m) = &
  \frac{1}{\Gamma^2(\frac{d}{2})} \left(\frac{\dl^2}{4}
    \frac{\dr^2}{4}\right)^\frac{d}{2} \int\limits_0^\infty dt\ t^{-(d/2+1)}\
  e^{-t m^2-\frac{(\dl^2/4)}{t}} \int\limits_0^\infty ds\ s^{-(d/2+1)}\ e^{-s
    m^2-\frac{(\dr^2/4)}{s}} \intertext{and a $d$ dimensional generalization
    of Eq.~\eqref{Fdef}} = & \frac{2^{2-d}}{\Gamma^2(\frac{d}{2})} (\dl
  m)^\frac{d}2 K_{\frac{d}{2}}(\dl m) (\dr m)^\frac{d}2 K_{\frac{d}{2}}(\dr m)
\ .
\end{align}
From here we obtain $B^T$ according to~\eqref{eq:Bdef}:
\begin{align}
  \label{eq:borel-via-F-resc}
  B^T(u,\vdl\mu,\vdr\mu) = & -{\rm e}^{\frac53 u} \frac{\sin \pi u}{\pi}\,
  \int_0^{\infty}\frac{dm^2}{m^2} \bigg(\frac{m^2}{\mu^2}\bigg)^{-u} [{\cal
    F}^T (\vdl m,\vdr m)-1]
\ .
\end{align}
We will first perform the $m^2$ integral by using~\eqref{eq:reg-calFT}:
\begin{align}
  \int\limits_0^{\infty}&\frac{dm^2}{m^2} \bigg(\frac{m^2}{\mu^2}\bigg)^{-u}
  [{\cal F}^T (\vdl m,\vdr m) -1] = \int_0^{\infty}\frac{dm^2}{m^2}
  \bigg(\frac{m^2}{\mu^2}\bigg)^{-u} \notag \\ & \times \left[
    \frac{1}{\Gamma^2(\frac{d}{2})} \left(\frac{\dl^2}{4}
      \frac{\dr^2}{4}\right)^\frac{d}{2} \int\limits_0^\infty dt_1\
    t_1^{-(d/2+1)}\ e^{-t_1 m^2-\frac{(\dl^2/4)}{t_1}} \int\limits_0^\infty
    dt_2\ t_2^{-(d/2+1)}\ e^{-t_2 m^2-\frac{(\dr^2/4)}{t_2}} -1 \right]
\ . \notag
  \intertext{The ${\cal F}-1$ structure serves to regulate at $m=0$. To
    capture its effect it is expedient to make use of
    $\frac{e^{-m^2(t_1+t_2)}-1}{m^2} = -\int\limits_0^{t_1+t_2}\!\! d\alpha\
    e^{-m^2 \alpha}$ to decouple the $m^2$ and $t_i$ integrals }
  \label{eq:intermediate-integral}
  = & - \frac{1}{\Gamma^2(\frac{d}{2})} \left(\frac{\dl^2}{4}
    \frac{\dr^2}{4}\right)^\frac{d}{2} \int\limits_0^\infty dt_1\
  t_1^{-(d/2+1)}\ e^{-\frac{(\dl^2/4)}{t_1}} \int\limits_0^\infty dt_2\
  t_2^{-(d/2+1)}\ e^{-\frac{(\dr^2/4)}{t_2}} \notag \\ &\times
  \int\limits_0^{t_1+t_2}\!\! d\alpha\ \underbrace{\int_0^{\infty}dm^2
    \bigg(\frac{m^2}{\mu^2}\bigg)^{-u}
    e^{-m^2\alpha}}_{\left(\mu^2\right)^u\alpha^{u-1}\Gamma(1-u)}
\ .
  \intertext{The $m^2$ integral converges only if $\text{Re}(u) < 1$.  $u=1$
    will be the location of the first renormalon pole. At this point we may
    set $d=2$ since all expressions are now explicitly finite: }  = &
  - \left(\mu^2\right)^u \frac{\dl^2}{4} \frac{\dr^2}{4} \int\limits_0^\infty
  dt_1\ t_1^{-2}\ e^{-\frac{(\dl^2/4)}{t_1}} \int\limits_0^\infty dt_2\
  t_2^{-2}\ e^{-\frac{(\dr^2/4)}{t_2}} \ \frac{(t_1+t_2)^u}{u} \Gamma(1-u)
\ .
  \label{eq:B-m-int}
\end{align}
One might guess that the poles of $\Gamma(1-u)$ reflect the location of the
renormalon poles, but this is not true: once one restores the pre-factor to
recover $B$, $\Gamma(1-u)$ is actually canceled by $\sin(\pi u)$):
\begin{align}
  \label{eq:B-poles-via-int}
  B^T(u,\vdl m,\vdr m) = & \frac{e^{\frac53 u}\left(\mu^2\right)^u
  }{\Gamma(u+1)} \left(\frac{\dl^2}{4} \frac{\dr^2}{4}\right)^\frac{d}{2}
  \int\limits_0^\infty dt_1\ t_1^{-(d/2+1)}\ e^{-\frac{(\dl^2/4)}{t_1}}
  \int\limits_0^\infty dt_2\ t_2^{-(d/2+1)}\ e^{-\frac{(\dr^2/4)}{t_2}} \notag
  \\ & \times (t_1+t_2)^u
\end{align}
which would be completely regular if not the $t$ integrations expose poles at
positive $u$.

This can be uncovered by using a Mellin-Barnes representation for the
$(t_1+t_2)^u$ factor in this expression (see~\cite{Smirnov:2004ym} for a
textbook on applications of the Mellin-Barnes technique in field theory). We
write
\begin{equation}
  \label{eq:MBfort}
  (t_1+t_2)^u = \frac1{2\pi i}\int\limits_\gamma d\nu
  \frac{\Gamma(-u+\nu)\Gamma(-\nu)}{\Gamma(-u)} \frac{t_1^\nu}{t_2^{-u+\nu}}
\ ,
\end{equation}
where the path $\gamma$ connects $-\eta-i\infty$ to $\eta+i\infty$ ($\eta$
real and infinitesimal) in such a way that it separates {\em all} left from
{\em all} right poles of the integrand. [To recall the definition: poles
originating from factors $\Gamma(a+\nu)$ and $\Gamma(a-\nu)$ are called left
and right poles respectively. The left hand side is recovered as a series in
either $t_1/t_2$ or $t_2/t_1$ by summing residues. The sign of $\ln(t_1/t_2)$
determines which way to close the contour and in turn the form of the series
by selecting the relevant residues.]  This decouples the $t$ integrations
which provide additional $\nu$ dependent $\Gamma$ functions allows us to write
$B^T$ as
\begin{align}
  \label{eq:doublepoleexpr}
  B^T(u,\vdl \mu,\vdr \mu) = -e^{\frac53 u}
  \frac{(\dr/4)^u}{\Gamma(u+1)\Gamma(-u)} \frac1{2\pi
    i}\int\limits_{-\eta-i\infty}^{-\eta+i\infty} d\nu\
  \left(\frac{\dl^2}{\dr^2}\right)^\nu \frac{\Gamma^2(1-u+\nu)}{-u+\nu}
  \frac{\Gamma^2(1-\nu)}{\nu}
\ .
\end{align}
The integral can now be done by summing residues. We have to distinguish two
cases: the sign of $\ln(\dl^2/\dr^2)$ determines where to close the contour.
It proves useful to emphasize the symmetry of this procedure by expressing the
result via $r_<$, $r_>$ and $\xi^2=r_<^2/r_>^2$:
\begin{itemize}
\item {\bf $\ln(\dl^2/\dr^2) < 0$, closing to the right:}
  \begin{align}
    \label{eq:rightpoles-><}
    \eqref{eq:doublepoleexpr} \to &
    \frac{\sin(\pi u)}{\pi} \left(\frac{4\,e^{-\frac53}}{r_>^2
        \mu^2}\right)^{-u}\Bigg\{ u\Gamma^2(-u) - \sum\limits_{n=0}^\infty
    \frac{(n+1)\left(\xi^2\right)^{n+1}\Gamma^2(2+n-u)}{ (n+1-u)\Gamma^2(n+2)}
    \notag \\ & \times \left[ -\frac{(2+2n-u)}{(n+1)(n+1-u)}
      +\ln\left(\xi^2\right) -2(\psi(n+1)-\psi(n+2-u)) \right] \Bigg\}
  \end{align}
\item {\bf $\ln(\dl^2/\dr^2) > 0$, closing to the left:}
  \begin{align}
    \label{eq:leftpoles-><}
    \eqref{eq:doublepoleexpr} \to & \frac{\sin(\pi u)}{\pi} \
    \left(\frac{4\,e^{-\frac53}}{r_<^2 \mu^2}\right)^{-u} \Bigg\{
    u\Gamma^2(-u)\left(\xi^2\right)^{-u} - \sum\limits_{n=0}^\infty
    \frac{(n+1)\left(\xi^2\right)^{(n+1)-u} \Gamma^2(2+n-u)}{
      (n+1-u)\Gamma^2(n+2)} \notag \\ & \times \left[
      -\frac{(2+2n-u)}{(n+1)(n+1-u)} +\ln\left(\xi^2\right)
      -2(\psi(n+1)-\psi(n+2-u)) \right] \Bigg\}
  \end{align}
  This is identical to~\eqref{eq:rightpoles-><}.
\end{itemize}
Using shift identities for the digamma functions, $2\psi(n+1)
=\psi(n+1)+\psi(n+2)-\frac1{n+1}$ and $2\psi(n+2-u)
=\psi(n+1-u)+\psi(n+2-u)+\frac1{n+1-u}$, we arrive at our final result
\begin{align}
  \label{eq:doublepoleexpr-exp-2}
  B^T(u,\vdl \mu,\vdr \mu) = & \frac{\sin(\pi u)}{\pi}
  \left(\frac{4\,e^{\frac53}}{r_>^2\mu^2}\right)^u\Bigg\{ u\Gamma^2(-u) -
  \sum\limits_{n=0}^\infty \frac{\left(\xi^2\right)^{(n+1)}
    \Gamma(n+1-u)\Gamma(n+2-u)}{ \Gamma(n+1)\Gamma(n+2)} \notag \\ & \times
  \left[ \psi(n+2-u)+\psi(n+1-u)-\psi(n+1)-\psi(n+2) +\ln\left(\xi^2\right)
  \right] \Bigg\}
\ .
\end{align}

\section{$B^L$ in coordinate space}
\label{sec:FT-of-BL}

The calculation of the coordinate expression for $B^L$ is slightly simpler
than that of $B^T$ and uses the same techniques.

We start from~\eqref{eq:Bdef} for the longitudinal contribution and note that
its contribution to the $m=0$ subtraction term vanishes:
\begin{align}
  \label{eq:Long-borel-def} {\cal K}_{\bm x z y}& \ B^L(u,\vdl\mu,\vdr\mu) =
  -{\rm e}^{\frac53 u} \frac{\sin \pi u}{\pi}\,
  \int_0^{\infty}\frac{dm^2}{m^2} \bigg(\frac{m^2}{\mu^2}\bigg)^{-u} {\cal
    K}_{\bm{x z y}}^{L,m} \ .  \intertext{Abbreviating $f(u)=-{\rm e}^{\frac53
      u} \frac{\sin \pi u}{\pi}$ and using the parameter the appropriate
    parameter representation for the $K_0$ factors this turns into} = & \,
  \frac{f(u)}{(2\pi)^2}\int_0^{\infty}\frac{dm^2}{m^2}
  \bigg(\frac{m^2}{\mu^2}\bigg)^{-u} (\minus m^2) \int\limits_0^\infty dt_1 e^{-t_2
    m^2-\frac{(\dl^2/4)}{t_1}} t_1^{-d/2} \pi^{d/2} \int\limits_0^\infty dt_2
  e^{-t_2 m^2-\frac{(\dl^2/4)}{t_2}} t_2^{-d/2} \pi^{d/2} \notag
  \intertext{
    \begin{minipage}{1.0\linewidth}
For the remainder of the caclulation we present only minimal comments.\\[1mm]
--separate off the $m$ integral--
    \end{minipage}
} = & \, \minus \frac{f(u)}{(2\pi)^2}
  \pi^d\int\limits_0^\infty dt_1 \int\limits_0^\infty dt_2 \
  e^{-\frac{(\dl^2/4)}{t_1}} t_1^{-d/2} e^{-\frac{(\dl^2/4)}{t_2}} t_2^{-d/2}
  \int_0^{\infty}\frac{dm^2}{m^2} \bigg(\frac{m^2}{\mu^2}\bigg)^{-u} m^2\
  e^{-(t_1+t_2) m^2} \notag \intertext{--the $m$ integral requires
    $\text{Re}(u) <1$--} = & \,
 \minus \frac{f(u)}{(2\pi)^2} (\mu^2)^u \pi^d\ \int\limits_0^\infty dt_1
  \int\limits_0^\infty dt_2 \ e^{-\frac{(\dl^2/4)}{t_1}} t_1^{-d/2}
  e^{-\frac{(\dl^2/4)}{t_2}} t_2^{-d/2} (t_1+t_2)^{u-1} \Gamma(1-u)
  \intertext{--use MB to factor $(t_1+t_2)^{u-1}$--} = & \,
 \minus \frac{f(u)}{(2\pi)^2} (\mu^2)^u \Gamma(1-u) \pi^d\ \int\limits_0^\infty dt_1
  \int\limits_0^\infty dt_2 \ e^{-\frac{(\dl^2/4)}{t_1}} t_1^{-d/2}
  e^{-\frac{(\dl^2/4)}{t_2}} t_2^{-d/2} \notag \\ & \times \frac1{2\pi
    i}\int\limits_\gamma d\nu
  \frac{\Gamma(-(u-1)+\nu)\Gamma(-\nu)}{\Gamma(-(u-1))}
  \frac{t_1^\nu}{t_2^{-(u-1)+\nu}} \intertext{--separate the $t$ integrals and
    perform them ($\text{Re}(-d/2+\nu) <-1$ and $\text{Re}(-d/2+u-1)-\nu <-1$)
    --} = & \, \minus \frac{f(u)}{(2\pi)^2} (\mu^2)^u \pi^d\ \frac1{2\pi
    i}\int\limits_\gamma d\nu \Gamma(-(u-1)+\nu)\Gamma(-\nu) \notag \\ &
  \times (\dl^2/4)^{1-\frac{d}{2}+\nu}\Gamma(\frac{d}{2}-1-\nu) \
  (\dr^2/4)^{-\frac{d}{2}+u-\nu}\Gamma(\frac{d}{2}-u+\nu)
  \intertext{--rearrange to expose convergence of the MB integral--} = & \,
  \minus \frac{f(u)}{(2\pi)^2} (\mu^2\ \dr^2/4)^u (\dl^2/4)^1 \pi^d\
  \left(\frac{\dl^2}{4}\frac{\dr^2}{4}\right)^{-\frac{d}{2}} \notag \\ &
  \times \frac1{2\pi i}\int\limits_\gamma d\nu\
  \Gamma(-(u-1)+\nu)\Gamma(\frac{d}{2}-u+\nu)
  \Gamma(-\nu)\Gamma(\frac{d}{2}-1-\nu)
  \left(\frac{\dr^2}{\dl^2}\right)^{-\nu} \intertext{--evaluate MB
    independently for both cases $\dr>\dl$ and $\dl>\dr$, find sums of
    residues to agree, set $d=2$ to obtain the final result--} = &
 \minus \frac{\sin(\pi u)}{\pi} \frac1{r_>^2}
  \left(\frac{4\, e^{-\frac53}}{\mu^2
      r_>^2}\right)^{-u} \sum\limits_{n=0}^\infty
  \left(\frac{\Gamma(n+1-u)}{\Gamma(n+1)}\right)^2 \left(\xi^2\right)^n \left(
    \ln(\xi^2) -2\psi(n+1)+2\psi(n+1-u) \right)
  \label{eq:KLBL-coord-sum}
\end{align}

\section{Pole and renormalon structure of $B^T$}
\label{sec:poles-BT}

To expose the pole structure of $B^T$ we find it easiest to start
from~\eqref{eq:BTb} and rewrite it as
\begin{align}
  \label{eq:transverse-1-xi2-poles}
  B^T(u,\vdl\mu,\vdr\mu) =& \left(\frac{4\,e^{-\frac53}}{r_>^2
      \mu^2}\right)^{-u} \frac{\sin(\pi u)}{\pi}\,
  \frac{u(1-u)\Big(\Gamma(1-u)\Gamma(-u)\Big)^2}{\Gamma(2(1-u))} \ \
  \qFp{2}{1}{1-u,-u}{2(1-u)}{1-\xi^2} \notag \\ = & -\sum\limits_{n=0}^\infty
  \frac{(1-\xi^2)^n}{\Gamma(n+1)} \left(\frac{4\,e^{-\frac53}}{r_>^2
      \mu^2}\right)^{-u} \frac{1}{\Gamma(u)}
  \frac{\Gamma(1-u+n)}{\Gamma(2(1-u)+n)} \Gamma(2-u)\Gamma(-u+n)
\end{align}
Here all is regular but the factor $\Gamma(2-u)\Gamma(n-u)$ which will cause
at most double poles at positive integers $u=m=1,2,\ldots$. In the following
we will encounter Laurent expansions around given pole locations and use the
abbreviation $\epsilon=u-m$.

\begin{itemize}
\item Double poles arise via combined divergence in
  $\Gamma(2-(m+\epsilon))\Gamma(n-(m+\epsilon))$. For a fixed pole location
  $u=m=2,3,\ldots,\infty$ the second factor appears to contribute as long as
  $n\le m$. The Laurent expansion around $u=m$ in these cases can generically
  be written as
  \begin{align}
    \label{eq:transv-gendoublepole}
    & 2(-1)^{m-n} \left(\frac{4\,e^{-\frac53}}{r_>^2 \mu^2}\right)^{-m}
    \frac{(1-\xi^2)^n}{\Gamma(n+1)} \frac{\Gamma(2m-(n+1))}{
      \Gamma(m)\Gamma(m-1)\Gamma(m-n)\Gamma(m+1-n)}
    \Bigg\{\frac{1}{\epsilon^2}
\notag \\ &
+\Bigg[-\ln\left(\frac{4\,e^{-\frac53}}{r_>^2
        \mu^2}\right)
    -\Psi(m-1)-\Psi(m)-\Psi(m-n)-\Psi(m+1-n)+2\Psi(2m-(n+1)) \Bigg]
    \frac{1}{\epsilon}
\notag \\ & +{\cal O}(\epsilon^0) \Bigg\}
  \end{align}
  We note that coefficient of the double pole also vanishes for $n=m$ so that
  the sum of the double pole terms may be expressed as
  \begin{align}
    \label{eq:BT-double-pole-app}
    B^T_{\text{double pole}} = &\, \sum\limits_{m=2}^\infty
    \sum\limits_{n=2}^{m-1} \frac{(1-\xi^2)^n}{\Gamma(n+1)}
\notag \\ & \times
 \left(\frac{4\,e^{-\frac53}}{r_>^2
        \mu^2}\right)^{-m}
    \frac{\Gamma(2m-(n+1))}{ \Gamma(m)\Gamma(m-1)\Gamma(m-n)\Gamma(m+1-n)}
    \frac{ 2(-1)^{m-n}}{(u-m)^2}
  \end{align}

\item As is obvious from the Laurent expansion
  in~\eqref{eq:transv-gendoublepole} we inherit single poles at the double
  pole locations. Additional single poles arise when
  \begin{itemize}
  \item only the first of the factors contributes a pole, i.e. for $n > m\ge
    2$, (the coefficients here vanish, however, where $n\le 2(m-1)$)
  \item or where only the second of the factors does contribute, i.e. for $2 >
    m \ge n$ (this is limited to $m=1, n=0,1$):
  \end{itemize}
  The Laurent expansion around any of these additional single poles is given
  by
  \begin{align}
    \label{eq:transv-gensinglepole}
    (-1)^m\left(\frac{4\,e^{-\frac53}}{r_>^2 \mu^2}\right)^{-m} (1-\xi^2)^n
    \frac{\Gamma(1-m+n)}{\Gamma(n+1)\Gamma(n+2-2m)\Gamma(m-1)}
    \frac{1}{\epsilon} +{\cal O}(\epsilon^0)
  \end{align}

  Adding all these leads to the following sum of single pole contributions
  which mark the power correction to to $B^T$
  \begin{align}
    \label{eq:BT-single-pole-app}
    B^T_{\text{single pole}} = & \left(\frac{4\,e^{-\frac53}}{r_>^2
        \mu^2}\right)^{-1}\frac{-2+(1-\xi^2)}{u-1} \notag \\ &
    +\sum\limits_{m=2}^\infty \sum\limits_{n=2(m-1)}^\infty
    \frac{(-1)^m}{u-m}\frac{(1-\xi^2)^n}{\Gamma(n+1)}
    \left(\frac{4\,e^{-\frac53}}{r_>^2 \mu^2}\right)^{-m}
    \frac{\Gamma(1-m+n)}{\Gamma(n-2(m-1))\Gamma(m-1)} \notag \\ & +
    \sum\limits_{m=2}^\infty \sum\limits_{n=2}^{m-1}
    \frac{(1-\xi^2)^n}{\Gamma(n+1)} \frac{ 2(-1)^{m-n}}{(u-m)} \notag \\ &
    \times \frac{d}{d m}\left\{ \left(\frac{4\,e^{-\frac53}}{r_>^2
          \mu^2}\right)^{-m} \frac{\Gamma(2m-(n+1))}{
        \Gamma(m)\Gamma(m-1)\Gamma(m-n)\Gamma(m+1-n)} \right\}
  \end{align}
  As a power correction each of the terms at fixed $m$ comes with an a priori
  unknown coefficient.
\end{itemize}

\section{Pole and renormalon structure of $B^L$}
\label{sec:poles-BL}

\begin{align}
  \label{eq:singlepole-1-xi-exp}
  {\cal K}_{\bm{x z y}} & B^L(u,\vdl\mu,\vdr\mu) = \plus \frac{\sin(\pi u)}{\pi}
  \frac{1}{{\bm r}_>^2} \left( \frac{4\,e^{-\frac53}}{{\bm r}_>^2\mu^2}
  \right)^{-u} \frac{\left(\Gamma(1-u)\right)^4}{\Gamma(2(1-u))}\ \
  \qFp{2}{1}{1-u,1-u}{2(1-u)}{1-\xi^2} \notag \\ = & \plus \sum_{n=0}^\infty
  \frac{(1-\xi^2)^n}{\Gamma(n+1)} \frac{1}{{\bm r}_>^2} \left(
    \frac{4\,e^{-\frac53}}{{\bm r}_>^2\mu^2} \right)^{-u}
  \frac{1}{\Gamma(u)}\frac{\Gamma(1+n-u)}{\Gamma(2(1-u)+n)} \
  \Gamma(1-u)\Gamma(1-u+n)
\end{align}
Renormalon poles again appear at $u=m=1,2,3,\ldots$, with everything regular
but the factor $\Gamma(1-u)\Gamma(1-u+n)$.
\begin{itemize}
\item Double poles at $m$ arise where both
  $\Gamma(1-(m+\epsilon))\Gamma(n+1-(m+\epsilon))$ diverge, i.e. for $1\le m
  \ge n+1$. The Laurent expansion around these pole locations $u=m$ is
  \begin{align}
    2 & (-1)^{m-n} \frac{1}{r_>^2} \left( \frac{4\,e^{-\frac53}}{{\bm
          r}_>^2\mu^2} \right)^{-m} \frac{(1-\xi^2)^n}{\Gamma(n+1)}
    \frac{\Gamma(2m-(n+1))}{(\Gamma(m))^2(\Gamma(m-n))^2} \notag \\ & \times
    \Bigg\{ \frac{1}{\epsilon^2} +\Bigg[ -\ln\left(
      \frac{4\,e^{-\frac53}}{{\bm r}_>^2\mu^2} \right)
    -2\Psi(m-1)-2\Psi(2(m-1))+2\Psi(m) \Bigg]\frac{1}{\epsilon} +{\cal
      O}(\epsilon^0) \Bigg\}
  \end{align}
  so that the double pole part of the longitudinal contribution may be written
  as
  \begin{align}
    \label{BL-double-pole-part}
    {\cal K}_{\bm{x z y}} B^L_{\text{double pole}} = \minus \sum\limits_{m=1}^\infty
    \sum\limits_{n=0}^{m-1} \frac{2 (-1)^{m-n}}{(u-m)^2}
    \frac{(1-\xi^2)^n}{\Gamma(n+1)} \frac{1}{r_>^2} \left(
      \frac{4\,e^{-\frac53}}{{\bm r}_>^2\mu^2} \right)^{-m}
    \frac{\Gamma(2m-(n+1))}{(\Gamma(m))^2(\Gamma(m-n))^2}
  \end{align}
\item As for $B^T$, we inherit single poles at the double pole locations.
  Additional single poles arise when only the first factor in
  $\Gamma(1-(m+\epsilon))\Gamma(n+1-(m+\epsilon))$ diverges, i.e. for $1\le m
  <n+1$. The Laurent expansion around these new pole locations reads
  \begin{align}
    \label{eq:singlepole-1-xi-res}
    \plus (-1)^m \frac{1}{r_>^2} \left( \frac{4\,e^{-\frac53}}{{\bm r}_>^2\mu^2}
    \right)^{-m} (1-\xi^2)^n
    \frac{(\Gamma(n+1-m))^2}{\Gamma(n+1)(\Gamma(m))^2\Gamma(n+2-2m)}
    \frac{1}{\epsilon} +{\cal O}(\epsilon^0)
  \end{align}
  Some of the residues vanish (where $2m > n+2$).

  Combining the two contributions leads to an expression for the single pole
  part of $B^L$ of the form
  \begin{align}
    \label{BL-single-pole-part-app}
    {\cal K}_{\bm{x z y}} & B^L_{\text{single pole}} =
    \minus \sum\limits_{m=1}^\infty \sum_{n=m}^\infty
    \frac{-(-1)^m}{u-m}\frac{1}{r_>^2} \left( \frac{4\,e^{-\frac53}}{{\bm
          r}_>^2\mu^2} \right)^{-m} \frac{(1-\xi^2)^n}{\Gamma(n+1)}
    \frac{(\Gamma(n+1-m))^2}{(\Gamma(m))^2\Gamma(n+2-2m)} \notag \\ &
\minus    \sum\limits_{m=1}^\infty \sum\limits_{n=0}^{m-1} \frac{2 (-1)^{m-n}}{u-m}
    \frac{(1-\xi^2)^n}{\Gamma(n+1)} \frac{1}{r_>^2} \frac{d}{d m}\left\{\left(
        \frac{4\,e^{-\frac53}}{{\bm r}_>^2\mu^2} \right)^{-m}
      \frac{\Gamma(2m-(n+1))}{(\Gamma(m))^2(\Gamma(m-n))^2} \right\}
  \end{align}

\end{itemize}

\section{Principal value definition of the perturbative sum}
\label{sec:PV-def-pert-sum}

The small $\xi^2$ series differ strongly from the $1-\xi^2$ series so that we
will deal with them separately.

\begin{itemize}
\item Closer inspection of~(\ref{eq:K-replacement}) via~\eqref{eq:BTa}
  and~(\ref{eq:BLb}) reveals, that the double poles in the obstructing terms
  have their origin in a $u$ derivative of a $\Gamma$ function, i.e. they
  arise from differentiating simple poles. To make this manifest we
  rewrite~\eqref{eq:BTa} and~\eqref{eq:BLa} as
  \begin{subequations}
    \begin{align}
      \label{eq:double-poles-as-deriv}
      B^T(u,\vdl \mu,\vdr \mu) = & \left(\frac{r_>^2 \mu^2 {\rm
            e}^{\frac53}}{4}\right)^{u}\,
      \bigg\{\frac{\Gamma(1-u)}{\Gamma(1+u)}
      \,\nonumber\\
      &+ \sum_{n=1}^{\infty} (-1)^{n+1}\ (\xi^2)^{n}\,
      \frac{\Gamma(1-u+n)}{n\Gamma^2(n)\ \Gamma(1+u-n)}\,
      \left(\Psi(1+n)+\Psi(n)-\ln \xi^2\right) \nonumber\\ & +
      \sum_{n=1}^{\infty} (-1)^{n+1}\ (\xi^2)^{n}\, \frac{
        2\frac{d}{du}\Gamma(1-u+n) +\Gamma(n-u)}{n\Gamma^2(n)\ \Gamma(1+u-n)}
      \bigg\}
      \\
      {\cal K}_{\bm{x z y}} B^L(u,\vdl \mu,\vdr \mu) = &
 \minus     \frac1{r_>^2}\left(\frac{\mu^2 r_>^2 e^{\frac53}}{4}\right)^u
      \sum\limits_{n=0}^\infty \frac{\left(\xi^2\right)^n}{(-1)^n
        \Gamma(u-n)\Big(\Gamma(n+1)\Big)^2} \notag \\ & \times
      \left[\Gamma(n+1-u)\left( \ln(\xi^2) +2\psi(n+1)\right)+2\frac{d}{d
          u}\Gamma(n+1-u) \right]
    \end{align}
  \end{subequations}
  The double pole contributions arise from factor $\frac{d}{du}\Gamma(1-u+n)$
  and as such they can be converted into single poles under the Borel integral
  by a partial integration in the expression for $R$,
  Eq.~\eqref{eq:K-replacement}.  For this, we note that boundary terms of the
  partial integration in~\eqref{eq:K-replacement} vanish to display the
  results separately for $R^T$ and $R^L$ as
  \begin{subequations}
    \label{eq:R-single-pole-version}
    \begin{align}
      \label{eq:RT-single-pole-version}
      R^T(\vdl &\Lambda,\vdr \Lambda,) = \frac{1}{\beta_0}
      \int\limits_0^{\infty}du \ T(u)
      \left(\frac{\mu^2}{\Lambda^2}\right)^{-u} B^T(u,\vdl \mu,\vdr \mu) \notag
      \displaybreak[0]\\ = & \frac{1}{\beta_0} \int\limits_0^{\infty}du \ T(u)
      \left(\frac{4\,e^{-\frac53}}{r_>^2\Lambda^2}\right)^{-u} \,
      \Bigg\{\frac{\Gamma(1-u)}{\Gamma(1+u)} \,\notag\displaybreak[0]\\ & +
      \sum_{k=1}^{\infty} (-1)^{k+1}\
      \frac{\left(\xi^2\right)^k}{k\Gamma^2(k)}\, \Bigg[
      \frac{\Gamma(1-u+k)}{\Gamma(1+u-k)}\, \left(\Psi(1+k)+\Psi(k)-\ln
        \xi^2\right) + \frac{ \Gamma(k-u)}{\Gamma(1+u-k)} \notag
      \displaybreak[0]\\ & \hspace{.5cm}+
      \frac{2\Gamma(1-u+k)}{\Gamma(1+u-k)}\left( \ln\left(\frac{4\,
            e^{-\frac53}}{r_>^2 \Lambda^2 }\right)
        -\frac{T'(u)}{T(u)}+\Psi(1+u-k)\right) \Bigg] \Bigg\} \intertext{and}
        \label{eq:RL-single-pole-version}
      {\cal K}_{\bm{x z y}} R^L(\vdl &\Lambda,\vdr \Lambda,) = \frac{1}{\beta_0}
      \int\limits_0^{\infty}du \ T(u)
      \left(\frac{\mu^2}{\Lambda^2}\right)^{-u} {\cal K}_{\bm{x z y}}
      B^L(u,\vdl \mu,\vdr \mu) \notag \\ = & \minus \frac{1}{\beta_0}
      \int\limits_0^{\infty}du \ T(u) \frac1{r_>^2} \left(\frac{4\,
          e^{-\frac53}}{r_>^2 \Lambda^2 }\right)^{-u} \sum\limits_{n=0}^\infty
      \frac{\Gamma(n+1-u) \left(\xi^2\right)^n}{ (-1)^n
        \Gamma(u-n)\Big(\Gamma(n+1)\Big)^2} \notag \\ & \hspace{1cm}\times
      \left[ \ln(\xi^2) -2\psi(n+1)+2\left( \ln\left(\frac{4\,
              e^{-\frac53}}{r_>^2 \Lambda^2 }\right)
          -\frac{T'(u)}{T(u)}-\psi(u-n) \right) \right] \ .
    \end{align}
  \end{subequations}

\item The single pole expressions given above converge only very slowly near
  $\xi=1$.  To circumvent this we also present single pole expressions based
  on~\eqref{eq:BTb} and~\eqref{eq:BLb}. For the transverse contributions we
  write
  \begin{subequations}
    \label{eq:RT-single-pole-version-2}
    \begin{align}
      \label{eq:RT-single-pole-version-2a}
      R^T(\vdl &\Lambda,\vdr \Lambda,) = \frac{1}{\beta_0}
      \int\limits_0^{\infty}du \ T(u)
      \left(\frac{\mu^2}{\Lambda^2}\right)^{-u} B^T(u,\vdl \mu,\vdr \mu)
      \notag \\
      = & \frac{1}{\beta_0} \int\limits_0^{\infty}du \ T(u)
      \left\{\left(\frac{\mu^2}{\Lambda^2}\right)^{-u} B^T(u,\vdl \mu,\vdr
        \mu)-\left[\left(\frac{\mu^2}{\Lambda^2}\right)^{-u} B^T(u,\vdl \mu,\vdr
          \mu) \right]_{\text{double pole}} \right\} \notag \\ &
      +\frac{1}{\beta_0} \int\limits_0^{\infty}du \ T(u)
      \left[\left(\frac{\mu^2}{\Lambda^2}\right)^{-u} B^T(u,\vdl \mu,\vdr \mu)
      \right]_{\text{double pole}}
    \end{align}
    where $B^T$ is taken from~\eqref{eq:BTb} and the double pole part of the
    integrand is derived in Appendix~\ref{sec:poles-BT}. It is given by
    \begin{align}
      \label{eq:BT-double-pole-part}
      \left[\left(\frac{\mu^2}{\Lambda^2}\right)^{-u} B^T(u,\vdl \mu,\vdr \mu)
      \right]_{\text{double pole}}= & \sum\limits_{m=2}^\infty
      \sum\limits_{n=2}^{m-1} \left(\frac{4\,e^{-\frac53}}{r_>^2
          \Lambda^2}\right)^{-m} \frac{(1-\xi^2)^n}{\Gamma(n+1)} \frac{
        2(-1)^{m-n}}{(u-m)^2} \notag \\ & \times \frac{\Gamma(2m-(n+1))}{
        \Gamma(m)\Gamma(m-1)\Gamma(m-n)\Gamma(m+1-n)}
    \end{align}
    The first term can now be integrated with a principal value prescription
    (it only contains single poles) while (for one loop running) the last term
    is given a meaning via integration by parts and gives only a boundary
    contribution. To this end we recall that for one loop running we may set
    $T(u)\to 1$, $T'(u)\to 0$ and identify $\ln\left(\mu^2/\Lambda^2\right) =
    \frac{\pi}{\beta_0\alpha_s(\mu^2)}$:
    \begin{align}
      \label{eq:BT-one-loop-boundary}
      \frac{1}{\beta_0} \int\limits_0^{\infty}du & \ T(u)
      \left[\left(\frac{\mu^2}{\Lambda^2}\right)^{-u} B^T(u,\vdl \mu,\vdr \mu)
      \right]_{\text{double pole}} \\ \notag \xrightarrow{T(u)\to 1} &
      -\sum\limits_{m=2}^\infty \sum\limits_{n=2}^{m-1}
      \left(\frac{4\,e^{-\frac53}}{r_>^2 \Lambda^2}\right)^{-m}
      \frac{(1-\xi^2)^n}{\Gamma(n+1)} \frac{ 2(-1)^{m-n}}{m}
      \frac{\Gamma(2m-(n+1))}{ \Gamma(m)\Gamma(m-1)\Gamma(m-n)\Gamma(m+1-n)}
    \end{align}
  \end{subequations}
  We have checked that this result matches with a very costly evaluation of
  $R^T$ via~\eqref{eq:R-single-pole-version}. $R^L$ can be treated analogously

  \begin{subequations}
    \label{eq:RL-single-pole-version-2}
    \begin{align}
      \label{eq:RL-single-pole-version-2a}
      R^L(\vdl &\Lambda,\vdr \Lambda,) = \frac{1}{\beta_0}
      \int\limits_0^{\infty}du \ T(u)
      \left(\frac{\mu^2}{\Lambda^2}\right)^{-u} B^L(u,\vdl \mu,\vdr \mu)
      \notag \\
      = & \frac{1}{\beta_0} \int\limits_0^{\infty}du \ T(u)
      \left\{\left(\frac{\mu^2}{\Lambda^2}\right)^{-u} B^L(u,\vdl \mu,\vdr
        \mu)-\left[\left(\frac{\mu^2}{\Lambda^2}\right)^{-u} B^L(u,\vdl \mu,\vdr
          \mu) \right]_{\text{double pole}} \right\} \notag \\ &
      +\frac{1}{\beta_0} \int\limits_0^{\infty}du \ T(u)
      \left[\left(\frac{\mu^2}{\Lambda^2}\right)^{-u} B^L(u,\vdl \mu,\vdr \mu)
      \right]_{\text{double pole}} \ ,
    \end{align}
    where now (see App.~\ref{sec:poles-BL})
    \begin{align}
      \label{eq:BL-double-pole-part}
      {\cal K}_{\bm{x z y}}\left[\left(\frac{\mu^2}{\Lambda^2}\right)^{-u}
        B^L(u,\vdl \mu,\vdr \mu) \right]_{\text{double pole}}= &
  \minus    \sum\limits_{m=1}^\infty \sum\limits_{n=0}^{m-1} \frac{2
        (-1)^{m-n}}{(u-m)^2} \frac{(1-\xi^2)^n}{\Gamma(n+1)} \frac{1}{r_>^2}
      \left( \frac{4\,e^{-\frac53}}{{\bm r}_>^2\Lambda^2} \right)^{-m} \notag
      \\ & \times \frac{\Gamma(2m-(n+1))}{(\Gamma(m))^2(\Gamma(m-n))^2}
    \end{align}
    and to one loop accuracy
    \begin{align}
      \label{eq:BL-one-loop-boundary}
      {\cal K}_{\bm{x z y}}& \frac{1}{\beta_0} \int\limits_0^{\infty}du \ T(u)
      \left[\left(\frac{\mu^2}{\Lambda^2}\right)^{-u} B^L(u,\vdl \mu,\vdr \mu)
      \right]_{\text{double pole}} \notag \\ & \xrightarrow{T(u)\to 0} \plus
      \sum\limits_{m=1}^\infty \sum\limits_{n=0}^{m-1} \frac{2 (-1)^{m-n}}{m}
      \frac{(1-\xi^2)^n}{\Gamma(n+1)} \frac{1}{r_>^2}
      \left( \frac{4\,e^{-\frac53}}{{\bm r}_>^2\Lambda^2} \right)^{-m}
      \frac{\Gamma(2m-(n+1))}{(\Gamma(m))^2(\Gamma(m-n))^2} \ .
    \end{align}
  \end{subequations}

\end{itemize}

The expressions on the r.h.s. of all the Borel integrals shown in
this section are in general well defined with any arbitrary contour
connecting $0$ to $\infty$ in the complex plane that avoids the
poles of the integrand on the real axis. All such expressions share
the same perturbative expansion (be they partially integrated or
not!). The uncertainty introduced by the this freedom of choice in
the contour is qualitatively determined by the size of residues of
these poles. An important consistency condition for using the
partially integrated version to study non-perturbative corrections
is therefore that the residues of the integrand remain the same as
in the original version~(\ref{eq:K-replacement}).  This is a generic
consequence of the fact that $\Gamma(2-u+k)$ (the function
contributing the poles) has no logarithmic terms in its Laurent
expansions.  Appendices~\ref{sec:poles-BT} and~\ref{sec:poles-BL}
contain an explicit discussion of the pole part of $B^T$ and $B^L$.


\providecommand{\href}[2]{#2}\begingroup\raggedright\endgroup

\end{document}